\documentclass[journal=jacsat,manuscript=article]{achemso}
\SectionsOn
\SectionNumbersOn
\usepackage{chemformula} %
\usepackage[T1]{fontenc} %
\usepackage[utf8]{inputenc}
\usepackage{textcomp}
\usepackage{amsmath,amssymb}
\usepackage{algorithm}
\usepackage{algpseudocode}

\DeclareUnicodeCharacter{2005}{}
\usepackage[utf8]{inputenc}
\DeclareUnicodeCharacter{03BC}{$\mu$}
\DeclareUnicodeCharacter{03C3}{$\sigma$}
\DeclareUnicodeCharacter{03B2}{$\beta$}
\usepackage{xr}
\usepackage{booktabs}
\usepackage{multirow}
\usepackage{comment}

\newcommand{\name}{MolPAL}

\newcommand{\errorbarsentence}[2]{Error bars {#1} and shaded regions {#2} denote $\pm$ one standard deviation across five runs. }

\newcommand{\topk}[1]{top ${\sim}{#1}$\%}
\newcommand{\capstopk}[1]{Top ${\sim}{#1}$\%}

\usepackage{xr}
\makeatletter

\newcommand*{\addFileDependency}[1]{%
\typeout{(#1)}%
\@addtofilelist{#1}
\IfFileExists{#1}{}{\typeout{No file #1.}}
}\makeatother

\addFileDependency{b_appendix_v1.tex}%
\addFileDependency{b_appendix_v1.aux}%

\author{Jenna C. Fromer}
\affiliation[Unknown University]
{Department of Chemical Engineering, MIT, Cambridge, MA 02139}
\author{David E. Graff}
\affiliation[Unknown University]
{Department of Chemical Engineering, MIT, Cambridge, MA 02139}
\alsoaffiliation{Department of Chemistry and Chemical Biology, Harvard University, Cambridge, MA 02138}
\author{Connor W. Coley}
\affiliation[Unknown University]
{Department of Chemical Engineering, MIT, Cambridge, MA 02139}
\alsoaffiliation[Second University]
{Department of Electrical Engineering and Computer Science, MIT, Cambridge, MA
02139}

\email{ccoley@mit.edu}
\title[]
  {Pareto Optimization to Accelerate Multi-Objective Virtual Screening}

\abbreviations{}

\begin{document}

\begin{abstract}
The discovery of therapeutic molecules is fundamentally a multi-objective optimization problem. One formulation of the problem is to identify molecules that simultaneously exhibit strong binding affinity for a target protein, minimal off-target interactions, and suitable pharmacokinetic properties. Inspired by prior work that uses active learning to accelerate the identification of strong binders, we implement multi-objective Bayesian optimization to reduce the computational cost of multi-property virtual screening and apply it to the identification of ligands predicted to be selective based on docking scores to on- and off-targets. We demonstrate the superiority of Pareto optimization over scalarization across three case studies. Further, we use the developed optimization tool to search a virtual library of over 4M molecules for those predicted to be selective dual inhibitors of EGFR and IGF1R, acquiring 100\% of the molecules that form the library's Pareto front after exploring only 8\% of the library. This workflow and associated open source software can reduce the screening burden of molecular design projects and is complementary to research aiming to improve the accuracy of binding predictions and other molecular properties. 
\end{abstract} 

\section{Introduction}

Molecular discovery aims to identify molecules that balance multiple, often competing, properties. The need to simultaneously optimize multiple properties is especially notable in drug discovery workflows. Small molecule drugs operating through direct single-target binding interactions must exhibit not only strong binding affinity for the target protein but also minimal off-target interactions and suitable pharmacokinetic properties \cite{hughes_principles_2011, kettle_standing_2016, beckers_25_2022}.
One formulation of small molecule drug discovery is to identify compounds that bind strongly to a protein of interest and subsequently modify them to fulfill remaining property constraints. %
\cite{keseru_hit_2006, hughes_principles_2011, sun_why_2022}. A candidate molecule with high activity but a poor pharmacokinetic profile may ultimately be abandoned, resulting in wasted time and resources \cite{segall_addressing_2014, beckers_25_2022}. %

Selectivity is one property that is often considered only after a hit with promising primary activity is identified \cite{hughes_principles_2011}. Selectivity may be measured with binding assays against off-targets that are structurally similar to the primary target or known to be associated with adverse side effects (e.g., cytochromes P450 and the hERG channel) \cite{van_vleet_screening_2019, beckers_25_2022}. %
Non-specific ligands that bind to many proteins in addition to the target may require additional optimization steps when compared to their selective counterparts \cite{bleicher_hit_2003, kettle_standing_2016}. Consideration of promiscuity early in a drug discovery project may aid in deprioritizing chemical series that are inherently nonselective  \cite{recanatini_silico_2004, macchiarulo_ligand_2004}. Kinases are protein targets for which binding selectivity is particularly relevant; by screening 367 small-molecule ATP-competitive kinase inhibitors against 224 recombinant kinases, \citet{elkins_comprehensive_2016} demonstrated the prevalence of unexpected cross-reactivity in identified hits: 39 of the tested compounds displayed >50\% inhibition of at least 10 of the screened kinases. In other settings, binding interactions with multiple targets can be advantageous \cite{raghavendra_dual_2018}. Therapeutics for Alzheimer's disease \cite{ibrahim_multitarget_2019, benek_perspective_2020} and thyroid cancer \cite{brassard_role_2012, okamoto_distinct_2015} have exhibited improved efficacy through affinity for multiple protein targets. Optimizing affinity to multiple targets is another goal that can be brought into earlier stages of hit discovery \cite{ma_-silico_2010, yousuf_structure-based_2017}.

Anticipating protein-ligand interactions that contribute to potency and selectivity is possible, albeit imperfectly, with structure-based drug design techniques that employ scoring functions to estimate energetic favorability. Docking to off-targets has been applied to improve the selectivity profiles of identified compounds \cite{chahal_combination_2023, schieferdecker_development_2023, matricon_structure-based_2023}. These demonstrations have revealed that falsely predicted non-binders may incorrectly be categorized as selective because scoring functions designed for hit-finding typically aim to minimize the false positive rate (i.e., weak binders predicted to bind strongly), not the false negative rate (i.e., binders predicted not to bind) \cite{weiss_selectivity_2018, matricon_structure-based_2023}. Although structure-based methods like docking are limited in their predictive accuracy \cite{chen_evaluating_2006, jain_bias_2008, cross_comparison_2009, irwin_docking_2016, boittier_assessing_2020, stanzione_chapter_2021}, docking-based virtual screens can still effectively enrich a virtual library for molecules that are more likely to exhibit target activity \cite{ling_use_2008, bajusz_discovery_2016, lyu_ultra-large_2019, gentile_automated_2021, alon_structures_2021}. 

The computational cost of virtual screening \cite{tingle_large-scale_2023} has motivated the development of model-guided optimization methods that reduce the total number of docking calculations required to recover top-performing molecules \cite{garnett_introducing_2015, smith_less_2018, gentile_deep_2020, graff_accelerating_2021, yang_efficient_2021,mehta_memes_2021, graff_self-focusing_2022}. As one example, \citet{graff_accelerating_2021} only require the docking scores of 2.4\% of a 100M member virtual library to identify over 90\% of the library's top-50,000 ligands. Similar principles apply when using more expensive evaluations such as relative binding free energy calculations \cite{thompson_optimizing_2022}, where a reduction of computational cost can be particularly beneficial. These methods are designed to optimize a single property and are inherently \emph{single-objective} optimizations.

The need for model-guided optimization methods in virtual screening is heightened when multiple properties are screened. The resources required for a multi-objective virtual screen scale linearly with the number of screened properties (``objectives'') and library size. In some settings, exhaustive screens of large virtual libraries (millions to billions) may be infeasible. Model-guided multi-objective optimization has the potential to reduce the computational cost of a multi-objective virtual screen without sacrificing performance. \citet{mehta_memes_2021} have previously applied %
multi-objective optimization with a scalarized acquisition function to identify molecules that simultaneously optimize the docking score to Tau Tubulin Kinase 1, calculated octanol-water partition coefficient (cLogP \cite{wildman_prediction_1999}), and synthetic accessibility score (SA\_Score \cite{ertl_estimation_2009}). Over 90\% of the most desirable molecules were recovered after scoring of only 6\% of the virtual library.  Multi-objective virtual screens involve multiple design choices primarily related to the acquisition strategy, which have not yet been compared in the context of virtual screening. Further, existing methods for multi-objective virtual screening do not implement Pareto optimization. Pareto optimization aims to identify the molecules that form or are close to the \textit{Pareto front}, for which an improvement in one objective necessitates a detriment to another. Molecules that form the Pareto front optimally balance multiple desired properties, illustrate which combinations of objective values are possible, and reveal trade-offs in the objective space; this is not possible with scalarization. %

In this work, we extend the molecular pool-based active learning tool MolPAL \cite{graff_accelerating_2021} to this setting of multi-objective virtual screening.  %
\name{} is publicly available, open source, and can be adopted for multi-objective virtual screening with any desired set of objective functions, including those beyond structure-based drug design. We demonstrate through three retrospective case studies that \name{} can efficiently search a virtual library for putative selective binders. %
We compare optimization performance across multi-objective acquisition functions and demonstrate the superiority of Pareto-based acquisition functions over scalarization ones. We also implement a diversity-enhanced acquisition strategy that increases the number of acquired scaffolds by 33\% %
with only a minor impact on optimization performance. Finally, we apply \name{} to efficiently search the Enamine Screening Collection \cite{noauthor_enamine_nodate} of over 4 million molecules for selective dual inhibitors of EGFR and IGF1R as an exemplary 3-objective optimization. After exploration of only 8\% of the virtual library, 100\% of the library's non-dominated points and over 60\% of the library's \topk{0.1}, defined by non-dominated sorting, are identified by \name{}.

\section{%
Multi-Objective Virtual Screening with \name{}}

\begin{figure}
    \centering
    \includegraphics{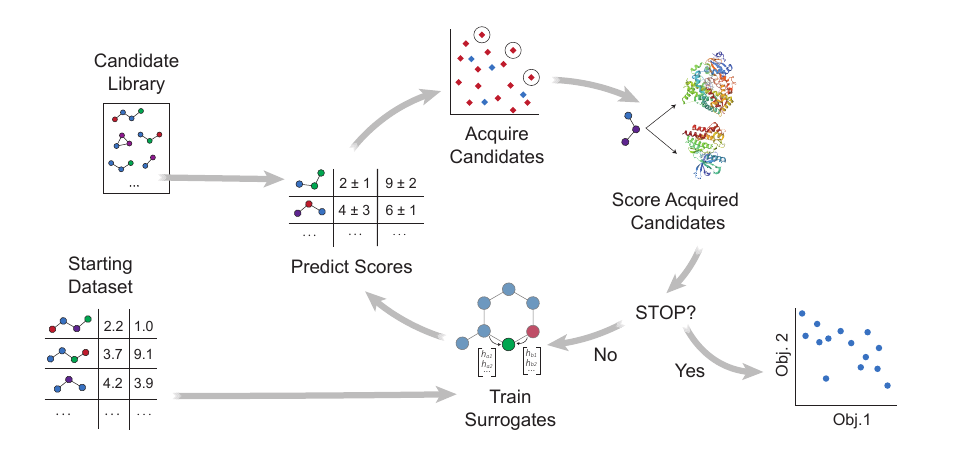}
    \caption{Overview of \name's workflow for multi-objective optimization. A surrogate model is first trained with an initial training set of randomly sampled molecules. Predictions and uncertainties from each surrogate model inform which molecules in the library to score next. After these acquired molecules are scored, each surrogate model is retrained, and the iterative loop continues. Once a stopping criterion is met, the set of observed points and their objective values can be analyzed. }
    \label{fig:workflow}
\end{figure}

\name{} applies the multi-objective pool-based workflow described in ref. \citenum{fromer_computer-aided_2023}, combining multi-objective Bayesian optimization and surrogate models to efficiently explore a virtual library for molecules that simultaneously optimize multiple properties (Figure \ref{fig:workflow}). Similar workflows have been demonstrated for the design of battery materials \cite{janet_accurate_2020, agarwal_discovery_2021} and other functional materials \cite{gopakumar_multi-objective_2018, del_rosario_assessing_2020}. As summarized in Algorithm \ref{S-alg:pareto-bo}, objective values are first calculated for a subset of the library, and surrogate models that predict each objective are trained on these initial observations. %
After objective values are predicted for all candidate molecules, an acquisition function selects a set of promising molecules for objective function evaluation. The surrogate models are then retrained with new observations, and the iterative loop repeats until a stopping criterion is met. 

Relative to its initial release in ref.~\citenum{graff_accelerating_2021}, MolPAL was extended primarily through modification of the acquisition strategy and handling of multiple surrogate models. The multi-objective extension of \name{} allows users to select between Pareto optimization and scalarization strategies. 

\begin{figure}
    \centering
    \includegraphics{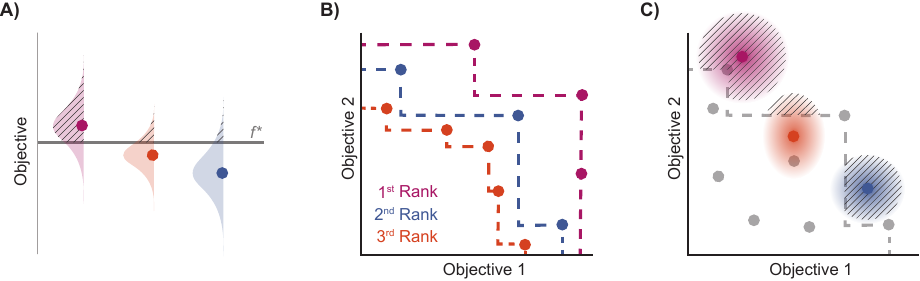}
    \caption{Depiction of acquisition functions in Bayesian optimization. (A) Probability of improvement (PI) estimates the likelihood that an as-yet unobserved objective function value exceeds the current maximum value $f^*$. Expected improvement (EI) estimates the amount that $f^*$ would increase if an unobserved point is acquired %
    \cite{keane_statistical_2006}. (B) Non-dominated sorting assigns integer ``Pareto ranks'' to each candidate. (C) Hypervolume-based acquisition functions \cite{keane_statistical_2006} extend the principles of PI and EI to multiple dimensions using the observed Pareto front (grey dashed line) rather than $f^*$. Probability of hypervolume improvement (PHI) represents the likelihood that acquisition of an unobserved point would increase the hypervolume by any amount. Expected hypervolume improvement (EHI) estimates the increase in hypervolume if the objective function value of such a point is scored \cite{keane_statistical_2006}. %
    Objectives are defined so that optimization corresponds to maximization.  }
    \label{fig:afs}
\end{figure}

Scalarization reduces a multi-objective optimization problem into a single-objective problem, often through a weighted sum: 
\begin{equation}
    f_{scal}(x) = \sum_i \lambda_i  f_i(x),  
    \label{eq:scal}
\end{equation}
with each $\lambda_i$ denoting the relative importance of objective $f_i$. Weighted sum scalarization requires the relative importance of objectives to be known before the optimization in order to assign weighting factors $\lambda$. Alternative scalarization strategies beyond a weighted sum include random scalarization \cite{paria_flexible_2020, zhang_random_2020} and Chebyshev scalarization \cite{steuer_interactive_1983, giagkiozis_methods_2015}, but these are not yet implemented in \name. Scalarization enables the use of single-objective acquisition functions, which 
include probability of improvement (PI) \cite{kushner_new_1964} or expected improvement (EI)\cite{mockus_bayesian_1975} (Figure \ref{fig:afs}A), greedy, and upper confidence bound (UCB)\cite{srinivas_information-theoretic_2012} (Table \ref{tab:afs}). %
Scalarization is implemented prior to surrogate model training, i.e., only one single-task surrogate model is necessary. Algorithm \ref{S-alg:scalarized-bo} summarizes \name's implementation of scalarized multi-objective optimization. 

Pareto optimization is a multi-objective optimization strategy that reveals the trade-offs between objectives and does not require any measure of the relative importance of objectives. %
Further, Pareto optimization aims to identify the entire Pareto front, a feature not guaranteed by single-objective methods such as weighted sum scalarization \cite{lin_three_1976, hu_revisiting_2023}. Common multi-objective acquisition functions include the probability of hypervolume improvement (PHI) \cite{keane_statistical_2006}, expected hypervolume improvement (EHI) \cite{keane_statistical_2006}, non-dominated sorting (NDS)\cite{srinivas_muiltiobjective_1994, deb_fast_2002}, and Pareto upper confidence bound (P-UCB) \cite{drugan_designing_2013}. These are natural extensions of single-objective acquisition functions (Table~\ref{tab:afs}) that instead aim to increase the region, or \textit{hypervolume}, dominated by the acquired points (Figure~\ref{fig:afs}). P-UCB and its single-objective analog UCB are not considered in this work. For model-guided Pareto optimization, either a multi-task surrogate model, multiple single-task surrogate models, or a combination thereof is needed to predict the set of objective function values. As outlined in Algorithm \ref{S-alg:pareto-bo}, \name{} trains $N$ single-task surrogate models to predict $N$ objectives to circumvent the challenge of loss function weighting in multi-task learning \cite{gong_comparison_2019}.

\begin{table}
\footnotesize
\def\arraystretch{1.5}%
    \centering
    \begin{tabular}{|c|c|c|c|} \hline 
         \multicolumn{2}{|c|}{\textbf{Single-objective}} & \multicolumn{2}{|c|}{\textbf{Multi-objective}} \\ \hline \hline 
         PI\cite{kushner_new_1964} & $\mathbb{P}_{f(x) \sim \mathcal{N}(\mu(x), \sigma(x))} [ f(x) > f^* ]$ & PHI\cite{keane_statistical_2006} & $\mathbb{P}_{\mathbf{f}(x) \sim \mathcal{N}(\boldsymbol{\mu}(x), \boldsymbol{\sigma}(x))} \, [\text{HV}(\mathcal{X}_{acq} \cup x) > \text{HV}(\mathcal{X}_{acq}) ]$ \\ \hline 
         EI\cite{mockus_bayesian_1975} & $\mathbb{E}_{f(x) \sim \mathcal{N}(\mu(x), \sigma(x))}[\max \{ f(x) - f^*, 0 \} ]$ & EHI\cite{keane_statistical_2006} &  $\mathbb{E}_{\mathbf{f}(x) \sim \mathcal{N}(\boldsymbol{\mu}(x), \boldsymbol{\sigma}(x))}\,[
         \text{HV}(\mathcal{X}_{acq} \cup x) - \text{HV}(\mathcal{X}_{acq}) ] %
         $\\ \hline 
         Greedy & $\mu(x)$ & NDS\cite{srinivas_muiltiobjective_1994, deb_fast_2002} &  $ \texttt{rank}(\boldsymbol{\mu}(x)) $\\ \hline UCB\cite{srinivas_information-theoretic_2012} & $\mu(x) + \beta  \sigma(x)$ & P-UCB\cite{drugan_designing_2013} & $ \texttt{rank}(\boldsymbol{\mu}(x) + \beta \boldsymbol{\sigma}(x))$ \\ \hline 
    \end{tabular}
    \caption{Common single-objective acquisition functions and their multi-objective analogs. $\mathbb{P}$ and $\mathbb{E}$ represent probability and expected value, respectively. $f^*$ is the current best objective value, HV is hypervolume, and $\texttt{rank}()$ is the Pareto rank. Surrogate models provide prediction means $\mu$ and standard deviations $\sigma$ for the objective value $f$ of candidate point $x$. $\mathcal{X}_{acq}$ is the set of points acquired in previous iterations. Bold variables are vectors. $\mathcal{N}(\boldsymbol{\mu}, \boldsymbol{\sigma})$ implies that the covariance matrix is treated as diagonal with entries $\sigma_i^2$, i.e., uncertainty is uncorrelated across objective functions.  }
    \label{tab:afs}
\end{table}

Acquiring a batch of $k$ points in a single iteration may be more efficient than sequential acquisition when objectives functions can be calculated in parallel. ``Top-$k$ batching'' naively selects the $k$ points with the highest acquisition scores \cite{bellamy_batched_2022}. More sophisticated batch acquisition strategies iteratively construct optimal batches one point at a time by hallucinating objective function values \cite{ginsbourger_kriging_2010, snoek_practical_2012, janusevskis_expected_2012, chevalier_fast_2013, jiang_efficient_2017,  tran_pbo-2gp-3b_2019}. %
Other strategies use heuristics to select batches that are diverse in the design space or objective space in order to improve the utility of a batch \cite{azimi_batch_2010, gonzalez_batch_2016, 
konakovic_lukovic_diversity-guided_2020, citovsky_batch_2021, maus_discovering_2023, gonzalez_new_2023}. 
\name{} implements both naive top-k batching and diversity-enhanced acquisition strategies that apply clustering in both the design space and in the objective space (Section \ref{section:clustering}).

\section{Results and Discussion}

\subsection{Description of Case Studies}

\begin{figure}[h]
    \centering
    \includegraphics{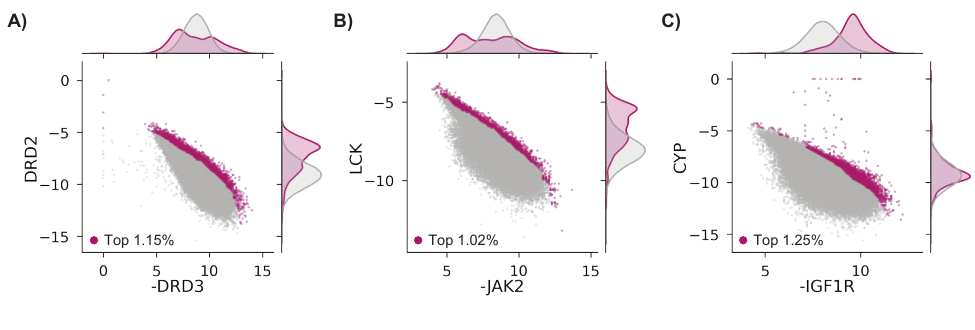}
    \caption{Docking scores in the DOCKSTRING dataset\cite{garcia-ortegon_dockstring_2022} for (A) Case 1, (B) Case 2, and (C) Case 3. All molecules considered in the \topk{1}, as determined by NDS rank (Section \ref{section:methods_metrics}), are shown as magenta points, with the remaining data in grey. There are 2986 (1.15\%), 2651 (1.02\%), and 3261 (1.25\%) molecules in the \topk{1} for Cases 1, 2, and 3, respectively. %
    }
    \label{fig:true-top-k}
\end{figure}

We test \name{} with three retrospective case studies with an emphasis on docking-predicted binding selectivity to compare optimization performance across acquisition functions. The pairs of objectives used in these case studies are exclusively docking scores even though the framework of \name{} is more general. Each case explores a virtual library of approximately 260k molecules and optimizes two competing docking score objectives from the DOCKSTRING benchmark \cite{garcia-ortegon_dockstring_2022}. Case 1's goal is modeled after identifying antagonists of dopamine receptor D$_3$ (DRD3) that are selective over dopamine receptor D$_2$ (DRD2), which may enable effective treatment of various conditions without the side effects triggered by DRD2 antagonists \cite{watson_selective_2012, williford_novel_2021, bonifazi_pharmacological_2023}. Case 2 treats  Janus Kinase 2 (JAK2), a lukemia target, as the on-target and lymphocyte-specific protein tyrosine kinase (LCK) as the off-target \cite{fridman_selective_2010, liu_design_2015}.  Finally, Case 3 aims to identify selective inhibitors of insulin-like growth factor 1 receptor (IGF1R) \cite{li_inhibition_2009, pasha_3d_2022} that do not bind to cytochrome P450 3A4 (CYP3A4) \cite{velaparthi_imidazole_2007, zimmermann_balancing_2008}, an off-target known to impact the pharmacokinetic properties of drugs though metabolism \cite{lin_inhibition_1998,  lynch_effect_2007,  cheng_insights_2011}. 
The property trade-offs for each case are shown in Figure \ref{fig:true-top-k}. %
Because both positive docking scores and scores of 0 should be interpreted as non-binders, we clip docking scores to 0. %

We employ single-task directed message-passing neural networks \cite{yang_analyzing_2019, heid_chemprop_2023} for each objective as surrogate models (Section \ref{section:methods_surrogate}). An initial set of 2,602 molecules is randomly sampled at the zeroth iteration, and 1\% of the library (2,602 molecules) is acquired in each subsequent iteration. Scalarization weighting factors ($\lambda_1$ and $\lambda_2$ in Eq. \ref{eq:scal}) were set to 0.5. Five trials with distinct initialization seeds were completed for each acquisition function. Section \ref{section:methods} contains full implementation details.

\subsection{Pareto Acquisition Functions Outperform Scalarization}
\label{section:acquisition_funcs}

We characterize optimization performance with four metrics: fraction of the \topk{1} acquired, inverted generational distance (IGD), hypervolume (HV), and fraction of non-dominated points acquired. We motivate the selection of these metrics and describe their implementation in Section \ref{section:methods_metrics}. %

\begin{figure}
    \centering
    \includegraphics{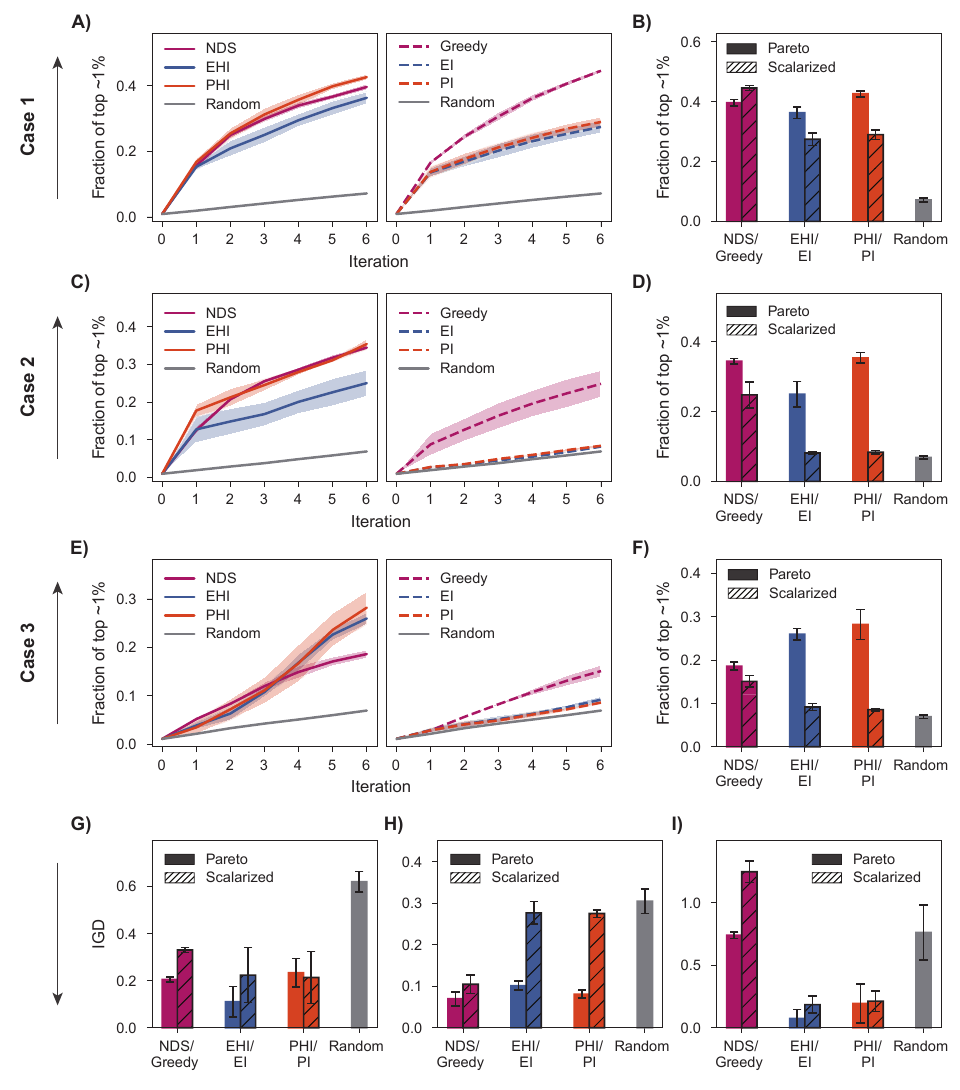}
    \caption{Fraction of the \topk{1} acquired and inverted generational distance (IGD) for Case 1, 2, and 3. \textbf{(A, C, E)} Fraction of \topk{1} using Pareto optimization acquisition functions (left) and scalarized acquisition functions (right). \textbf{(B, D, F)} Fraction of the \topk{1} acquired after 6 iterations. \textbf{(G-I)} IGD after 6 iterations for Case 1 (G), Case 2 (H), and Case 3 (I).  \errorbarsentence{(B, D, F, G-I)}{(A, C, E)} }
    \label{fig:top-k-profiles}
\end{figure}

All Pareto optimization acquisition functions show substantial improvement over random acquisition according to the \topk{1} metric, with PHI performing most consistently across cases (Figure \ref{fig:top-k-profiles}A-F). Analyzing the same metric, greedy is clearly the best scalarized acquisition function despite its poor performance in Case 3 relative to PHI and EHI. %
Only in Case 1 does the best scalarization acquisition function outperform the best Pareto acquisition function. 

\begin{figure}
    \centering
    \includegraphics{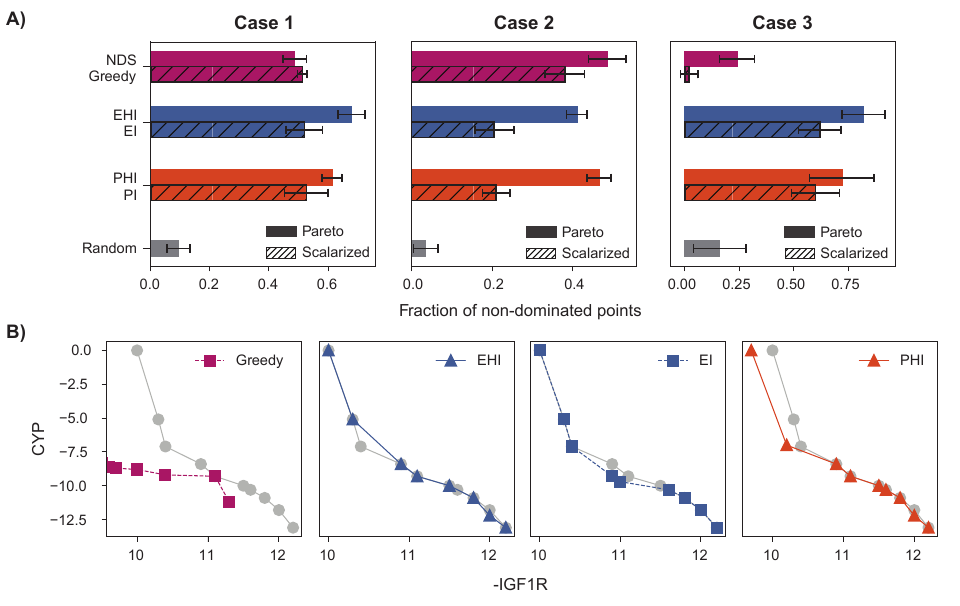}
    \caption{Identification of molecules on the true Pareto front. \textbf{(A)} Fraction of non-dominated points acquired for Cases 1, 2, and 3. \textbf{(B)} Final Pareto front acquired in Case 3 using probability of greedy, expected hypervolume improvement (EHI), expected improvement (EI), and hypervolume improvement (PHI) acquisition, with the true Pareto front shown in grey. The Pareto front identified with greedy acquisition poorly reflects the shape of the true Pareto front, a discrepancy captured well by the fraction of non-dominated points metric. Each plot represents individual runs, all initialized with the same model seed and starting acquired set. Error bars denote $\pm$ one standard deviation across five runs. 
    }
    \label{fig:nd-points}
\end{figure}

There is better differentiation among acquisition functions in terms of the IGD (Figure \ref{fig:top-k-profiles}G-I). %
EHI achieves the smallest IGD in Cases 1 and 3, while all three Pareto acquisition functions perform similarly in Case 2. When performance is measured by the fraction of non-dominated points acquired, EHI and PHI outperform scalarization acquisition functions (Figure \ref{fig:nd-points}) even though the degree of improvement varies across cases. The sensitivity of the hypervolume metric to outliers on the Pareto front leads to noisy hypervolume profiles that follow the same trends as the IGD metric (Figure \ref{S-fig:hv_si}). Overall, Pareto optimization acquisition functions strongly outperform or match the performance of scalarization for the pairs of objectives and virtual library considered.

The variation in relative performance of different acquisition functions across evaluation metrics is noteworthy. For example, greedy scalarization performs quite well in the \topk{1} metric for all cases but worse than random acquisition according to IGD, hypervolume, and fraction of non-dominated points in Case 3. %
The results shown here highlight the importance of assessing the performance across different sets of objectives. Despite the variation across cases and metrics, EHI and PHI consistently perform as well as scalarization or substantially better than scalarization. These acquisition functions are suitable choices for new sets of objectives. Alternatively, when a previously unexplored set of objectives is to be optimized, a retrospective study on a scored subset of the library can inform the selection of a suitable acquisition strategy.  %
Each reported metric corresponds to distinct optimization goals (Section \ref{section:methods_metrics}); the primary application of a multi-objective virtual screen should inform which metric to use for acquisition function comparison and selection. 

\subsection{Clustering-based Acquisition Improves Molecular Diversity } %
\label{section:cluster_results}
Scoring functions used in structure-based virtual screening can exhibit systematic errors that bias selection toward specific interactions \cite{bender_practical_2021}. %
This poses a risk for experimental validation if specific scaffolds are overrepresented in the top-scoring molecules. 
Selecting a structurally diverse set of candidates is one strategy to mitigate this risk and can be achieved via a diversity-enhanced acquisition strategy (Section \ref{section:clustering}). %
First, a set of molecules larger than the target batch size is selected according to the acquisition function and is then partitioned into a number of clusters equal to the batch size in feature (e.g., molecular fingerprint) space. %
The molecule with the best acquisition score in each cluster is acquired. %
We analyzed the performance of diversity-enhanced acquisition for Case 3, using PHI as the acquisition function. All hyperparameters were the same as those used for previous experiments (Section \ref{section:methods}).

\begin{figure}
    \centering
    \includegraphics{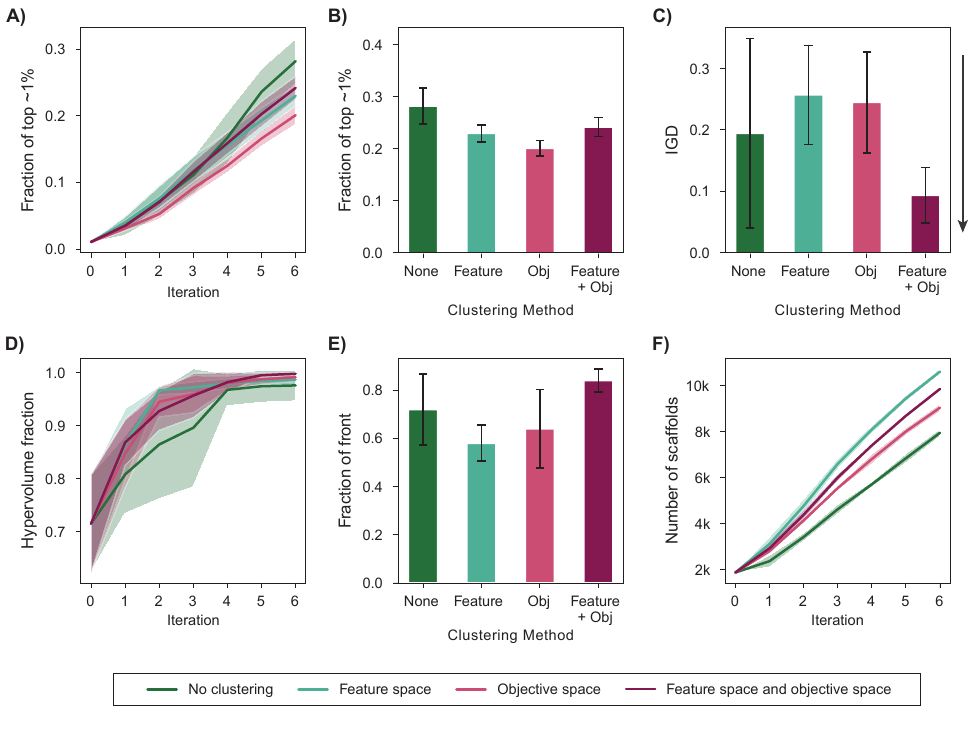}
    \caption{Impact of clustering in fingerprint and objective space on optimization performance. \textbf{(A)} Fraction of \topk{1} acquired. \textbf{(B)} Fraction of \topk{1} acquired acquired after 6 iterations. \textbf{(C)} Inverted generational distance (IGD) after 6 iterations. \textbf{(D)} Hypervolume profiles. \textbf{(E)} Fraction of the library's non-dominated points acquired after 6 iterations. \textbf{(F)} Number of distinct graph-based Bemis-Murcko scaffolds acquired. \errorbarsentence{(B, C, E)}{(A, D, F)} }
    \label{fig:cluster_metrics}
\end{figure}

Feature space clustering slightly hinders optimization performance according to all four performance metrics (Figure \ref{fig:cluster_metrics}A-E), but it also increases the number of graph-based Bemis-Murcko scaffolds \cite{bemis_properties_1996} acquired by 33\% when compared to a naïve batch construction strategy. This degradation in measured performance is expected given that the performance metrics do not consider the overall diversity of the selected molecules. The increased structural diversity of the selected molecules can be qualitatively visualized via dimensionality reduction through UMAP \cite{mcinnes_umap_2018} (Figure \ref{S-fig:umap}). 

Certain multi-objective optimization methods such as NSGA-II \cite{deb_fast_2002} also incorporate \textit{objective space} diversity, i.e., the selection of points better distributed along the Pareto front \cite{fromer_computer-aided_2023}. We find that clustering in the objective space during acquisition (Section \ref{section:clustering}) mildly hinders performance in all optimization metrics (Figure \ref{fig:cluster_metrics}A-E).  
An acquisition strategy that considers diversity in both the objective space and feature space (Section \ref{section:clustering}) performs similarly to the standard acquisition strategy across most optimization metrics (Figure \ref{fig:cluster_metrics}A-D) while acquiring a more structurally diverse set of molecules when compared to standard acquisition (Figure \ref{fig:cluster_metrics}F and \ref{S-fig:umap}). Overall, we recommend the use of feature space clustering if having structurally distinct candidates is a priority for experimental validation.

\subsection{\name{} Scales to 3 Objectives and Larger Libraries} %
\label{section:large_scale_results}

\begin{figure}
    \centering
    \includegraphics{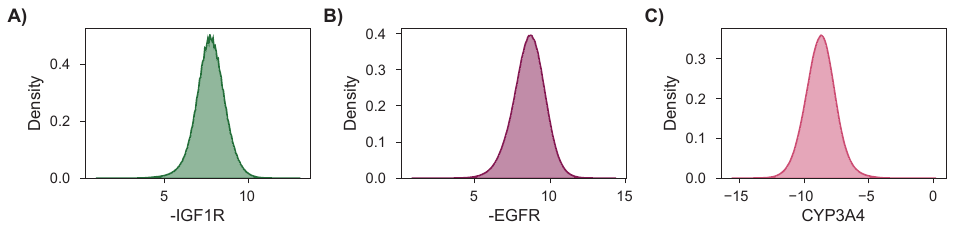}
    \caption{Distributions of objective function values to be maximized for a retrospective 3-objective virtual screen. \textbf{(A,B)} Docking scores for targets IGF1R and EGFR, respectively. \textbf{(C)} Docking scores for off-target CYP3A4. } %
    \label{fig:3obj_distributions}
\end{figure}

As a final demonstration, we show that \name{} scales well to larger virtual libraries and more than two objectives by searching the Enamine Screening Library \cite{noauthor_enamine_nodate} of over 4 million molecules for those that optimize three docking objectives. The objectives were defined to identify putative dual inhibitors of IGF1R and EGFR \cite{tandon_rbx10080307_2013, hu_dual_2022,  kang_dual_2022} with selectivity over CYP3A4 \cite{abourehab_globally_2021, lin_inhibition_1998,  lynch_effect_2007,  cheng_insights_2011}, which could in principle serve as starting points for esophageal cancer therapeutics. To analyze performance according to the four considered evaluation metrics, we perform this search retrospectively after docking the entire library using DOCKSTRING's protocol for each target \cite{garcia-ortegon_dockstring_2022} (Section \ref{section:methods_data}). The distributions of individual objectives for the entire library are shown in Figure \ref{fig:3obj_distributions}. We use PHI acquisition without clustering and acquire 1\% of the library at each iteration, repeating each experiment three times. 

\begin{table}
    \centering
    \begin{tabular}{llllll}
    \toprule
    Acquisition & \capstopk{0.1} $\uparrow$ & \capstopk{0.5} $\uparrow$ & HV $\uparrow$  & IGD $\downarrow$ & Fraction of \\
    Function & & & & & True Front $\uparrow$ \\
    \midrule
PHI & $0.65 \pm 0.06$ & $0.50 \pm 0.09$ & $1.00 \pm 0.00$ & $0.00 \pm 0.00$ & $1.00 \pm 0.00$ \\
Random & $0.10 \pm 0.00$ & $0.10 \pm 0.00$ & $0.81 \pm 0.05$ & $0.88 \pm 0.04$ & $0.10 \pm 0.04$ \\
    \bottomrule
    \end{tabular}
    \caption{Performance metrics after acquisition of 10\% of the library. Means and standard deviation across three trials are shown. Metrics include hypervolume (HV) and inverted generational distance (IGD), as well as the fraction of the library's top 0.12\% (4829 molecules), top 0.524\% (21015 molecules), and Pareto front points (39 molecules) acquired. }
    \label{tab:enamine_scores}
\end{table}

\begin{figure}
    \centering
    \includegraphics{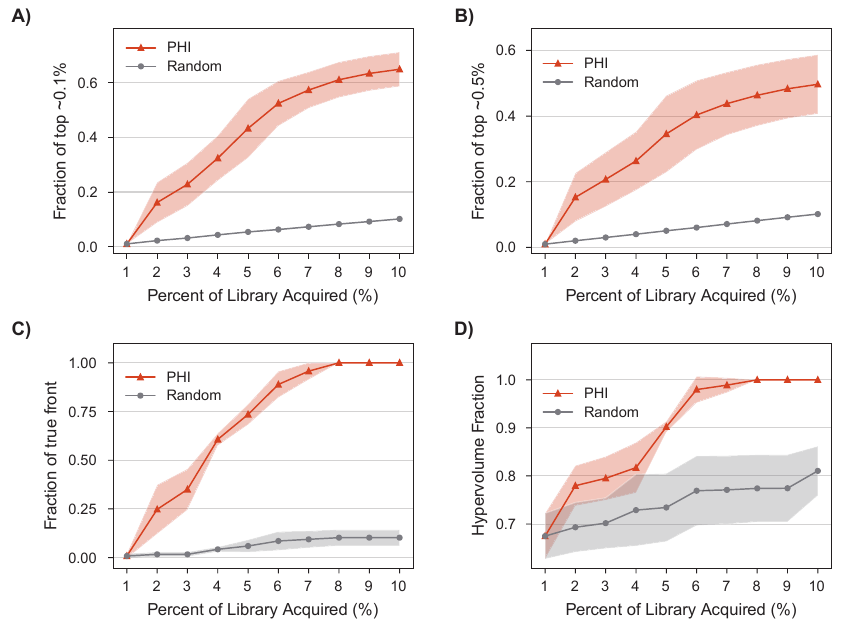}
    \caption{Performance of \name{} for the identification of selective dual IGF1R/EGFR inhibitors from the 4M-member Enamine screening library as an exemplary three-objective optimization. Profiles are depicted for the fraction of \textbf{(A)} \topk{0.1}, \textbf{(B)} \topk{0.5}, and \textbf{(C)} non-dominated points acquired, as well as the \textbf{(D)} hypervolume.  }
    \label{fig:three_obj_metrics}
\end{figure}

\name{} succeeds in acquiring 100\% of the library's non-dominated points in all three replicates after exploring only 8\% of the search space (Figure \ref{fig:three_obj_metrics}C), a 9X improvement of in the fraction of acquired non-dominated points over random acquisition. %
At this same degree of exploration (8\% of the library), over 60\% of the library's \topk{0.1} molecules have been identified (Figure \ref{fig:three_obj_metrics}A). Improvements in IGD and hypervolume with \name{} over random acquisition are also notable (Figure \ref{fig:three_obj_metrics}, Table \ref{tab:enamine_scores}). The results of this retrospective run indicate that \name{} can substantially reduce the computational resources required to identify molecules that simultaneously optimize multiple properties from a virtual library of millions of molecules. %

\begin{figure}
    \centering
    \includegraphics{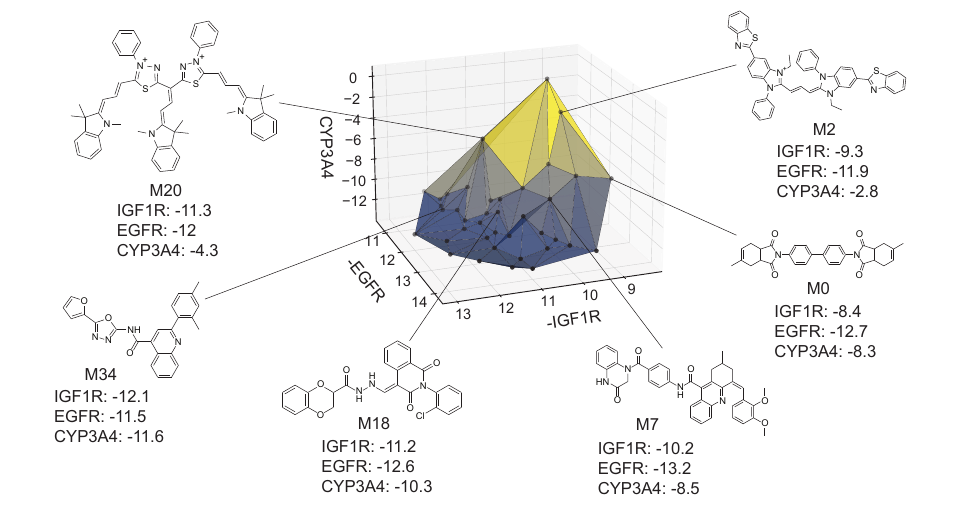}
    \caption{Visualization of the Pareto front for the 3-objective optimization for the identification of putative selective dual inhibitors. All non-dominated points depicted were acquired by \name{} after scoring only 8\% of the virtual library. Structures for some molecules on the Pareto front are shown. Docking scores to targets EGFR and IGF1R and off-target CYP3A4 were calculated with the AutoDock Vina wrapper DOCKSTRING \cite{garcia-ortegon_dockstring_2022}. }
    \label{fig:pf_annotated}
\end{figure}

The virtual library's true Pareto front of 39 molecules, all of which were recovered by \name{}, is visualized in Figure \ref{fig:pf_annotated}. The structures of all non-dominated molecules are shown in Figures \ref{S-fig:pf_all_1}-\ref{S-fig:pf_all_3}. %
The structures of these molecules expose some remaining challenges of docking for selectivity prediction. Molecules predicted to be non-binders to CYP3A4, such as M2 and M20 in Figure \ref{fig:pf_annotated}, are relatively large molecules that do not fit inside the pocket of CYP3A4, leading to steric clashes and less favorable binding energetics. The ability for such molecules to score well against the IGF1R and EGFR but poorly against CYP3A4 is exemplified by the computed docking poses and protein-ligand interactions of M2 (Figure \ref{S-fig:poses}). Although steric clashes may be a valid reason for hypothetical selectivity towards EGFR/IGF1R over CYP3A4, the molecules predicted to be most selective (e.g., M2 and M20) are not attractive candidates for experimental validation. In practice, dominated molecules that are close to the identified Pareto front should also be considered for experimental validation or follow-up studies. A ligand efficiency score may also be used as an objective function to penalize very large molecules \cite{pan_consideration_2003}. Nevertheless, given the imperfections of docking as an oracle function, \name{} perfectly identifies the Pareto front at a reduced computational cost and scores well in terms of all evaluation metrics. We do not intend to nominate the visualized molecules as starting points for a drug discovery project but instead aim to demonstrate the ability for \name{} to efficiently identify molecules that optimize any set of oracles such as those that predict binding affinity.

\section{Conclusion}

We have introduced an open source multi-objective extension of the pool-based optimization tool MolPAL \cite{graff_accelerating_2021} and demonstrated its ability to accelerate multi-objective virtual screening. 
\name{} provides a flexible Pareto optimization framework that allows users to systematically vary key design choices like acquisition strategy. %
\name{} is most appropriate for optimizing objectives that are more expensive to calculate than surrogate molecular property prediction models. Beyond docking, these include objectives that require binding free energy calculations, %
quantum mechanical simulations, or experiments to measure. Objectives that are calculable in CPU milliseconds, such as SA\_Score or ClogP, can be screened exhaustively and do not warrant model-guided optimization tools. \name{} could also be applied to consensus docking by optimizing multiple scoring functions that predict binding affinity to the same target \cite{li_effective_2009}.

We first assessed \name{} on three two-objective case studies that aim to identify putative selective binders. We found that expected hypervolume improvement and probability of hypervolume improvement, both Pareto optimization acquisition functions, consistently performed as well as or better than scalarization. %
A diversity-enhanced acquisition strategy that applies clustering in molecular fingerprint space was found to increase the number of Bemis-Murcko scaffolds observed by 33\% when compared to standard acquisition. Finally, we demonstrated that \name{} can efficiently search large virtual libraries and optimize three objectives simultaneously through a case study aiming to identify putative selective dual inhibitors from the Enamine Screening Library of over 4 million molecules; \name{} acquired all of the library's non-dominated molecules after exploring only 8\% of the library in all three replicates.

Exploration of other multi-objective optimization approaches, such as random scalarization and Chebyshev scalarization, in the context of molecular optimization could expose benefits of strategies not explored in this work. Other published diversity-enhanced acquisition strategies \cite{kirsch_batchbald_2019, gonzalez_new_2023} may have the potential to improve both molecular diversity \textit{and} optimization performance. %

While the use of docking to off-targets as a proxy for selectivity is well precedented \cite{chahal_combination_2023, schieferdecker_development_2023, matricon_structure-based_2023}, the high false negative rates of docking screens (i.e., binders predicted to be nonbinding) pose a risk for experimental validation of molecules predicted to be selective \cite{huggins_rational_2012, weiss_selectivity_2018, matricon_structure-based_2023}. Pharmacophore models and scoring functions that are designed for off-target activity \cite{recanatini_silico_2004, klabunde_gpcr_2005, schieferdecker_development_2023} could be more appropriate for cross-docking for selectivity in the future. Because \name{} is a general multi-objective optimization strategy that can be applied to any combination of oracle functions, it will maintain its relevance as these oracle functions improve in time. Future efforts applying \name{} to hit discovery and early-stage molecular design are necessary to validate the benefits of considering multiple objectives at early stages of molecular discovery.

\section{Methods}
All code required to reproduce the reported results can be found in the \texttt{multiobj} branch of MolPAL at \url{https://github.com/coleygroup/molpal/tree/multiobj}. 

\label{section:methods}

\subsection{Data Collection}
\label{section:methods_data}
All docking scores used in Sections \ref{section:acquisition_funcs} and \ref{section:cluster_results} were used without reprocessing from the DOCKSTRING benchmark\cite{garcia-ortegon_dockstring_2022}. For the larger scale study in Section \ref{section:large_scale_results}, the Enamine Screening library of over 4 million compounds \cite{noauthor_enamine_nodate} was docked against IGF1R, EGFR, and CYP3A4 using DOCKSTRING \cite{garcia-ortegon_dockstring_2022} with default settings for each target. Before docking, molecules were stripped of salts using RDKit's SaltRemover module \cite{noauthor_rdkit_nodate}. Of the 4,032,152 in the library (at the time of download), 4,010,199 were docked successfully to CYP3A4, 4,010,191 to IGF1R, and 4,010,187 to EGFR. The full set of docking scores are available at \url{https://figshare.com/articles/dataset/Enamine_screen_CYP3A4_EGFR_IGF1R_zip/23978547}. 

\subsection{Objectives}
\label{section:objective_cap}
Docking scores to on-targets were minimized objectives, and scores to off-targets were maximized objectives. A positive docking score is not meaningfully different from a docking score of zero. We therefore adjust docking objectives to be: 
\begin{equation}
    f_{dock}(x) = \min(0, f(x))
\end{equation}
where $f(x)$ is the raw docking score and $f_{\text {dock}}(x)$ is the minimized (for on-targets) or maximized (for off-targets) objective. Clipping docking scores in this manner also mitigates the effect of outliers on Pareto optimization metrics like hypervolume. %

\subsection{Acquisition Functions}
\label{section:methods_afs}
We modify PHYSBO's implementation \cite{motoyama_bayesian_2022} of expected hypervolume improvement and probability of hypervolume improvement, which applies the algorithm proposed by \citet{couckuyt_fast_2014}. \name{} uses the \texttt{non\_dominated\_front\_2d()} utility in pygmo \cite{biscani_parallel_2020} for non-dominated sorting in two dimensions. In more than three dimensions, \name{} iteratively identifies non-dominated points using the Pareto class in PHYSBO \cite{motoyama_bayesian_2022} and removes those points from the unranked set until enough points have been ranked. 

In the scalarization runs, $f$ was calculated according to Eq. \ref{eq:scal} with $\lambda_1 = \lambda_2 = 0.5$. For prediction mean $\mu$, prediction standard deviation $\sigma$, and current maximum value $f^*$, expected improvement and probability of improvement in the scalarization runs were calculated as: 
\begin{equation}
    \begin{aligned}
    & \text{EI}(x, f^*) & = &\;\; (\mu(x) - f^* + \xi) \cdot \Phi \left( \frac{\mu(x) - f^* + \xi}{\sigma(x)} \right) + \sigma(x) \cdot \phi \left( \frac{\mu(x) - f^* + \xi}{\sigma(x)} \right) \\
    & \text{PI}(x, f^*) & = &\;\; \Phi \left( \frac{\mu(x) - f^* + \xi}{\sigma(x)} \right) \\
    & \text{Greedy}(x) & = &\;\; \mu(x)  \\
    \end{aligned}
\end{equation}
where $\Phi$ is the cumulative distribution function of the standard normal distribution and $\phi$ is the probability density function of the standard normal distribution function. $\xi$ is a hyperparameter that controls the competitiveness of the sampling \cite{kushner_new_1964}, which we set to 0.01. Higher values of $\xi$ encourages exploration of uncertain points, while low values prioritize prediction means \cite{kushner_new_1964}.

\subsection{Surrogate Model}
\label{section:methods_surrogate}
Because the optimization campaigns explored in this study comprise batches of thousands of molecules, we elect to use a directed message-passing neural network architecture \cite{yang_analyzing_2019} as the surrogate model for all runs. \name{} contains a PyTorch \cite{paszke_pytorch_2019} implementation of ChemProp \cite{yang_analyzing_2019, heid_chemprop_2023} as in the original publication \cite{graff_accelerating_2021}. We use an encoder depth of 3, directed edge messages only (no atom messages), a hidden size of 300, ReLU activation, and two layers in the feed-forward neural network as default parameters. The model was trained with an initial learning rate of $10^{-4}$ using a Noam learning rate scheduler and an Adam optimizer. In each iteration, the model is retrained from scratch, as this was found to provide a benefit over fine-tuning \cite{graff_accelerating_2021}.     

We use mean-variance estimation \cite{nix_estimating_1994, hirschfeld_uncertainty_2020} for surrogate model uncertainty quantification. The surrogate model is trained to predict both the mean $f(x)$ and the variance of the estimation $\sigma^2(x)$ via a negative log likelihood loss function: %
\begin{equation}
    L(f(x), \mu(x),  \sigma(x)) = \frac{\ln{2\pi}}{2} + \frac{1}{2} \ln{\sigma^2(x)} + \frac{\left( \mu(x) - y(x) \right)^2}{2\sigma^2(x)}
\end{equation}
where $f(x)$ is the true objective value, $\mu$ is the predicted mean, and $\sigma^2(x)$ is the predicted variance. %

\subsection{Clustering}
\label{section:clustering}
We implement three diversity-enhancing acquisition strategies using clustering: in the feature space, in the objective space, and in both (Section \ref{section:cluster_results}). Feature space clustering is performed according to 2048-bit atom-pair fingerprints \cite{carhart_atom_1985} calculated using the \\ \noindent \texttt{GetHashedAtomPairFingerprintAsBitVect} function in RDKit \cite{noauthor_rdkit_nodate} with the minimum length and maximum length of paths between pairs to 1 and 3, respectively. Atom pair fingerprints have been shown to outperform other extended connectivity fingerprints in recovering 3D-shape analogs \cite{awale_atom_2014} and ranking close analogs by structural similarity \cite{oboyle_comparing_2016}. The predicted objective means were used for clustering in the objective space.  %

The acquisition strategy first selects a superset of molecules according to the acquisition scores as in a standard acquisition strategy. In our trials, the size of the superset is 10X the batch size, $b$. Then, the superset is clustered according to molecular fingerprints or predicted objective values. We use the implementation of \texttt{MiniBatchKMeans} \cite{sculley_web-scale_2010} in scikit-learn \cite{pedregosa_scikit-learn_2011} with $b$ clusters, 10 random initializations, an initialization size of $3b$, and a reassignment ratio of 0 for batch size $b$. Although $b$ clusters are specified, some clusters may be returned as empty in the case of a low-dimensional cluster basis. We ensure that $b$ points are acquired by iteratively looping through all clusters, beginning with the largest clusters, and acquiring the molecule in the cluster with the highest acquisition score. This process is continued until $b$ molecules are acquired. %

When clustering in \textit{both} the feature and objective spaces, the superset is first clustered in the objective space into $\mathtt{ceil}(b/2)$ clusters, and $\mathtt{ceil}(b/2)$ points are acquired by iterating through the clusters and selecting the molecule with the highest acquisition score. The superset is then supplemented with the $b/2$ unacquired candidate points that have the highest acquisition scores. Then, the superset is clustered in the atom-pair fingerprint space into $b-\mathtt{ceil}(b/2)$ clusters, and the point with the highest acquisition score in each cluster is acquired. Thus, when both forms of clustering are used, effectively $b/2$ points are selected to improve objective space diversity, and $b/2$ points are selected to improve feature space diversity.  

\subsection{Performance Metrics}
\label{section:methods_metrics}

Four performance metrics designed for Pareto optimization were measured to holistically assess the performance of \name. The underlying goal of a multi-property virtual screen (e.g., understanding the trade-offs between objectives, identifying a set of potentially promising selective inhibitors) can guide which evaluation metric to be deemed most important. Reported metrics and the motivation for selecting each is described below. %

\subsubsection*{Fraction of \topk{k}}

Virtual screening workflows based on active learning are often evaluated using a top-$k$ metric: the fraction of the true top-$k$ molecules that have been acquired. Using a top-$k$ metric captures the goal of separating promising molecules from the bulk while acknowledging that docking scores are imperfect and imprecise predictors of binding affinity. The best observed performance of a single compound is a less useful metric in this context, as there is a high likelihood it would not validate as a binder experimentally. %

For multi-property virtual screening, the set of top-$k$ molecules is not well-defined. We opt to define the \topk{k} through non-dominated sorting of the library (Figure \ref{S-fig:metrics}).
The non-dominated points in the library are iteratively selected and removed from the candidate pool until $\ge k$\% of the library has been selected. These points then serve as the \topk{k}. %
Because multiple points are selected as part of the \topk{k} %
in each iteration (i.e., all points of a certain Pareto rank), the fraction is slightly larger than $k\%$. %

The \topk{k} metric best describes virtual screens aiming to identify many molecules roughly close to the Pareto front. Compared to other metrics that consider points on or much closer to the Pareto front, it best captures the expectation that not all top-performing molecules will validate experimentally. 

\subsubsection*{Fraction of non-dominated points}
In some multi-property screens, the aim may be to elucidate as many points on the true Pareto front as possible. This enables the shape of the Pareto front and the inherent trade-off between objectives to be well-understood. A metric related to only the true non-dominated points is most appropriate in these cases. The library's non-dominated points are identified using non-dominated sorting (Section \ref{section:methods_afs}), and this metric is simply the fraction of the non-dominated molecules that were acquired. 

\subsubsection*{Hypervolume}

Hypervolume is a common metric used to assess the performance of multi-objective optimization methods \cite{tanabe_analysis_2020}. It measures the size of the region dominated by the Pareto front; in a bi-objective optimization, the hypervolume is the area dominated by the acquired points (Figure \ref{S-fig:metrics}B). Importantly, the hypervolume can be very sensitive to outliers, leading to high variability across repeat experiments. Because outliers may fail to be experimentally validated, the strong sensitivity of hypervolume to specific points makes it slightly misleading. We use the python package pygmo \cite{biscani_parallel_2020} to calculate hypervolume and report it as the fraction of the virtual library's total hypervolume. In contrast to the other evaluated metrics, the absolute hypervolume can be measured in prospective studies. 

\subsubsection*{Inverted generational distance}
The inverted generational distance (IGD), another widely adopted metric for Pareto optimization performance \cite{tanabe_analysis_2020, bosman_balance_2003}, quantifies the distance between the true Pareto front and the Pareto front of the acquired points. Therefore, IGD is a useful metric for understanding the extent to which the shape of the true Pareto front is captured by the acquired points. The IGD is computed as the average of the shortest distance between each point on the true Pareto front and the set of acquired points (Figure \ref{S-fig:metrics}): 
\begin{equation}
    \text{IGD}(\mathcal{X}_{acq},Z) = \frac{1}{|Z|} \sum_{z \in Z} \min_{x \in \mathcal{X}_{acq}} d(x, z),
\end{equation}
where $\mathcal{X}_{acq}$ is the set of acquired points, $Z$ is the set of non-dominated points, and $d(x,z)$ is the Euclidian distance between acquired point $x$ and non-dominated point $z$ in the objective space. IGD is calculated with the python package pymoo\cite{blank_pymoo_2020}.

\subsection{Scaffold Analysis}
To quantify the improvement in molecular diversity using diversity-enhanced acquisition, we count the number of observed Bemis-Murcko scaffolds \cite{bemis_properties_1996} using RDKit's \texttt{MurckoScaffoldSmilesFromSmiles} and \texttt{MakeScaffoldGeneric} functions.

\section*{Data Availability}
All code required to reproduce the reported results can be found in the \texttt{multiobj} branch of MolPAL at \url{https://github.com/coleygroup/molpal/tree/multiobj}. Docking scores from the DOCKSTRING benchmark are available at \url{https://figshare.com/articles/dataset/dockstring_dataset/16511577}. Docking scores for IGF1R, EGFR, and CYP3A4 computed as a part of this work can be found at \url{https://figshare.com/articles/dataset/Enamine_screen_CYP3A4_EGFR_IGF1R_zip/23978547}. 

\section*{Author Contributions}
J. C. F. extended MolPAL to perform multi-objective optimization and conducted the experiments; J. C. F, D. E. G., and C. W. C. designed the experiments; J. C. F., D. E. G., and C. W. C. wrote the manuscript. C. W. C. supervised the work.

\section*{Conflicts of Interest}
There are no conflicts to declare.

\section*{Acknowledgment}
This work was funded by the DARPA Accelerated Molecular Discovery program under contract number HR00111920025. J.C.F. received additional support from the National Science Foundation Graduate Research Fellowship under Grant No. 2141064. The authors acknowledge the MIT SuperCloud \cite{reuther_interactive_2018}, Lincoln Laboratory Supercomputing Center, and the MIT Engaging Cluster for providing HPC resources that have contributed to the research results reported within this paper. The authors thank Samuel Goldman for helpful discussions about the content of this manuscript.

\bibliography{references.bib}

\providecommand{\latin}[1]{#1}
\makeatletter
\providecommand{\doi}
  {\begingroup\let\do\@makeother\dospecials
  \catcode`\{=1 \catcode`\}=2 \doi@aux}
\providecommand{\doi@aux}[1]{\endgroup\texttt{#1}}
\makeatother
\providecommand*\mcitethebibliography{\thebibliography}
\csname @ifundefined\endcsname{endmcitethebibliography}
  {\let\endmcitethebibliography\endthebibliography}{}
\begin{mcitethebibliography}{121}
\providecommand*\natexlab[1]{#1}
\providecommand*\mciteSetBstSublistMode[1]{}
\providecommand*\mciteSetBstMaxWidthForm[2]{}
\providecommand*\mciteBstWouldAddEndPuncttrue
  {\def\EndOfBibitem{\unskip.}}
\providecommand*\mciteBstWouldAddEndPunctfalse
  {\let\EndOfBibitem\relax}
\providecommand*\mciteSetBstMidEndSepPunct[3]{}
\providecommand*\mciteSetBstSublistLabelBeginEnd[3]{}
\providecommand*\EndOfBibitem{}
\mciteSetBstSublistMode{f}
\mciteSetBstMaxWidthForm{subitem}{(\alph{mcitesubitemcount})}
\mciteSetBstSublistLabelBeginEnd
  {\mcitemaxwidthsubitemform\space}
  {\relax}
  {\relax}

\bibitem[Hughes \latin{et~al.}(2011)Hughes, Rees, Kalindjian, and
  Philpott]{hughes_principles_2011}
Hughes,~J.; Rees,~S.; Kalindjian,~S.; Philpott,~K. Principles of {Early} {Drug}
  {Discovery}. \emph{British Journal of Pharmacology} \textbf{2011},
  \emph{162}, 1239--1249\relax
\mciteBstWouldAddEndPuncttrue
\mciteSetBstMidEndSepPunct{\mcitedefaultmidpunct}
{\mcitedefaultendpunct}{\mcitedefaultseppunct}\relax
\EndOfBibitem
\bibitem[Kettle and Wilson(2016)Kettle, and Wilson]{kettle_standing_2016}
Kettle,~J.~G.; Wilson,~D.~M. Standing on the {Shoulders} of {Giants}: {A}
  {Retrospective} {Analysis} of {Kinase} {Drug} {Discovery} at {Astrazeneca}.
  \emph{Drug Discovery Today} \textbf{2016}, \emph{21}, 1596--1608\relax
\mciteBstWouldAddEndPuncttrue
\mciteSetBstMidEndSepPunct{\mcitedefaultmidpunct}
{\mcitedefaultendpunct}{\mcitedefaultseppunct}\relax
\EndOfBibitem
\bibitem[Beckers \latin{et~al.}(2022)Beckers, Fechner, and
  Stiefl]{beckers_25_2022}
Beckers,~M.; Fechner,~N.; Stiefl,~N. 25 {Years} of {Small}-{Molecule}
  {Optimization} at {Novartis}: {A} {Retrospective} {Analysis} of {Chemical}
  {Series} {Evolution}. \emph{Journal of Chemical Information and Modeling}
  \textbf{2022}, \emph{62}, 6002--6021\relax
\mciteBstWouldAddEndPuncttrue
\mciteSetBstMidEndSepPunct{\mcitedefaultmidpunct}
{\mcitedefaultendpunct}{\mcitedefaultseppunct}\relax
\EndOfBibitem
\bibitem[Keserű and Makara(2006)Keserű, and Makara]{keseru_hit_2006}
Keserű,~G.~M.; Makara,~G.~M. Hit {Discovery} and {Hit}-to-{Lead} {Approaches}.
  \emph{Drug Discovery Today} \textbf{2006}, \emph{11}, 741--748\relax
\mciteBstWouldAddEndPuncttrue
\mciteSetBstMidEndSepPunct{\mcitedefaultmidpunct}
{\mcitedefaultendpunct}{\mcitedefaultseppunct}\relax
\EndOfBibitem
\bibitem[Sun \latin{et~al.}(2022)Sun, Gao, Hu, and Zhou]{sun_why_2022}
Sun,~D.; Gao,~W.; Hu,~H.; Zhou,~S. Why 90\% of {Clinical} {Drug} {Development}
  {Fails} and {How} to {Improve} {It}? \emph{Acta Pharmaceutica Sinica B}
  \textbf{2022}, \emph{12}, 3049--3062\relax
\mciteBstWouldAddEndPuncttrue
\mciteSetBstMidEndSepPunct{\mcitedefaultmidpunct}
{\mcitedefaultendpunct}{\mcitedefaultseppunct}\relax
\EndOfBibitem
\bibitem[Segall and Barber(2014)Segall, and Barber]{segall_addressing_2014}
Segall,~M.~D.; Barber,~C. Addressing {Toxicity} {Risk} {When} {Designing} and
  {Selecting} {Compounds} in {Early} {Drug} {Discovery}. \emph{Drug Discovery
  Today} \textbf{2014}, \emph{19}, 688--693\relax
\mciteBstWouldAddEndPuncttrue
\mciteSetBstMidEndSepPunct{\mcitedefaultmidpunct}
{\mcitedefaultendpunct}{\mcitedefaultseppunct}\relax
\EndOfBibitem
\bibitem[Van~Vleet \latin{et~al.}(2019)Van~Vleet, Liguori, Lynch, Rao, and
  Warder]{van_vleet_screening_2019}
Van~Vleet,~T.~R.; Liguori,~M.~J.; Lynch,~I.,~James~J.; Rao,~M.; Warder,~S.
  Screening {Strategies} and {Methods} for {Better} {Off}-{Target} {Liability}
  {Prediction} and {Identification} of {Small}-{Molecule} {Pharmaceuticals}.
  \emph{SLAS Discovery} \textbf{2019}, \emph{24}, 1--24\relax
\mciteBstWouldAddEndPuncttrue
\mciteSetBstMidEndSepPunct{\mcitedefaultmidpunct}
{\mcitedefaultendpunct}{\mcitedefaultseppunct}\relax
\EndOfBibitem
\bibitem[Bleicher \latin{et~al.}(2003)Bleicher, Böhm, Müller, and
  Alanine]{bleicher_hit_2003}
Bleicher,~K.~H.; Böhm,~H.-J.; Müller,~K.; Alanine,~A.~I. Hit and {Lead}
  {Generation}: {Beyond} {High}-{Throughput} {Screening}. \emph{Nature Reviews
  Drug Discovery} \textbf{2003}, \emph{2}, 369--378\relax
\mciteBstWouldAddEndPuncttrue
\mciteSetBstMidEndSepPunct{\mcitedefaultmidpunct}
{\mcitedefaultendpunct}{\mcitedefaultseppunct}\relax
\EndOfBibitem
\bibitem[Recanatini \latin{et~al.}(2004)Recanatini, Bottegoni, and
  Cavalli]{recanatini_silico_2004}
Recanatini,~M.; Bottegoni,~G.; Cavalli,~A. In {Silico} {Antitarget}
  {Screening}. \emph{Drug Discovery Today: Technologies} \textbf{2004},
  \emph{1}, 209--215\relax
\mciteBstWouldAddEndPuncttrue
\mciteSetBstMidEndSepPunct{\mcitedefaultmidpunct}
{\mcitedefaultendpunct}{\mcitedefaultseppunct}\relax
\EndOfBibitem
\bibitem[Macchiarulo \latin{et~al.}(2004)Macchiarulo, Nobeli, and
  Thornton]{macchiarulo_ligand_2004}
Macchiarulo,~A.; Nobeli,~I.; Thornton,~J.~M. Ligand {Selectivity} and
  {Competition} {Between} {Enzymes} in {Silico}. \emph{Nature Biotechnology}
  \textbf{2004}, \emph{22}, 1039--1045\relax
\mciteBstWouldAddEndPuncttrue
\mciteSetBstMidEndSepPunct{\mcitedefaultmidpunct}
{\mcitedefaultendpunct}{\mcitedefaultseppunct}\relax
\EndOfBibitem
\bibitem[Elkins \latin{et~al.}(2016)Elkins, Fedele, Szklarz, Abdul~Azeez,
  Salah, Mikolajczyk, Romanov, Sepetov, Huang, Roth, Al~Haj~Zen, Fourches,
  Muratov, Tropsha, Morris, Teicher, Kunkel, Polley, Lackey, Atkinson,
  Overington, Bamborough, Müller, Price, Willson, Drewry, Knapp, and
  Zuercher]{elkins_comprehensive_2016}
Elkins,~J.~M. \latin{et~al.}  Comprehensive {Characterization} of the
  {Published} {Kinase} {Inhibitor} {Set}. \emph{Nature Biotechnology}
  \textbf{2016}, \emph{34}, 95--103\relax
\mciteBstWouldAddEndPuncttrue
\mciteSetBstMidEndSepPunct{\mcitedefaultmidpunct}
{\mcitedefaultendpunct}{\mcitedefaultseppunct}\relax
\EndOfBibitem
\bibitem[Raghavendra \latin{et~al.}(2018)Raghavendra, Pingili, Kadasi, Mettu,
  and Prasad]{raghavendra_dual_2018}
Raghavendra,~N.~M.; Pingili,~D.; Kadasi,~S.; Mettu,~A.; Prasad,~S. V. U.~M.
  Dual or {Multi}-{Targeting} {Inhibitors}: {The} {Next} {Generation}
  {Anticancer} {Agents}. \emph{European Journal of Medicinal Chemistry}
  \textbf{2018}, \emph{143}, 1277--1300\relax
\mciteBstWouldAddEndPuncttrue
\mciteSetBstMidEndSepPunct{\mcitedefaultmidpunct}
{\mcitedefaultendpunct}{\mcitedefaultseppunct}\relax
\EndOfBibitem
\bibitem[Ibrahim and Gabr(2019)Ibrahim, and Gabr]{ibrahim_multitarget_2019}
Ibrahim,~M.~M.; Gabr,~M.~T. Multitarget {Therapeutic} {Strategies} for
  {Alzheimer}’s {Disease}. \emph{Neural Regeneration Research} \textbf{2019},
  \emph{14}, 437--440\relax
\mciteBstWouldAddEndPuncttrue
\mciteSetBstMidEndSepPunct{\mcitedefaultmidpunct}
{\mcitedefaultendpunct}{\mcitedefaultseppunct}\relax
\EndOfBibitem
\bibitem[Benek \latin{et~al.}(2020)Benek, Korabecny, and
  Soukup]{benek_perspective_2020}
Benek,~O.; Korabecny,~J.; Soukup,~O. A {Perspective} on {Multi}-target {Drugs}
  for {Alzheimer}’s {Disease}. \emph{Trends in Pharmacological Sciences}
  \textbf{2020}, \emph{41}, 434--445\relax
\mciteBstWouldAddEndPuncttrue
\mciteSetBstMidEndSepPunct{\mcitedefaultmidpunct}
{\mcitedefaultendpunct}{\mcitedefaultseppunct}\relax
\EndOfBibitem
\bibitem[Brassard and Rondeau(2012)Brassard, and Rondeau]{brassard_role_2012}
Brassard,~M.; Rondeau,~G. Role of {Vandetanib} in the {Management} of
  {Medullary} {Thyroid} {Cancer}. \emph{Biologics : Targets \& Therapy}
  \textbf{2012}, \emph{6}, 59--66\relax
\mciteBstWouldAddEndPuncttrue
\mciteSetBstMidEndSepPunct{\mcitedefaultmidpunct}
{\mcitedefaultendpunct}{\mcitedefaultseppunct}\relax
\EndOfBibitem
\bibitem[Okamoto \latin{et~al.}(2015)Okamoto, Ikemori-Kawada, Jestel, von
  König, Funahashi, Matsushima, Tsuruoka, Inoue, and
  Matsui]{okamoto_distinct_2015}
Okamoto,~K.; Ikemori-Kawada,~M.; Jestel,~A.; von König,~K.; Funahashi,~Y.;
  Matsushima,~T.; Tsuruoka,~A.; Inoue,~A.; Matsui,~J. Distinct {Binding} {Mode}
  of {Multikinase} {Inhibitor} {Lenvatinib} {Revealed} by {Biochemical}
  {Characterization}. \emph{ACS Medicinal Chemistry Letters} \textbf{2015},
  \emph{6}, 89--94\relax
\mciteBstWouldAddEndPuncttrue
\mciteSetBstMidEndSepPunct{\mcitedefaultmidpunct}
{\mcitedefaultendpunct}{\mcitedefaultseppunct}\relax
\EndOfBibitem
\bibitem[Ma \latin{et~al.}(2010)Ma, Shi, Tan, Jiang, Go, Low, and
  Chen]{ma_-silico_2010}
Ma,~X.~H.; Shi,~Z.; Tan,~C.; Jiang,~Y.; Go,~M.~L.; Low,~B.~C.; Chen,~Y.~Z.
  In-{Silico} {Approaches} to {Multi}-target {Drug} {Discovery}.
  \emph{Pharmaceutical Research} \textbf{2010}, \emph{27}, 739--749\relax
\mciteBstWouldAddEndPuncttrue
\mciteSetBstMidEndSepPunct{\mcitedefaultmidpunct}
{\mcitedefaultendpunct}{\mcitedefaultseppunct}\relax
\EndOfBibitem
\bibitem[Yousuf \latin{et~al.}(2017)Yousuf, Iman, Iftikhar, and
  Mirza]{yousuf_structure-based_2017}
Yousuf,~Z.; Iman,~K.; Iftikhar,~N.; Mirza,~M.~U. Structure-{Based} {Virtual}
  {Screening} and {Molecular} {Docking} for the {Identification} of {Potential}
  {Multi}-{Targeted} {Inhibitors} {Against} {Breast} {Cancer}. \emph{Breast
  Cancer: Targets and Therapy} \textbf{2017}, \emph{9}, 447--459\relax
\mciteBstWouldAddEndPuncttrue
\mciteSetBstMidEndSepPunct{\mcitedefaultmidpunct}
{\mcitedefaultendpunct}{\mcitedefaultseppunct}\relax
\EndOfBibitem
\bibitem[Chahal and Kakkar(2023)Chahal, and Kakkar]{chahal_combination_2023}
Chahal,~V.; Kakkar,~R. A {Combination} {Strategy} of {Structure}-{Based}
  {Virtual} {Screening}, {MM}-{GBSA}, {Cross} {Docking}, {Molecular} {Dynamics}
  and {Metadynamics} {Simulations} {Used} to {Investigate} {Natural}
  {Compounds} as {Potent} and {Specific} {Inhibitors} of {Tumor} {Linked}
  {Human} {Carbonic} {Anhydrase} {Ix}. \emph{Journal of Biomolecular Structure
  and Dynamics} \textbf{2023}, \emph{41}, 5465--5480\relax
\mciteBstWouldAddEndPuncttrue
\mciteSetBstMidEndSepPunct{\mcitedefaultmidpunct}
{\mcitedefaultendpunct}{\mcitedefaultseppunct}\relax
\EndOfBibitem
\bibitem[Schieferdecker and Vock(2023)Schieferdecker, and
  Vock]{schieferdecker_development_2023}
Schieferdecker,~S.; Vock,~E. Development of {Pharmacophore} {Models} for the
  {Important} {Off}-{Target} 5-{HT2B} {Receptor}. \emph{Journal of Medicinal
  Chemistry} \textbf{2023}, \emph{66}, 1509--1521\relax
\mciteBstWouldAddEndPuncttrue
\mciteSetBstMidEndSepPunct{\mcitedefaultmidpunct}
{\mcitedefaultendpunct}{\mcitedefaultseppunct}\relax
\EndOfBibitem
\bibitem[Matricon \latin{et~al.}(2023)Matricon, Nguyen, Vo, Baltos, Jaiteh,
  Luttens, Kampen, Christopoulos, Kihlberg, May, and
  Carlsson]{matricon_structure-based_2023}
Matricon,~P.; Nguyen,~A.~T.; Vo,~D.~D.; Baltos,~J.-A.; Jaiteh,~M.; Luttens,~A.;
  Kampen,~S.; Christopoulos,~A.; Kihlberg,~J.; May,~L.~T.; Carlsson,~J.
  Structure-{Based} {Virtual} {Screening} {Discovers} {Potent} and {Selective}
  {Adenosine} {A1} {Receptor} {Antagonists}. \emph{European Journal of
  Medicinal Chemistry} \textbf{2023}, \emph{257}, 115419\relax
\mciteBstWouldAddEndPuncttrue
\mciteSetBstMidEndSepPunct{\mcitedefaultmidpunct}
{\mcitedefaultendpunct}{\mcitedefaultseppunct}\relax
\EndOfBibitem
\bibitem[Weiss \latin{et~al.}(2018)Weiss, Karpiak, Huang, Sassano, Lyu, Roth,
  and Shoichet]{weiss_selectivity_2018}
Weiss,~D.~R.; Karpiak,~J.; Huang,~X.-P.; Sassano,~M.~F.; Lyu,~J.; Roth,~B.~L.;
  Shoichet,~B.~K. Selectivity {Challenges} in {Docking} {Screens} for {GPCR}
  {Targets} and {Antitargets}. \emph{Journal of Medicinal Chemistry}
  \textbf{2018}, \emph{61}, 6830--6845\relax
\mciteBstWouldAddEndPuncttrue
\mciteSetBstMidEndSepPunct{\mcitedefaultmidpunct}
{\mcitedefaultendpunct}{\mcitedefaultseppunct}\relax
\EndOfBibitem
\bibitem[Chen \latin{et~al.}(2006)Chen, Lyne, Giordanetto, Lovell, and
  Li]{chen_evaluating_2006}
Chen,~H.; Lyne,~P.~D.; Giordanetto,~F.; Lovell,~T.; Li,~J. On {Evaluating}
  {Molecular}-{Docking} {Methods} for {Pose} {Prediction} and {Enrichment}
  {Factors}. \emph{Journal of Chemical Information and Modeling} \textbf{2006},
  \emph{46}, 401--415\relax
\mciteBstWouldAddEndPuncttrue
\mciteSetBstMidEndSepPunct{\mcitedefaultmidpunct}
{\mcitedefaultendpunct}{\mcitedefaultseppunct}\relax
\EndOfBibitem
\bibitem[Jain(2008)]{jain_bias_2008}
Jain,~A.~N. Bias, {Reporting}, and {Sharing}: {Computational} {Evaluations} of
  {Docking} {Methods}. \emph{Journal of Computer-Aided Molecular Design}
  \textbf{2008}, \emph{22}, 201--212\relax
\mciteBstWouldAddEndPuncttrue
\mciteSetBstMidEndSepPunct{\mcitedefaultmidpunct}
{\mcitedefaultendpunct}{\mcitedefaultseppunct}\relax
\EndOfBibitem
\bibitem[Cross \latin{et~al.}(2009)Cross, Thompson, Rai, Baber, Fan, Hu, and
  Humblet]{cross_comparison_2009}
Cross,~J.~B.; Thompson,~D.~C.; Rai,~B.~K.; Baber,~J.~C.; Fan,~K.~Y.; Hu,~Y.;
  Humblet,~C. Comparison of {Several} {Molecular} {Docking} {Programs}: {Pose}
  {Prediction} and {Virtual} {Screening} {Accuracy}. \emph{Journal of Chemical
  Information and Modeling} \textbf{2009}, \emph{49}, 1455--1474\relax
\mciteBstWouldAddEndPuncttrue
\mciteSetBstMidEndSepPunct{\mcitedefaultmidpunct}
{\mcitedefaultendpunct}{\mcitedefaultseppunct}\relax
\EndOfBibitem
\bibitem[Irwin and Shoichet(2016)Irwin, and Shoichet]{irwin_docking_2016}
Irwin,~J.~J.; Shoichet,~B.~K. Docking {Screens} for {Novel} {Ligands}
  {Conferring} {New} {Biology}. \emph{Journal of Medicinal Chemistry}
  \textbf{2016}, \emph{59}, 4103--4120\relax
\mciteBstWouldAddEndPuncttrue
\mciteSetBstMidEndSepPunct{\mcitedefaultmidpunct}
{\mcitedefaultendpunct}{\mcitedefaultseppunct}\relax
\EndOfBibitem
\bibitem[Boittier \latin{et~al.}(2020)Boittier, Tang, Buckley, Schuurs,
  Richard, and Gandhi]{boittier_assessing_2020}
Boittier,~E.~D.; Tang,~Y.~Y.; Buckley,~M.~E.; Schuurs,~Z.~P.; Richard,~D.~J.;
  Gandhi,~N.~S. Assessing {Molecular} {Docking} {Tools} to {Guide} {Targeted}
  {Drug} {Discovery} of {CD38} {Inhibitors}. \emph{International Journal of
  Molecular Sciences} \textbf{2020}, \emph{21}, 5183\relax
\mciteBstWouldAddEndPuncttrue
\mciteSetBstMidEndSepPunct{\mcitedefaultmidpunct}
{\mcitedefaultendpunct}{\mcitedefaultseppunct}\relax
\EndOfBibitem
\bibitem[Stanzione \latin{et~al.}(2021)Stanzione, Giangreco, and
  Cole]{stanzione_chapter_2021}
Stanzione,~F.; Giangreco,~I.; Cole,~J.~C. In \emph{Progress in {Medicinal}
  {Chemistry}}; Witty,~D.~R., Cox,~B., Eds.; Elsevier, 2021; Vol.~60; pp
  273--343\relax
\mciteBstWouldAddEndPuncttrue
\mciteSetBstMidEndSepPunct{\mcitedefaultmidpunct}
{\mcitedefaultendpunct}{\mcitedefaultseppunct}\relax
\EndOfBibitem
\bibitem[Ling \latin{et~al.}(2008)Ling, Singh, Chuaqui, Boriack-Sjodin,
  Corbley, Lepage, Silverio, Sun, Papadatos, Shan, Pontz, Cheung, Zhang,
  Arduini, Mead, Newman, Bowes, Josiah, and Lee]{ling_use_2008}
Ling,~L.~E. \latin{et~al.}  In \emph{Transforming {Growth} {Factor}-β in
  {Cancer} {Therapy}, {Volume} {II}: {Cancer} {Treatment} and {Therapy}};
  Jakowlew,~S.~B., Ed.; Cancer {Drug} {Discovery} and {Development}; Humana
  Press: Totowa, NJ, 2008; pp 685--696\relax
\mciteBstWouldAddEndPuncttrue
\mciteSetBstMidEndSepPunct{\mcitedefaultmidpunct}
{\mcitedefaultendpunct}{\mcitedefaultseppunct}\relax
\EndOfBibitem
\bibitem[Bajusz \latin{et~al.}(2016)Bajusz, Ferenczy, and
  Keserű]{bajusz_discovery_2016}
Bajusz,~D.; Ferenczy,~G.~G.; Keserű,~G.~M. Discovery of {Subtype} {Selective}
  {Janus} {Kinase} ({JAK}) {Inhibitors} by {Structure}-{Based} {Virtual}
  {Screening}. \emph{Journal of Chemical Information and Modeling}
  \textbf{2016}, \emph{56}, 234--247\relax
\mciteBstWouldAddEndPuncttrue
\mciteSetBstMidEndSepPunct{\mcitedefaultmidpunct}
{\mcitedefaultendpunct}{\mcitedefaultseppunct}\relax
\EndOfBibitem
\bibitem[Lyu \latin{et~al.}(2019)Lyu, Wang, Balius, Singh, Levit, Moroz,
  O’Meara, Che, Algaa, Tolmachova, Tolmachev, Shoichet, Roth, and
  Irwin]{lyu_ultra-large_2019}
Lyu,~J.; Wang,~S.; Balius,~T.~E.; Singh,~I.; Levit,~A.; Moroz,~Y.~S.;
  O’Meara,~M.~J.; Che,~T.; Algaa,~E.; Tolmachova,~K.; Tolmachev,~A.~A.;
  Shoichet,~B.~K.; Roth,~B.~L.; Irwin,~J.~J. Ultra-{Large} {Library} {Docking}
  for {Discovering} {New} {Chemotypes}. \emph{Nature} \textbf{2019},
  \emph{566}, 224--229\relax
\mciteBstWouldAddEndPuncttrue
\mciteSetBstMidEndSepPunct{\mcitedefaultmidpunct}
{\mcitedefaultendpunct}{\mcitedefaultseppunct}\relax
\EndOfBibitem
\bibitem[Gentile \latin{et~al.}(2021)Gentile, Fernandez, Ban, Ton, Mslati,
  Perez, Leblanc, Yaacoub, Gleave, Stern, Wong, Jean, Strynadka, and
  Cherkasov]{gentile_automated_2021}
Gentile,~F.; Fernandez,~M.; Ban,~F.; Ton,~A.-T.; Mslati,~H.; Perez,~C.~F.;
  Leblanc,~E.; Yaacoub,~J.~C.; Gleave,~J.; Stern,~A.; Wong,~B.; Jean,~F.;
  Strynadka,~N.; Cherkasov,~A. Automated {Discovery} of {Noncovalent}
  {Inhibitors} of {Sars}-{Cov}-2 {Main} {Protease} by {Consensus} {Deep}
  {Docking} of 40 {Billion} {Small} {Molecules}. \emph{Chemical Science}
  \textbf{2021}, \emph{12}, 15960--15974\relax
\mciteBstWouldAddEndPuncttrue
\mciteSetBstMidEndSepPunct{\mcitedefaultmidpunct}
{\mcitedefaultendpunct}{\mcitedefaultseppunct}\relax
\EndOfBibitem
\bibitem[Alon \latin{et~al.}(2021)Alon, Lyu, Braz, Tummino, Craik, O’Meara,
  Webb, Radchenko, Moroz, Huang, Liu, Roth, Irwin, Basbaum, Shoichet, and
  Kruse]{alon_structures_2021}
Alon,~A. \latin{et~al.}  Structures of the σ2 {Receptor} {Enable} {Docking}
  for {Bioactive} {Ligand} {Discovery}. \emph{Nature} \textbf{2021},
  \emph{600}, 759--764\relax
\mciteBstWouldAddEndPuncttrue
\mciteSetBstMidEndSepPunct{\mcitedefaultmidpunct}
{\mcitedefaultendpunct}{\mcitedefaultseppunct}\relax
\EndOfBibitem
\bibitem[Tingle and Irwin(2023)Tingle, and Irwin]{tingle_large-scale_2023}
Tingle,~B.~I.; Irwin,~J.~J. Large-{Scale} {Docking} in the {Cloud}.
  \emph{Journal of Chemical Information and Modeling} \textbf{2023}, \emph{63},
  2735--2741\relax
\mciteBstWouldAddEndPuncttrue
\mciteSetBstMidEndSepPunct{\mcitedefaultmidpunct}
{\mcitedefaultendpunct}{\mcitedefaultseppunct}\relax
\EndOfBibitem
\bibitem[Garnett \latin{et~al.}(2015)Garnett, Gärtner, Vogt, and
  Bajorath]{garnett_introducing_2015}
Garnett,~R.; Gärtner,~T.; Vogt,~M.; Bajorath,~J. Introducing the ‘{Active}
  {Search}’ {Method} for {Iterative} {Virtual} {Screening}. \emph{Journal of
  Computer-Aided Molecular Design} \textbf{2015}, \emph{29}, 305--314\relax
\mciteBstWouldAddEndPuncttrue
\mciteSetBstMidEndSepPunct{\mcitedefaultmidpunct}
{\mcitedefaultendpunct}{\mcitedefaultseppunct}\relax
\EndOfBibitem
\bibitem[Smith \latin{et~al.}(2018)Smith, Nebgen, Lubbers, Isayev, and
  Roitberg]{smith_less_2018}
Smith,~J.~S.; Nebgen,~B.; Lubbers,~N.; Isayev,~O.; Roitberg,~A.~E. Less {Is}
  {More}: {Sampling} {Chemical} {Space} with {Active} {Learning}. \emph{The
  Journal of Chemical Physics} \textbf{2018}, \emph{148}, 241733\relax
\mciteBstWouldAddEndPuncttrue
\mciteSetBstMidEndSepPunct{\mcitedefaultmidpunct}
{\mcitedefaultendpunct}{\mcitedefaultseppunct}\relax
\EndOfBibitem
\bibitem[Gentile \latin{et~al.}(2020)Gentile, Agrawal, Hsing, Ton, Ban,
  Norinder, Gleave, and Cherkasov]{gentile_deep_2020}
Gentile,~F.; Agrawal,~V.; Hsing,~M.; Ton,~A.-T.; Ban,~F.; Norinder,~U.;
  Gleave,~M.~E.; Cherkasov,~A. Deep {Docking}: {A} {Deep} {Learning} {Platform}
  for {Augmentation} of {Structure} {Based} {Drug} {Discovery}. \emph{ACS
  Central Science} \textbf{2020}, \emph{6}, 939--949\relax
\mciteBstWouldAddEndPuncttrue
\mciteSetBstMidEndSepPunct{\mcitedefaultmidpunct}
{\mcitedefaultendpunct}{\mcitedefaultseppunct}\relax
\EndOfBibitem
\bibitem[Graff \latin{et~al.}(2021)Graff, Shakhnovich, and
  Coley]{graff_accelerating_2021}
Graff,~D.~E.; Shakhnovich,~E.~I.; Coley,~C.~W. Accelerating {High}-{Throughput}
  {Virtual} {Screening} {Through} {Molecular} {Pool}-{Based} {Active}
  {Learning}. \emph{Chemical Science} \textbf{2021}, \emph{12},
  7866--7881\relax
\mciteBstWouldAddEndPuncttrue
\mciteSetBstMidEndSepPunct{\mcitedefaultmidpunct}
{\mcitedefaultendpunct}{\mcitedefaultseppunct}\relax
\EndOfBibitem
\bibitem[Yang \latin{et~al.}(2021)Yang, Yao, Repasky, Leswing, Abel, Shoichet,
  and Jerome]{yang_efficient_2021}
Yang,~Y.; Yao,~K.; Repasky,~M.~P.; Leswing,~K.; Abel,~R.; Shoichet,~B.~K.;
  Jerome,~S.~V. Efficient {Exploration} of {Chemical} {Space} with {Docking}
  and {Deep} {Learning}. \emph{Journal of Chemical Theory and Computation}
  \textbf{2021}, \emph{17}, 7106--7119\relax
\mciteBstWouldAddEndPuncttrue
\mciteSetBstMidEndSepPunct{\mcitedefaultmidpunct}
{\mcitedefaultendpunct}{\mcitedefaultseppunct}\relax
\EndOfBibitem
\bibitem[Mehta \latin{et~al.}(2021)Mehta, Laghuvarapu, Pathak, Sethi, Alvala,
  and Priyakumar]{mehta_memes_2021}
Mehta,~S.; Laghuvarapu,~S.; Pathak,~Y.; Sethi,~A.; Alvala,~M.;
  Priyakumar,~U.~D. {MEMES}: {Machine} {Learning} {Framework} for {Enhanced}
  {Molecular} {Screening}. \emph{Chemical Science} \textbf{2021}, \emph{12},
  11710--11721\relax
\mciteBstWouldAddEndPuncttrue
\mciteSetBstMidEndSepPunct{\mcitedefaultmidpunct}
{\mcitedefaultendpunct}{\mcitedefaultseppunct}\relax
\EndOfBibitem
\bibitem[Graff \latin{et~al.}(2022)Graff, Aldeghi, Morrone, Jordan,
  Pyzer-Knapp, and Coley]{graff_self-focusing_2022}
Graff,~D.~E.; Aldeghi,~M.; Morrone,~J.~A.; Jordan,~K.~E.; Pyzer-Knapp,~E.~O.;
  Coley,~C.~W. Self-{Focusing} {Virtual} {Screening} with {Active} {Design}
  {Space} {Pruning}. \emph{Journal of Chemical Information and Modeling}
  \textbf{2022}, \emph{62}, 3854--3862\relax
\mciteBstWouldAddEndPuncttrue
\mciteSetBstMidEndSepPunct{\mcitedefaultmidpunct}
{\mcitedefaultendpunct}{\mcitedefaultseppunct}\relax
\EndOfBibitem
\bibitem[Thompson \latin{et~al.}(2022)Thompson, Walters, Feng, Pabon, Xu,
  Maser, Goldman, Moustakas, Schmidt, and York]{thompson_optimizing_2022}
Thompson,~J.; Walters,~W.~P.; Feng,~J.~A.; Pabon,~N.~A.; Xu,~H.; Maser,~M.;
  Goldman,~B.~B.; Moustakas,~D.; Schmidt,~M.; York,~F. Optimizing {Active}
  {Learning} for {Free} {Energy} {Calculations}. \emph{Artificial Intelligence
  in the Life Sciences} \textbf{2022}, \emph{2}, 100050\relax
\mciteBstWouldAddEndPuncttrue
\mciteSetBstMidEndSepPunct{\mcitedefaultmidpunct}
{\mcitedefaultendpunct}{\mcitedefaultseppunct}\relax
\EndOfBibitem
\bibitem[Wildman and Crippen(1999)Wildman, and
  Crippen]{wildman_prediction_1999}
Wildman,~S.~A.; Crippen,~G.~M. Prediction of {Physicochemical} {Parameters} by
  {Atomic} {Contributions}. \emph{Journal of Chemical Information and Computer
  Sciences} \textbf{1999}, \emph{39}, 868--873\relax
\mciteBstWouldAddEndPuncttrue
\mciteSetBstMidEndSepPunct{\mcitedefaultmidpunct}
{\mcitedefaultendpunct}{\mcitedefaultseppunct}\relax
\EndOfBibitem
\bibitem[Ertl and Schuffenhauer(2009)Ertl, and
  Schuffenhauer]{ertl_estimation_2009}
Ertl,~P.; Schuffenhauer,~A. Estimation of {Synthetic} {Accessibility} {Score}
  of {Drug}-like {Molecules} {Based} on {Molecular} {Complexity} and {Fragment}
  {Contributions}. \emph{Journal of Cheminformatics} \textbf{2009}, \emph{1},
  8\relax
\mciteBstWouldAddEndPuncttrue
\mciteSetBstMidEndSepPunct{\mcitedefaultmidpunct}
{\mcitedefaultendpunct}{\mcitedefaultseppunct}\relax
\EndOfBibitem
\bibitem[noa()]{noauthor_enamine_nodate}
Enamine {Screening} {Collections}.
  \url{https://enamine.net/compound-collections/screening-collection}\relax
\mciteBstWouldAddEndPuncttrue
\mciteSetBstMidEndSepPunct{\mcitedefaultmidpunct}
{\mcitedefaultendpunct}{\mcitedefaultseppunct}\relax
\EndOfBibitem
\bibitem[Fromer and Coley(2023)Fromer, and Coley]{fromer_computer-aided_2023}
Fromer,~J.~C.; Coley,~C.~W. Computer-{Aided} {Multi}-{Objective} {Optimization}
  in {Small} {Molecule} {Discovery}. \emph{Patterns} \textbf{2023}, \emph{4},
  100678\relax
\mciteBstWouldAddEndPuncttrue
\mciteSetBstMidEndSepPunct{\mcitedefaultmidpunct}
{\mcitedefaultendpunct}{\mcitedefaultseppunct}\relax
\EndOfBibitem
\bibitem[Janet \latin{et~al.}(2020)Janet, Ramesh, Duan, and
  Kulik]{janet_accurate_2020}
Janet,~J.~P.; Ramesh,~S.; Duan,~C.; Kulik,~H.~J. Accurate {Multiobjective}
  {Design} in a {Space} of {Millions} of {Transition} {Metal} {Complexes} with
  {Neural}-{Network}-{Driven} {Efficient} {Global} {Optimization}. \emph{ACS
  Central Science} \textbf{2020}, \emph{6}, 513--524\relax
\mciteBstWouldAddEndPuncttrue
\mciteSetBstMidEndSepPunct{\mcitedefaultmidpunct}
{\mcitedefaultendpunct}{\mcitedefaultseppunct}\relax
\EndOfBibitem
\bibitem[Agarwal \latin{et~al.}(2021)Agarwal, Doan, Robertson, Zhang, and
  Assary]{agarwal_discovery_2021}
Agarwal,~G.; Doan,~H.~A.; Robertson,~L.~A.; Zhang,~L.; Assary,~R.~S. Discovery
  of {Energy} {Storage} {Molecular} {Materials} {Using} {Quantum}
  {Chemistry}-{Guided} {Multiobjective} {Bayesian} {Optimization}.
  \emph{Chemistry of Materials} \textbf{2021}, \emph{33}, 8133--8144\relax
\mciteBstWouldAddEndPuncttrue
\mciteSetBstMidEndSepPunct{\mcitedefaultmidpunct}
{\mcitedefaultendpunct}{\mcitedefaultseppunct}\relax
\EndOfBibitem
\bibitem[Gopakumar \latin{et~al.}(2018)Gopakumar, Balachandran, Xue,
  Gubernatis, and Lookman]{gopakumar_multi-objective_2018}
Gopakumar,~A.~M.; Balachandran,~P.~V.; Xue,~D.; Gubernatis,~J.~E.; Lookman,~T.
  Multi-objective {Optimization} for {Materials} {Discovery} via {Adaptive}
  {Design}. \emph{Scientific Reports} \textbf{2018}, \emph{8}, 3738\relax
\mciteBstWouldAddEndPuncttrue
\mciteSetBstMidEndSepPunct{\mcitedefaultmidpunct}
{\mcitedefaultendpunct}{\mcitedefaultseppunct}\relax
\EndOfBibitem
\bibitem[del Rosario \latin{et~al.}(2020)del Rosario, Rupp, Kim, Antono, and
  Ling]{del_rosario_assessing_2020}
del Rosario,~Z.; Rupp,~M.; Kim,~Y.; Antono,~E.; Ling,~J. Assessing the
  {Frontier}: {Active} {Learning}, {Model} {Accuracy}, and {Multi}-{Objective}
  {Candidate} {Discovery} and {Optimization}. \emph{The Journal of Chemical
  Physics} \textbf{2020}, \emph{153}, 024112\relax
\mciteBstWouldAddEndPuncttrue
\mciteSetBstMidEndSepPunct{\mcitedefaultmidpunct}
{\mcitedefaultendpunct}{\mcitedefaultseppunct}\relax
\EndOfBibitem
\bibitem[Keane(2006)]{keane_statistical_2006}
Keane,~A.~J. Statistical {Improvement} {Criteria} for {Use} in {Multiobjective}
  {Design} {Optimization}. \emph{AIAA Journal} \textbf{2006}, \emph{44},
  879--891\relax
\mciteBstWouldAddEndPuncttrue
\mciteSetBstMidEndSepPunct{\mcitedefaultmidpunct}
{\mcitedefaultendpunct}{\mcitedefaultseppunct}\relax
\EndOfBibitem
\bibitem[Paria \latin{et~al.}(2020)Paria, Kandasamy, and
  Póczos]{paria_flexible_2020}
Paria,~B.; Kandasamy,~K.; Póczos,~B. A {Flexible} {Framework} for
  {Multi}-{Objective} {Bayesian} {Optimization} using {Random}
  {Scalarizations}. Proceedings of {The} 35th {Uncertainty} in {Artificial}
  {Intelligence} {Conference}. 2020; pp 766--776\relax
\mciteBstWouldAddEndPuncttrue
\mciteSetBstMidEndSepPunct{\mcitedefaultmidpunct}
{\mcitedefaultendpunct}{\mcitedefaultseppunct}\relax
\EndOfBibitem
\bibitem[Zhang and Golovin(2020)Zhang, and Golovin]{zhang_random_2020}
Zhang,~R.; Golovin,~D. Random {Hypervolume} {Scalarizations} for {Provable}
  {Multi}-{Objective} {Black} {Box} {Optimization}. Proceedings of the 37th
  {International} {Conference} on {Machine} {Learning}. 2020; pp
  11096--11105\relax
\mciteBstWouldAddEndPuncttrue
\mciteSetBstMidEndSepPunct{\mcitedefaultmidpunct}
{\mcitedefaultendpunct}{\mcitedefaultseppunct}\relax
\EndOfBibitem
\bibitem[Steuer and Choo(1983)Steuer, and Choo]{steuer_interactive_1983}
Steuer,~R.~E.; Choo,~E.-U. An {Interactive} {Weighted} {Tchebycheff}
  {Procedure} for {Multiple} {Objective} {Programming}. \emph{Mathematical
  Programming} \textbf{1983}, \emph{26}, 326--344\relax
\mciteBstWouldAddEndPuncttrue
\mciteSetBstMidEndSepPunct{\mcitedefaultmidpunct}
{\mcitedefaultendpunct}{\mcitedefaultseppunct}\relax
\EndOfBibitem
\bibitem[Giagkiozis and Fleming(2015)Giagkiozis, and
  Fleming]{giagkiozis_methods_2015}
Giagkiozis,~I.; Fleming,~P.~J. Methods for {Multi}-{Objective} {Optimization}:
  {An} {Analysis}. \emph{Information Sciences} \textbf{2015}, \emph{293},
  338--350\relax
\mciteBstWouldAddEndPuncttrue
\mciteSetBstMidEndSepPunct{\mcitedefaultmidpunct}
{\mcitedefaultendpunct}{\mcitedefaultseppunct}\relax
\EndOfBibitem
\bibitem[Kushner(1964)]{kushner_new_1964}
Kushner,~H.~J. A {New} {Method} of {Locating} the {Maximum} {Point} of an
  {Arbitrary} {Multipeak} {Curve} in the {Presence} of {Noise}. \emph{Journal
  of Basic Engineering} \textbf{1964}, \emph{86}, 97--106\relax
\mciteBstWouldAddEndPuncttrue
\mciteSetBstMidEndSepPunct{\mcitedefaultmidpunct}
{\mcitedefaultendpunct}{\mcitedefaultseppunct}\relax
\EndOfBibitem
\bibitem[Močkus(1975)]{mockus_bayesian_1975}
Močkus,~J. On {Bayesian} {Methods} for {Seeking} the {Extremum}. Optimization
  {Techniques} {IFIP} {Technical} {Conference} {Novosibirsk}, {July} 1–7,
  1974. Berlin, Heidelberg, 1975; pp 400--404\relax
\mciteBstWouldAddEndPuncttrue
\mciteSetBstMidEndSepPunct{\mcitedefaultmidpunct}
{\mcitedefaultendpunct}{\mcitedefaultseppunct}\relax
\EndOfBibitem
\bibitem[Srinivas \latin{et~al.}(2012)Srinivas, Krause, Kakade, and
  Seeger]{srinivas_information-theoretic_2012}
Srinivas,~N.; Krause,~A.; Kakade,~S.~M.; Seeger,~M.~W. Information-{Theoretic}
  {Regret} {Bounds} for {Gaussian} {Process} {Optimization} in the {Bandit}
  {Setting}. \emph{IEEE Transactions on Information Theory} \textbf{2012},
  \emph{58}, 3250--3265\relax
\mciteBstWouldAddEndPuncttrue
\mciteSetBstMidEndSepPunct{\mcitedefaultmidpunct}
{\mcitedefaultendpunct}{\mcitedefaultseppunct}\relax
\EndOfBibitem
\bibitem[Lin(1976)]{lin_three_1976}
Lin,~J.~G. In \emph{Directions in {Large}-{Scale} {Systems}: {Many}-{Person}
  {Optimization} and {Decentralized} {Control}}; Ho,~Y.~C., Mitter,~S.~K.,
  Eds.; Springer US: Boston, MA, 1976; pp 117--138\relax
\mciteBstWouldAddEndPuncttrue
\mciteSetBstMidEndSepPunct{\mcitedefaultmidpunct}
{\mcitedefaultendpunct}{\mcitedefaultseppunct}\relax
\EndOfBibitem
\bibitem[Hu \latin{et~al.}(2023)Hu, Xian, Wu, Fan, Yin, and
  Zhao]{hu_revisiting_2023}
Hu,~Y.; Xian,~R.; Wu,~Q.; Fan,~Q.; Yin,~L.; Zhao,~H. Revisiting {Scalarization}
  in {Multi}-{Task} {Learning}: {A} {Theoretical} {Perspective}. 2023;
  \url{http://arxiv.org/abs/2308.13985}\relax
\mciteBstWouldAddEndPuncttrue
\mciteSetBstMidEndSepPunct{\mcitedefaultmidpunct}
{\mcitedefaultendpunct}{\mcitedefaultseppunct}\relax
\EndOfBibitem
\bibitem[Srinivas and Deb(1994)Srinivas, and
  Deb]{srinivas_muiltiobjective_1994}
Srinivas,~N.; Deb,~K. Muiltiobjective {Optimization} {Using} {Nondominated}
  {Sorting} in {Genetic} {Algorithms}. \emph{Evolutionary Computation}
  \textbf{1994}, \emph{2}, 221--248\relax
\mciteBstWouldAddEndPuncttrue
\mciteSetBstMidEndSepPunct{\mcitedefaultmidpunct}
{\mcitedefaultendpunct}{\mcitedefaultseppunct}\relax
\EndOfBibitem
\bibitem[Deb \latin{et~al.}(2002)Deb, Pratap, Agarwal, and
  Meyarivan]{deb_fast_2002}
Deb,~K.; Pratap,~A.; Agarwal,~S.; Meyarivan,~T. A {Fast} and {Elitist}
  {Multiobjective} {Genetic} {Algorithm}: {NSGA}-{II}. \emph{IEEE Transactions
  on Evolutionary Computation} \textbf{2002}, \emph{6}, 182--197\relax
\mciteBstWouldAddEndPuncttrue
\mciteSetBstMidEndSepPunct{\mcitedefaultmidpunct}
{\mcitedefaultendpunct}{\mcitedefaultseppunct}\relax
\EndOfBibitem
\bibitem[Drugan and Nowe(2013)Drugan, and Nowe]{drugan_designing_2013}
Drugan,~M.~M.; Nowe,~A. Designing {Multi}-{Objective} {Multi}-{Armed} {Bandits}
  {Algorithms}: {A} {Study}. The 2013 {International} {Joint} {Conference} on
  {Neural} {Networks} ({IJCNN}). 2013; pp 1--8\relax
\mciteBstWouldAddEndPuncttrue
\mciteSetBstMidEndSepPunct{\mcitedefaultmidpunct}
{\mcitedefaultendpunct}{\mcitedefaultseppunct}\relax
\EndOfBibitem
\bibitem[Gong \latin{et~al.}(2019)Gong, Lee, Stephenson, Renduchintala, Padhy,
  Ndirango, Keskin, and Elibol]{gong_comparison_2019}
Gong,~T.; Lee,~T.; Stephenson,~C.; Renduchintala,~V.; Padhy,~S.; Ndirango,~A.;
  Keskin,~G.; Elibol,~O.~H. A {Comparison} of {Loss} {Weighting} {Strategies}
  for {Multi} task {Learning} in {Deep} {Neural} {Networks}. \emph{IEEE Access}
  \textbf{2019}, \emph{7}, 141627--141632\relax
\mciteBstWouldAddEndPuncttrue
\mciteSetBstMidEndSepPunct{\mcitedefaultmidpunct}
{\mcitedefaultendpunct}{\mcitedefaultseppunct}\relax
\EndOfBibitem
\bibitem[Bellamy \latin{et~al.}(2022)Bellamy, Rehim, Orhobor, and
  King]{bellamy_batched_2022}
Bellamy,~H.; Rehim,~A.~A.; Orhobor,~O.~I.; King,~R. Batched {Bayesian}
  {Optimization} for {Drug} {Design} in {Noisy} {Environments}. \emph{Journal
  of Chemical Information and Modeling} \textbf{2022}, \emph{62},
  3970--3981\relax
\mciteBstWouldAddEndPuncttrue
\mciteSetBstMidEndSepPunct{\mcitedefaultmidpunct}
{\mcitedefaultendpunct}{\mcitedefaultseppunct}\relax
\EndOfBibitem
\bibitem[Ginsbourger \latin{et~al.}(2010)Ginsbourger, Le~Riche, and
  Carraro]{ginsbourger_kriging_2010}
Ginsbourger,~D.; Le~Riche,~R.; Carraro,~L. In \emph{Computational
  {Intelligence} in {Expensive} {Optimization} {Problems}}; Tenne,~Y.,
  Goh,~C.-K., Eds.; Adaptation {Learning} and {Optimization}; Springer: Berlin,
  Heidelberg, 2010; pp 131--162\relax
\mciteBstWouldAddEndPuncttrue
\mciteSetBstMidEndSepPunct{\mcitedefaultmidpunct}
{\mcitedefaultendpunct}{\mcitedefaultseppunct}\relax
\EndOfBibitem
\bibitem[Snoek \latin{et~al.}(2012)Snoek, Larochelle, and
  Adams]{snoek_practical_2012}
Snoek,~J.; Larochelle,~H.; Adams,~R.~P. Practical {Bayesian} {Optimization} of
  {Machine} {Learning} {Algorithms}. Advances in {Neural} {Information}
  {Processing} {Systems}. 2012\relax
\mciteBstWouldAddEndPuncttrue
\mciteSetBstMidEndSepPunct{\mcitedefaultmidpunct}
{\mcitedefaultendpunct}{\mcitedefaultseppunct}\relax
\EndOfBibitem
\bibitem[Janusevskis \latin{et~al.}(2012)Janusevskis, Le~Riche, Ginsbourger,
  and Girdziusas]{janusevskis_expected_2012}
Janusevskis,~J.; Le~Riche,~R.; Ginsbourger,~D.; Girdziusas,~R. Expected
  {Improvements} for the {Asynchronous} {Parallel} {Global} {Optimization} of
  {Expensive} {Functions}: {Potentials} and {Challenges}. Learning and
  {Intelligent} {Optimization}. Berlin, Heidelberg, 2012; pp 413--418\relax
\mciteBstWouldAddEndPuncttrue
\mciteSetBstMidEndSepPunct{\mcitedefaultmidpunct}
{\mcitedefaultendpunct}{\mcitedefaultseppunct}\relax
\EndOfBibitem
\bibitem[Chevalier and Ginsbourger(2013)Chevalier, and
  Ginsbourger]{chevalier_fast_2013}
Chevalier,~C.; Ginsbourger,~D. Fast {Computation} of the {Multi}-{Points}
  {Expected} {Improvement} with {Applications} in {Batch} {Selection}. Learning
  and {Intelligent} {Optimization}. Berlin, Heidelberg, 2013; pp 59--69\relax
\mciteBstWouldAddEndPuncttrue
\mciteSetBstMidEndSepPunct{\mcitedefaultmidpunct}
{\mcitedefaultendpunct}{\mcitedefaultseppunct}\relax
\EndOfBibitem
\bibitem[Jiang \latin{et~al.}(2017)Jiang, Malkomes, Converse, Shofner, Moseley,
  and Garnett]{jiang_efficient_2017}
Jiang,~S.; Malkomes,~G.; Converse,~G.; Shofner,~A.; Moseley,~B.; Garnett,~R.
  Efficient {Nonmyopic} {Active} {Search}. Proceedings of the 34th
  {International} {Conference} on {Machine} {Learning}. 2017; pp
  1714--1723\relax
\mciteBstWouldAddEndPuncttrue
\mciteSetBstMidEndSepPunct{\mcitedefaultmidpunct}
{\mcitedefaultendpunct}{\mcitedefaultseppunct}\relax
\EndOfBibitem
\bibitem[Tran \latin{et~al.}(2019)Tran, Sun, Furlan, Pagalthivarthi,
  Visintainer, and Wang]{tran_pbo-2gp-3b_2019}
Tran,~A.; Sun,~J.; Furlan,~J.~M.; Pagalthivarthi,~K.~V.; Visintainer,~R.~J.;
  Wang,~Y. {pBO}-{2GP}-{3B}: {A} {Batch} {Parallel} {Known}/{Unknown}
  {Constrained} {Bayesian} {Optimization} with {Feasibility} {Classification}
  and {Its} {Applications} in {Computational} {Fluid} {Dynamics}.
  \emph{Computer Methods in Applied Mechanics and Engineering} \textbf{2019},
  \emph{347}, 827--852\relax
\mciteBstWouldAddEndPuncttrue
\mciteSetBstMidEndSepPunct{\mcitedefaultmidpunct}
{\mcitedefaultendpunct}{\mcitedefaultseppunct}\relax
\EndOfBibitem
\bibitem[Azimi \latin{et~al.}(2010)Azimi, Fern, and Fern]{azimi_batch_2010}
Azimi,~J.; Fern,~A.; Fern,~X. Batch {Bayesian} {Optimization} via {Simulation}
  {Matching}. Advances in {Neural} {Information} {Processing} {Systems}.
  2010\relax
\mciteBstWouldAddEndPuncttrue
\mciteSetBstMidEndSepPunct{\mcitedefaultmidpunct}
{\mcitedefaultendpunct}{\mcitedefaultseppunct}\relax
\EndOfBibitem
\bibitem[Gonzalez \latin{et~al.}(2016)Gonzalez, Dai, Hennig, and
  Lawrence]{gonzalez_batch_2016}
Gonzalez,~J.; Dai,~Z.; Hennig,~P.; Lawrence,~N. Batch {Bayesian} {Optimization}
  via {Local} {Penalization}. Proceedings of the 19th {International}
  {Conference} on {Artificial} {Intelligence} and {Statistics}. 2016; pp
  648--657\relax
\mciteBstWouldAddEndPuncttrue
\mciteSetBstMidEndSepPunct{\mcitedefaultmidpunct}
{\mcitedefaultendpunct}{\mcitedefaultseppunct}\relax
\EndOfBibitem
\bibitem[Konakovic~Lukovic \latin{et~al.}(2020)Konakovic~Lukovic, Tian, and
  Matusik]{konakovic_lukovic_diversity-guided_2020}
Konakovic~Lukovic,~M.; Tian,~Y.; Matusik,~W. Diversity-{Guided}
  {Multi}-{Objective} {Bayesian} {Optimization} {With} {Batch} {Evaluations}.
  Advances in {Neural} {Information} {Processing} {Systems}. 2020; pp
  17708--17720\relax
\mciteBstWouldAddEndPuncttrue
\mciteSetBstMidEndSepPunct{\mcitedefaultmidpunct}
{\mcitedefaultendpunct}{\mcitedefaultseppunct}\relax
\EndOfBibitem
\bibitem[Citovsky \latin{et~al.}(2021)Citovsky, DeSalvo, Gentile, Karydas,
  Rajagopalan, Rostamizadeh, and Kumar]{citovsky_batch_2021}
Citovsky,~G.; DeSalvo,~G.; Gentile,~C.; Karydas,~L.; Rajagopalan,~A.;
  Rostamizadeh,~A.; Kumar,~S. Batch {Active} {Learning} at {Scale}. Advances in
  {Neural} {Information} {Processing} {Systems}. 2021; pp 11933--11944\relax
\mciteBstWouldAddEndPuncttrue
\mciteSetBstMidEndSepPunct{\mcitedefaultmidpunct}
{\mcitedefaultendpunct}{\mcitedefaultseppunct}\relax
\EndOfBibitem
\bibitem[Maus \latin{et~al.}(2023)Maus, Wu, Eriksson, and
  Gardner]{maus_discovering_2023}
Maus,~N.; Wu,~K.; Eriksson,~D.; Gardner,~J. Discovering {Many} {Diverse}
  {Solutions} with {Bayesian} {Optimization}. Proceedings of the 26th
  {International} {Conference} on {Artificial} {Intelligence} and {Statistics}
  ({AISTATS}). Valencia, Spain, 2023\relax
\mciteBstWouldAddEndPuncttrue
\mciteSetBstMidEndSepPunct{\mcitedefaultmidpunct}
{\mcitedefaultendpunct}{\mcitedefaultseppunct}\relax
\EndOfBibitem
\bibitem[González and Zavala(2023)González, and Zavala]{gonzalez_new_2023}
González,~L.~D.; Zavala,~V.~M. New {Paradigms} for {Exploiting} {Parallel}
  {Experiments} in {Bayesian} {Optimization}. \emph{Computers \& Chemical
  Engineering} \textbf{2023}, \emph{170}, 108110\relax
\mciteBstWouldAddEndPuncttrue
\mciteSetBstMidEndSepPunct{\mcitedefaultmidpunct}
{\mcitedefaultendpunct}{\mcitedefaultseppunct}\relax
\EndOfBibitem
\bibitem[García-Ortegón \latin{et~al.}(2022)García-Ortegón, Simm, Tripp,
  Hernández-Lobato, Bender, and Bacallado]{garcia-ortegon_dockstring_2022}
García-Ortegón,~M.; Simm,~G. N.~C.; Tripp,~A.~J.; Hernández-Lobato,~J.~M.;
  Bender,~A.; Bacallado,~S. {DOCKSTRING}: {Easy} {Molecular} {Docking} {Yields}
  {Better} {Benchmarks} for {Ligand} {Design}. \emph{Journal of Chemical
  Information and Modeling} \textbf{2022}, \emph{62}, 3486--3502\relax
\mciteBstWouldAddEndPuncttrue
\mciteSetBstMidEndSepPunct{\mcitedefaultmidpunct}
{\mcitedefaultendpunct}{\mcitedefaultseppunct}\relax
\EndOfBibitem
\bibitem[Watson \latin{et~al.}(2012)Watson, Loiseau, Ingallinesi, Millan,
  Marsden, and Fone]{watson_selective_2012}
Watson,~D.~J.; Loiseau,~F.; Ingallinesi,~M.; Millan,~M.~J.; Marsden,~C.~A.;
  Fone,~K.~C. Selective {Blockade} of {Dopamine} {D3} {Receptors} {Enhances}
  while {D2} {Receptor} {Antagonism} {Impairs} {Social} {Novelty}
  {Discrimination} and {Novel} {Object} {Recognition} in {Rats}: {A} {Key}
  {Role} for the {Prefrontal} {Cortex}. \emph{Neuropsychopharmacology}
  \textbf{2012}, \emph{37}, 770--786\relax
\mciteBstWouldAddEndPuncttrue
\mciteSetBstMidEndSepPunct{\mcitedefaultmidpunct}
{\mcitedefaultendpunct}{\mcitedefaultseppunct}\relax
\EndOfBibitem
\bibitem[Williford \latin{et~al.}(2021)Williford, Libby, Ayokanmbi, Otamias,
  Gordillo, Gordon, Cooper, Redmann, Li, Griguer, Zhang, Napierala, Ananthan,
  and Hjelmeland]{williford_novel_2021}
Williford,~S.~E.; Libby,~C.~J.; Ayokanmbi,~A.; Otamias,~A.; Gordillo,~J.~J.;
  Gordon,~E.~R.; Cooper,~S.~J.; Redmann,~M.; Li,~Y.; Griguer,~C.; Zhang,~J.;
  Napierala,~M.; Ananthan,~S.; Hjelmeland,~A.~B. Novel {Dopamine} {Receptor} 3
  {Antagonists} {Inhibit} the {Growth} of {Primary} and {Temozolomide}
  {Resistant} {Glioblastoma} {Cells}. \emph{PLoS ONE} \textbf{2021}, \emph{16},
  e0250649\relax
\mciteBstWouldAddEndPuncttrue
\mciteSetBstMidEndSepPunct{\mcitedefaultmidpunct}
{\mcitedefaultendpunct}{\mcitedefaultseppunct}\relax
\EndOfBibitem
\bibitem[Bonifazi \latin{et~al.}(2023)Bonifazi, Saab, Sanchez, Nazarova, Zaidi,
  Jahan, Katritch, Canals, Lane, and Newman]{bonifazi_pharmacological_2023}
Bonifazi,~A.; Saab,~E.; Sanchez,~J.; Nazarova,~A.~L.; Zaidi,~S.~A.; Jahan,~K.;
  Katritch,~V.; Canals,~M.; Lane,~J.~R.; Newman,~A.~H. Pharmacological and
  {Physicochemical} {Properties} {Optimization} for {Dual}-{Target} {Dopamine}
  {D3} ({D3R}) and μ-{Opioid} ({MOR}) {Receptor} {Ligands} as {Potentially}
  {Safer} {Analgesics}. \emph{Journal of Medicinal Chemistry} \textbf{2023},
  \relax
\mciteBstWouldAddEndPunctfalse
\mciteSetBstMidEndSepPunct{\mcitedefaultmidpunct}
{}{\mcitedefaultseppunct}\relax
\EndOfBibitem
\bibitem[Fridman \latin{et~al.}(2010)Fridman, Scherle, Collins, Burn, Li, Li,
  Covington, Thomas, Collier, Favata, Wen, Shi, McGee, Haley, Shepard, Rodgers,
  Yeleswaram, Hollis, Newton, Metcalf, Friedman, and
  Vaddi]{fridman_selective_2010}
Fridman,~J.~S. \latin{et~al.}  Selective {Inhibition} of {JAK1} and {JAK2} {Is}
  {Efficacious} in {Rodent} {Models} of {Arthritis}: {Preclinical}
  {Characterization} of {INCB028050}. \emph{The Journal of Immunology}
  \textbf{2010}, \emph{184}, 5298--5307\relax
\mciteBstWouldAddEndPuncttrue
\mciteSetBstMidEndSepPunct{\mcitedefaultmidpunct}
{\mcitedefaultendpunct}{\mcitedefaultseppunct}\relax
\EndOfBibitem
\bibitem[Liu \latin{et~al.}(2015)Liu, Batt, Lippy, Surti, Tebben, Muckelbauer,
  Chen, An, Chang, Pokross, Yang, Wang, Burke, Carter, and
  Tino]{liu_design_2015}
Liu,~Q.; Batt,~D.~G.; Lippy,~J.~S.; Surti,~N.; Tebben,~A.~J.;
  Muckelbauer,~J.~K.; Chen,~L.; An,~Y.; Chang,~C.; Pokross,~M.; Yang,~Z.;
  Wang,~H.; Burke,~J.~R.; Carter,~P.~H.; Tino,~J.~A. Design and {Synthesis} of
  {Carbazole} {Carboxamides} as {Promising} {Inhibitors} of {Bruton}’s
  {Tyrosine} {Kinase} ({BTK}) and {Janus} {Kinase} 2 ({JAK2}). \emph{Bioorganic
  \& Medicinal Chemistry Letters} \textbf{2015}, \emph{25}, 4265--4269\relax
\mciteBstWouldAddEndPuncttrue
\mciteSetBstMidEndSepPunct{\mcitedefaultmidpunct}
{\mcitedefaultendpunct}{\mcitedefaultseppunct}\relax
\EndOfBibitem
\bibitem[Li \latin{et~al.}(2009)Li, Pourpak, and Morris]{li_inhibition_2009}
Li,~R.; Pourpak,~A.; Morris,~S.~W. Inhibition of the {Insulin}-like {Growth}
  {Factor}-1 {Receptor} ({IGF1R}) {Tyrosine} {Kinase} as a {Novel} {Cancer}
  {Therapy} {Approach}. \emph{Journal of Medicinal Chemistry} \textbf{2009},
  \emph{52}, 4981--5004\relax
\mciteBstWouldAddEndPuncttrue
\mciteSetBstMidEndSepPunct{\mcitedefaultmidpunct}
{\mcitedefaultendpunct}{\mcitedefaultseppunct}\relax
\EndOfBibitem
\bibitem[Pasha \latin{et~al.}(2022)Pasha, Jabeen, and
  Samarasinghe]{pasha_3d_2022}
Pasha,~M.~K.; Jabeen,~I.; Samarasinghe,~S. {3D} {QSAR} and {Pharmacophore}
  {Studies} on {Inhibitors} of {Insuline} {Like} {Growth} {Factor} 1 {Receptor}
  ({IGF}-{1R}) and {Insulin} {Receptor} ({IR}) as {Potential} {Anti}-{Cancer}
  {Agents}. \emph{Current Research in Chemical Biology} \textbf{2022},
  \emph{2}, 100019\relax
\mciteBstWouldAddEndPuncttrue
\mciteSetBstMidEndSepPunct{\mcitedefaultmidpunct}
{\mcitedefaultendpunct}{\mcitedefaultseppunct}\relax
\EndOfBibitem
\bibitem[Velaparthi \latin{et~al.}(2007)Velaparthi, Liu, Balasubramanian,
  Carboni, Attar, Gottardis, Li, Greer, Zoeckler, Wittman, and
  Vyas]{velaparthi_imidazole_2007}
Velaparthi,~U.; Liu,~P.; Balasubramanian,~B.; Carboni,~J.; Attar,~R.;
  Gottardis,~M.; Li,~A.; Greer,~A.; Zoeckler,~M.; Wittman,~M.~D.; Vyas,~D.
  Imidazole {Moiety} {Replacements} in the
  3-({1H}-{Benzo}[d]imidazol-2-{Yl})pyridin-2({1H})-{One} {Inhibitors} of
  {Insulin}-{Like} {Growth} {Factor} {Receptor}-1 ({IGF}-{1R}) to {Improve}
  {Cytochrome} {P450} {Profile}. \emph{Bioorganic \& Medicinal Chemistry
  Letters} \textbf{2007}, \emph{17}, 3072--3076\relax
\mciteBstWouldAddEndPuncttrue
\mciteSetBstMidEndSepPunct{\mcitedefaultmidpunct}
{\mcitedefaultendpunct}{\mcitedefaultseppunct}\relax
\EndOfBibitem
\bibitem[Zimmermann \latin{et~al.}(2008)Zimmermann, Wittman, Saulnier,
  Velaparthi, Langley, Sang, Frennesson, Carboni, Li, Greer, Gottardis, Attar,
  Yang, Balimane, Discenza, and Vyas]{zimmermann_balancing_2008}
Zimmermann,~K. \latin{et~al.}  Balancing {Oral} {Exposure} with {CYP3A4}
  {Inhibition} in {Benzimidazole}-{Based} {IGF}-{IR} {Inhibitors}.
  \emph{Bioorganic \& Medicinal Chemistry Letters} \textbf{2008}, \emph{18},
  4075--4080\relax
\mciteBstWouldAddEndPuncttrue
\mciteSetBstMidEndSepPunct{\mcitedefaultmidpunct}
{\mcitedefaultendpunct}{\mcitedefaultseppunct}\relax
\EndOfBibitem
\bibitem[Lin and Lu(1998)Lin, and Lu]{lin_inhibition_1998}
Lin,~J.~H.; Lu,~A. Y.~H. Inhibition and {Induction} of {Cytochrome} {P450} and
  the {Clinical} {Implications}. \emph{Clinical Pharmacokinetics}
  \textbf{1998}, \emph{35}, 361--390\relax
\mciteBstWouldAddEndPuncttrue
\mciteSetBstMidEndSepPunct{\mcitedefaultmidpunct}
{\mcitedefaultendpunct}{\mcitedefaultseppunct}\relax
\EndOfBibitem
\bibitem[Lynch and Price(2007)Lynch, and Price]{lynch_effect_2007}
Lynch,~T.; Price,~A. The {Effect} of {Cytochrome} {P450} {Metabolism} on {Drug}
  {Response}, {Interactions}, and {Adverse} {Effects}. \emph{American Family
  Physician} \textbf{2007}, \emph{76}, 391--396\relax
\mciteBstWouldAddEndPuncttrue
\mciteSetBstMidEndSepPunct{\mcitedefaultmidpunct}
{\mcitedefaultendpunct}{\mcitedefaultseppunct}\relax
\EndOfBibitem
\bibitem[Cheng \latin{et~al.}(2011)Cheng, Yu, Zhou, Shen, Xiao, Liu, Li, Lee,
  and Tang]{cheng_insights_2011}
Cheng,~F.; Yu,~Y.; Zhou,~Y.; Shen,~Z.; Xiao,~W.; Liu,~G.; Li,~W.; Lee,~P.~W.;
  Tang,~Y. Insights into {Molecular} {Basis} of {Cytochrome} {P450}
  {Inhibitory} {Promiscuity} of {Compounds}. \emph{Journal of Chemical
  Information and Modeling} \textbf{2011}, \emph{51}, 2482--2495\relax
\mciteBstWouldAddEndPuncttrue
\mciteSetBstMidEndSepPunct{\mcitedefaultmidpunct}
{\mcitedefaultendpunct}{\mcitedefaultseppunct}\relax
\EndOfBibitem
\bibitem[Yang \latin{et~al.}(2019)Yang, Swanson, Jin, Coley, Eiden, Gao,
  Guzman-Perez, Hopper, Kelley, Mathea, Palmer, Settels, Jaakkola, Jensen, and
  Barzilay]{yang_analyzing_2019}
Yang,~K.; Swanson,~K.; Jin,~W.; Coley,~C.; Eiden,~P.; Gao,~H.;
  Guzman-Perez,~A.; Hopper,~T.; Kelley,~B.; Mathea,~M.; Palmer,~A.;
  Settels,~V.; Jaakkola,~T.; Jensen,~K.; Barzilay,~R. Analyzing {Learned}
  {Molecular} {Representations} for {Property} {Prediction}. \emph{Journal of
  Chemical Information and Modeling} \textbf{2019}, \emph{59}, 3370--3388\relax
\mciteBstWouldAddEndPuncttrue
\mciteSetBstMidEndSepPunct{\mcitedefaultmidpunct}
{\mcitedefaultendpunct}{\mcitedefaultseppunct}\relax
\EndOfBibitem
\bibitem[Heid \latin{et~al.}(2023)Heid, Greenman, Chung, Li, Graff, Vermeire,
  Wu, Green, and McGill]{heid_chemprop_2023}
Heid,~E.; Greenman,~K.~P.; Chung,~Y.; Li,~S.-C.; Graff,~D.~E.; Vermeire,~F.~H.;
  Wu,~H.; Green,~W.~H.; McGill,~C.~J. Chemprop: {A} {Machine} {Learning}
  {Package} for {Chemical} {Property} {Prediction}. 2023;
  \url{https://chemrxiv.org/engage/chemrxiv/article-details/64d1f13d4a3f7d0c0dcd836b}\relax
\mciteBstWouldAddEndPuncttrue
\mciteSetBstMidEndSepPunct{\mcitedefaultmidpunct}
{\mcitedefaultendpunct}{\mcitedefaultseppunct}\relax
\EndOfBibitem
\bibitem[Bender \latin{et~al.}(2021)Bender, Gahbauer, Luttens, Lyu, Webb,
  Stein, Fink, Balius, Carlsson, Irwin, and Shoichet]{bender_practical_2021}
Bender,~B.~J.; Gahbauer,~S.; Luttens,~A.; Lyu,~J.; Webb,~C.~M.; Stein,~R.~M.;
  Fink,~E.~A.; Balius,~T.~E.; Carlsson,~J.; Irwin,~J.~J.; Shoichet,~B.~K. A
  {Practical} {Guide} to {Large}-{Scale} {Docking}. \emph{Nature Protocols}
  \textbf{2021}, \emph{16}, 4799--4832\relax
\mciteBstWouldAddEndPuncttrue
\mciteSetBstMidEndSepPunct{\mcitedefaultmidpunct}
{\mcitedefaultendpunct}{\mcitedefaultseppunct}\relax
\EndOfBibitem
\bibitem[Bemis and Murcko(1996)Bemis, and Murcko]{bemis_properties_1996}
Bemis,~G.~W.; Murcko,~M.~A. The {Properties} of {Known} {Drugs}. 1. {Molecular}
  {Frameworks}. \emph{Journal of Medicinal Chemistry} \textbf{1996}, \emph{39},
  2887--2893\relax
\mciteBstWouldAddEndPuncttrue
\mciteSetBstMidEndSepPunct{\mcitedefaultmidpunct}
{\mcitedefaultendpunct}{\mcitedefaultseppunct}\relax
\EndOfBibitem
\bibitem[McInnes \latin{et~al.}(2018)McInnes, Healy, Saul, and
  Großberger]{mcinnes_umap_2018}
McInnes,~L.; Healy,~J.; Saul,~N.; Großberger,~L. {UMAP}: {Uniform} {Manifold}
  {Approximation} and {Projection}. \emph{Journal of Open Source Software}
  \textbf{2018}, \emph{3}, 861\relax
\mciteBstWouldAddEndPuncttrue
\mciteSetBstMidEndSepPunct{\mcitedefaultmidpunct}
{\mcitedefaultendpunct}{\mcitedefaultseppunct}\relax
\EndOfBibitem
\bibitem[Tandon \latin{et~al.}(2013)Tandon, Senthil, Nithya, Pamidiboina,
  Kumar, Malik, Chaira, Diwan, Gupta, Venkataramanan, Malik, Das, Dastidar,
  Cliffe, Ray, and Bhatnagar]{tandon_rbx10080307_2013}
Tandon,~R. \latin{et~al.}  {RBx10080307}, a {Dual} {EGFR}/{IGF}-{1R}
  {Inhibitor} for {Anticancer} {Therapy}. \emph{European Journal of
  Pharmacology} \textbf{2013}, \emph{711}, 19--26\relax
\mciteBstWouldAddEndPuncttrue
\mciteSetBstMidEndSepPunct{\mcitedefaultmidpunct}
{\mcitedefaultendpunct}{\mcitedefaultseppunct}\relax
\EndOfBibitem
\bibitem[Hu \latin{et~al.}(2022)Hu, Fan, Shi, Song, Wang, He, and
  Qi]{hu_dual_2022}
Hu,~L.; Fan,~M.; Shi,~S.; Song,~X.; Wang,~F.; He,~H.; Qi,~B. Dual {Target}
  {Inhibitors} {Based} on {EGFR}: {Promising} {Anticancer} {Agents} for the
  {Treatment} of {Cancers} (2017-). \emph{European Journal of Medicinal
  Chemistry} \textbf{2022}, \emph{227}, 113963\relax
\mciteBstWouldAddEndPuncttrue
\mciteSetBstMidEndSepPunct{\mcitedefaultmidpunct}
{\mcitedefaultendpunct}{\mcitedefaultseppunct}\relax
\EndOfBibitem
\bibitem[Kang \latin{et~al.}(2022)Kang, Guo, Zhang, Guo, Zhu, and
  Guo]{kang_dual_2022}
Kang,~J.; Guo,~Z.; Zhang,~H.; Guo,~R.; Zhu,~X.; Guo,~X. Dual {Inhibition} of
  {EGFR} and {IGF}-{1R} {Signaling} {Leads} to {Enhanced} {Antitumor}
  {Efficacy} against {Esophageal} {Squamous} {Cancer}. \emph{International
  Journal of Molecular Sciences} \textbf{2022}, \emph{23}, 10382\relax
\mciteBstWouldAddEndPuncttrue
\mciteSetBstMidEndSepPunct{\mcitedefaultmidpunct}
{\mcitedefaultendpunct}{\mcitedefaultseppunct}\relax
\EndOfBibitem
\bibitem[Abourehab \latin{et~al.}(2021)Abourehab, Alqahtani, Youssif, and
  Gouda]{abourehab_globally_2021}
Abourehab,~M. A.~S.; Alqahtani,~A.~M.; Youssif,~B. G.~M.; Gouda,~A.~M. Globally
  {Approved} {EGFR} {Inhibitors}: {Insights} into {Their} {Syntheses}, {Target}
  {Kinases}, {Biological} {Activities}, {Receptor} {Interactions}, and
  {Metabolism}. \emph{Molecules} \textbf{2021}, \emph{26}, 6677\relax
\mciteBstWouldAddEndPuncttrue
\mciteSetBstMidEndSepPunct{\mcitedefaultmidpunct}
{\mcitedefaultendpunct}{\mcitedefaultseppunct}\relax
\EndOfBibitem
\bibitem[Pan \latin{et~al.}(2003)Pan, Huang, Cho, and
  MacKerell]{pan_consideration_2003}
Pan,~Y.; Huang,~N.; Cho,~S.; MacKerell,~A.~D. Consideration of molecular weight
  during compound selection in virtual target-based database screening.
  \emph{Journal of Chemical Information and Computer Sciences} \textbf{2003},
  \emph{43}, 267--272\relax
\mciteBstWouldAddEndPuncttrue
\mciteSetBstMidEndSepPunct{\mcitedefaultmidpunct}
{\mcitedefaultendpunct}{\mcitedefaultseppunct}\relax
\EndOfBibitem
\bibitem[Li \latin{et~al.}(2009)Li, Zhang, Zheng, Luo, Kang, Liu, Wang, and
  Jiang]{li_effective_2009}
Li,~H.; Zhang,~H.; Zheng,~M.; Luo,~J.; Kang,~L.; Liu,~X.; Wang,~X.; Jiang,~H.
  An effective docking strategy for virtual screening based on multi-objective
  optimization algorithm. \emph{BMC Bioinformatics} \textbf{2009}, \emph{10},
  58\relax
\mciteBstWouldAddEndPuncttrue
\mciteSetBstMidEndSepPunct{\mcitedefaultmidpunct}
{\mcitedefaultendpunct}{\mcitedefaultseppunct}\relax
\EndOfBibitem
\bibitem[Kirsch \latin{et~al.}(2019)Kirsch, van Amersfoort, and
  Gal]{kirsch_batchbald_2019}
Kirsch,~A.; van Amersfoort,~J.; Gal,~Y. {BatchBALD}: {Efficient} and {Diverse}
  {Batch} {Acquisition} for {Deep} {Bayesian} {Active} {Learning}. Advances in
  {Neural} {Information} {Processing} {Systems}. 2019\relax
\mciteBstWouldAddEndPuncttrue
\mciteSetBstMidEndSepPunct{\mcitedefaultmidpunct}
{\mcitedefaultendpunct}{\mcitedefaultseppunct}\relax
\EndOfBibitem
\bibitem[Huggins \latin{et~al.}(2012)Huggins, Sherman, and
  Tidor]{huggins_rational_2012}
Huggins,~D.~J.; Sherman,~W.; Tidor,~B. Rational {Approaches} to {Improving}
  {Selectivity} in {Drug} {Design}. \emph{Journal of Medicinal Chemistry}
  \textbf{2012}, \emph{55}, 1424--1444\relax
\mciteBstWouldAddEndPuncttrue
\mciteSetBstMidEndSepPunct{\mcitedefaultmidpunct}
{\mcitedefaultendpunct}{\mcitedefaultseppunct}\relax
\EndOfBibitem
\bibitem[Klabunde and Evers(2005)Klabunde, and Evers]{klabunde_gpcr_2005}
Klabunde,~T.; Evers,~A. {GPCR} {Antitarget} {Modeling}: {Pharmacophore}
  {Models} for {Biogenic} {Amine} {Binding} {GPCRs} to {Avoid}
  {GPCR}-{Mediated} {Side} {Effects}. \emph{ChemBioChem} \textbf{2005},
  \emph{6}, 876--889\relax
\mciteBstWouldAddEndPuncttrue
\mciteSetBstMidEndSepPunct{\mcitedefaultmidpunct}
{\mcitedefaultendpunct}{\mcitedefaultseppunct}\relax
\EndOfBibitem
\bibitem[noa()]{noauthor_rdkit_nodate}
{RDKit}: {Open}-{Source} {Cheminformatics} {Software}.
  \url{http://www.rdkit.org/}\relax
\mciteBstWouldAddEndPuncttrue
\mciteSetBstMidEndSepPunct{\mcitedefaultmidpunct}
{\mcitedefaultendpunct}{\mcitedefaultseppunct}\relax
\EndOfBibitem
\bibitem[Motoyama \latin{et~al.}(2022)Motoyama, Tamura, Yoshimi, Terayama,
  Ueno, and Tsuda]{motoyama_bayesian_2022}
Motoyama,~Y.; Tamura,~R.; Yoshimi,~K.; Terayama,~K.; Ueno,~T.; Tsuda,~K.
  Bayesian {Optimization} {Package}: {PHYSBO}. \emph{Computer Physics
  Communications} \textbf{2022}, \emph{278}, 108405\relax
\mciteBstWouldAddEndPuncttrue
\mciteSetBstMidEndSepPunct{\mcitedefaultmidpunct}
{\mcitedefaultendpunct}{\mcitedefaultseppunct}\relax
\EndOfBibitem
\bibitem[Couckuyt \latin{et~al.}(2014)Couckuyt, Deschrijver, and
  Dhaene]{couckuyt_fast_2014}
Couckuyt,~I.; Deschrijver,~D.; Dhaene,~T. Fast {Calculation} of
  {Multiobjective} {Probability} of {Improvement} and {Expected} {Improvement}
  {Criteria} for {Pareto} {Optimization}. \emph{Journal of Global Optimization}
  \textbf{2014}, \emph{60}, 575--594\relax
\mciteBstWouldAddEndPuncttrue
\mciteSetBstMidEndSepPunct{\mcitedefaultmidpunct}
{\mcitedefaultendpunct}{\mcitedefaultseppunct}\relax
\EndOfBibitem
\bibitem[Biscani and Izzo(2020)Biscani, and Izzo]{biscani_parallel_2020}
Biscani,~F.; Izzo,~D. A {Parallel} {Global} {Multiobjective} {Framework} for
  {Optimization}: {Pagmo}. \emph{Journal of Open Source Software}
  \textbf{2020}, \emph{5}, 2338\relax
\mciteBstWouldAddEndPuncttrue
\mciteSetBstMidEndSepPunct{\mcitedefaultmidpunct}
{\mcitedefaultendpunct}{\mcitedefaultseppunct}\relax
\EndOfBibitem
\bibitem[Paszke \latin{et~al.}(2019)Paszke, Gross, Massa, Lerer, Bradbury,
  Chanan, Killeen, Lin, Gimelshein, Antiga, Desmaison, Kopf, Yang, DeVito,
  Raison, Tejani, Chilamkurthy, Steiner, Fang, Bai, and
  Chintala]{paszke_pytorch_2019}
Paszke,~A. \latin{et~al.}  {PyTorch}: {An} {Imperative} {Style},
  {High}-{Performance} {Deep} {Learning} {Library}. Advances in {Neural}
  {Information} {Processing} {Systems}. 2019\relax
\mciteBstWouldAddEndPuncttrue
\mciteSetBstMidEndSepPunct{\mcitedefaultmidpunct}
{\mcitedefaultendpunct}{\mcitedefaultseppunct}\relax
\EndOfBibitem
\bibitem[Nix and Weigend(1994)Nix, and Weigend]{nix_estimating_1994}
Nix,~D.; Weigend,~A. Estimating the {Mean} and {Variance} of the {Target}
  {Probability} {Distribution}. Proceedings of 1994 {IEEE} {International}
  {Conference} on {Neural} {Networks} ({ICNN}'94). 1994; pp 55--60 vol.1\relax
\mciteBstWouldAddEndPuncttrue
\mciteSetBstMidEndSepPunct{\mcitedefaultmidpunct}
{\mcitedefaultendpunct}{\mcitedefaultseppunct}\relax
\EndOfBibitem
\bibitem[Hirschfeld \latin{et~al.}(2020)Hirschfeld, Swanson, Yang, Barzilay,
  and Coley]{hirschfeld_uncertainty_2020}
Hirschfeld,~L.; Swanson,~K.; Yang,~K.; Barzilay,~R.; Coley,~C.~W. Uncertainty
  {Quantification} {Using} {Neural} {Networks} for {Molecular} {Property}
  {Prediction}. \emph{Journal of Chemical Information and Modeling}
  \textbf{2020}, \emph{60}, 3770--3780\relax
\mciteBstWouldAddEndPuncttrue
\mciteSetBstMidEndSepPunct{\mcitedefaultmidpunct}
{\mcitedefaultendpunct}{\mcitedefaultseppunct}\relax
\EndOfBibitem
\bibitem[Carhart \latin{et~al.}(1985)Carhart, Smith, and
  Venkataraghavan]{carhart_atom_1985}
Carhart,~R.~E.; Smith,~D.~H.; Venkataraghavan,~R. Atom {Pairs} as {Molecular}
  {Features} in {Structure}-{Activity} {Studies}: {Definition} and
  {Applications}. \emph{Journal of Chemical Information and Computer Sciences}
  \textbf{1985}, \emph{25}, 64--73\relax
\mciteBstWouldAddEndPuncttrue
\mciteSetBstMidEndSepPunct{\mcitedefaultmidpunct}
{\mcitedefaultendpunct}{\mcitedefaultseppunct}\relax
\EndOfBibitem
\bibitem[Awale and Reymond(2014)Awale, and Reymond]{awale_atom_2014}
Awale,~M.; Reymond,~J.-L. Atom {Pair} {2D}-{Fingerprints} {Perceive}
  {3D}-{Molecular} {Shape} and {Pharmacophores} for {Very} {Fast} {Virtual}
  {Screening} of {ZINC} and {GDB}-17. \emph{Journal of Chemical Information and
  Modeling} \textbf{2014}, \emph{54}, 1892--1907\relax
\mciteBstWouldAddEndPuncttrue
\mciteSetBstMidEndSepPunct{\mcitedefaultmidpunct}
{\mcitedefaultendpunct}{\mcitedefaultseppunct}\relax
\EndOfBibitem
\bibitem[O’Boyle and Sayle(2016)O’Boyle, and Sayle]{oboyle_comparing_2016}
O’Boyle,~N.~M.; Sayle,~R.~A. Comparing structural fingerprints using a
  literature-based similarity benchmark. \emph{Journal of Cheminformatics}
  \textbf{2016}, \emph{8}, 36\relax
\mciteBstWouldAddEndPuncttrue
\mciteSetBstMidEndSepPunct{\mcitedefaultmidpunct}
{\mcitedefaultendpunct}{\mcitedefaultseppunct}\relax
\EndOfBibitem
\bibitem[Sculley(2010)]{sculley_web-scale_2010}
Sculley,~D. Web-{Scale} k-{Means} {Clustering}. Proceedings of the 19th
  international conference on {World} wide web. Raleigh North Carolina USA,
  2010; pp 1177--1178\relax
\mciteBstWouldAddEndPuncttrue
\mciteSetBstMidEndSepPunct{\mcitedefaultmidpunct}
{\mcitedefaultendpunct}{\mcitedefaultseppunct}\relax
\EndOfBibitem
\bibitem[Pedregosa \latin{et~al.}(2011)Pedregosa, Varoquaux, Gramfort, Michel,
  Thirion, Grisel, Blondel, Prettenhofer, Weiss, Dubourg, Vanderplas, Passos,
  Cournapeau, Brucher, Perrot, and Duchesnay]{pedregosa_scikit-learn_2011}
Pedregosa,~F. \latin{et~al.}  Scikit-learn: {Machine} {Learning} in {Python}.
  \emph{Journal of Machine Learning Research} \textbf{2011}, \emph{12},
  2825--2830\relax
\mciteBstWouldAddEndPuncttrue
\mciteSetBstMidEndSepPunct{\mcitedefaultmidpunct}
{\mcitedefaultendpunct}{\mcitedefaultseppunct}\relax
\EndOfBibitem
\bibitem[Tanabe and Ishibuchi(2020)Tanabe, and Ishibuchi]{tanabe_analysis_2020}
Tanabe,~R.; Ishibuchi,~H. An {Analysis} of {Quality} {Indicators} {Using}
  {Approximated} {Optimal} {Distributions} in a 3-{D} {Objective} {Space}.
  \emph{IEEE Transactions on Evolutionary Computation} \textbf{2020},
  \emph{24}, 853--867\relax
\mciteBstWouldAddEndPuncttrue
\mciteSetBstMidEndSepPunct{\mcitedefaultmidpunct}
{\mcitedefaultendpunct}{\mcitedefaultseppunct}\relax
\EndOfBibitem
\bibitem[Bosman and Thierens(2003)Bosman, and Thierens]{bosman_balance_2003}
Bosman,~P.; Thierens,~D. The {Balance} {Between} {Proximity} and {Diversity} in
  {Multiobjective} {Evolutionary} {Algorithms}. \emph{IEEE Transactions on
  Evolutionary Computation} \textbf{2003}, \emph{7}, 174--188\relax
\mciteBstWouldAddEndPuncttrue
\mciteSetBstMidEndSepPunct{\mcitedefaultmidpunct}
{\mcitedefaultendpunct}{\mcitedefaultseppunct}\relax
\EndOfBibitem
\bibitem[Blank and Deb(2020)Blank, and Deb]{blank_pymoo_2020}
Blank,~J.; Deb,~K. Pymoo: {Multi}-{Objective} {Optimization} in {Python}.
  \emph{IEEE Access} \textbf{2020}, \emph{8}, 89497--89509\relax
\mciteBstWouldAddEndPuncttrue
\mciteSetBstMidEndSepPunct{\mcitedefaultmidpunct}
{\mcitedefaultendpunct}{\mcitedefaultseppunct}\relax
\EndOfBibitem
\bibitem[Reuther \latin{et~al.}(2018)Reuther, Kepner, Byun, Samsi, Arcand,
  Bestor, Bergeron, Gadepally, Houle, Hubbell, Jones, Klein, Milechin, Mullen,
  Prout, Rosa, Yee, and Michaleas]{reuther_interactive_2018}
Reuther,~A. \latin{et~al.}  Interactive {Supercomputing} on 40,000 {Cores} for
  {Machine} {Learning} and {Data} {Analysis}. 2018 {IEEE} {High} {Performance}
  extreme {Computing} {Conference} ({HPEC}). 2018; pp 1--6\relax
\mciteBstWouldAddEndPuncttrue
\mciteSetBstMidEndSepPunct{\mcitedefaultmidpunct}
{\mcitedefaultendpunct}{\mcitedefaultseppunct}\relax
\EndOfBibitem
\end{mcitethebibliography}


\providecommand{\latin}[1]{#1}
\makeatletter
\providecommand{\doi}
  {\begingroup\let\do\@makeother\dospecials
  \catcode`\{=1 \catcode`\}=2 \doi@aux}
\providecommand{\doi@aux}[1]{\endgroup\texttt{#1}}
\makeatother
\providecommand*\mcitethebibliography{\thebibliography}
\csname @ifundefined\endcsname{endmcitethebibliography}
  {\let\endmcitethebibliography\endthebibliography}{}
\begin{mcitethebibliography}{5}
\providecommand*\natexlab[1]{#1}
\providecommand*\mciteSetBstSublistMode[1]{}
\providecommand*\mciteSetBstMaxWidthForm[2]{}
\providecommand*\mciteBstWouldAddEndPuncttrue
  {\def\EndOfBibitem{\unskip.}}
\providecommand*\mciteBstWouldAddEndPunctfalse
  {\let\EndOfBibitem\relax}
\providecommand*\mciteSetBstMidEndSepPunct[3]{}
\providecommand*\mciteSetBstSublistLabelBeginEnd[3]{}
\providecommand*\EndOfBibitem{}
\mciteSetBstSublistMode{f}
\mciteSetBstMaxWidthForm{subitem}{(\alph{mcitesubitemcount})}
\mciteSetBstSublistLabelBeginEnd
  {\mcitemaxwidthsubitemform\space}
  {\relax}
  {\relax}

\bibitem[García-Ortegón \latin{et~al.}(2022)García-Ortegón, Simm, Tripp,
  Hernández-Lobato, Bender, and Bacallado]{garcia-ortegon_dockstring_2022}
García-Ortegón,~M.; Simm,~G. N.~C.; Tripp,~A.~J.; Hernández-Lobato,~J.~M.;
  Bender,~A.; Bacallado,~S. {DOCKSTRING}: {Easy} {Molecular} {Docking} {Yields}
  {Better} {Benchmarks} for {Ligand} {Design}. \emph{Journal of Chemical
  Information and Modeling} \textbf{2022}, \emph{62}, 3486--3502\relax
\mciteBstWouldAddEndPuncttrue
\mciteSetBstMidEndSepPunct{\mcitedefaultmidpunct}
{\mcitedefaultendpunct}{\mcitedefaultseppunct}\relax
\EndOfBibitem
\bibitem[McInnes \latin{et~al.}(2018)McInnes, Healy, Saul, and
  Großberger]{mcinnes_umap_2018}
McInnes,~L.; Healy,~J.; Saul,~N.; Großberger,~L. {UMAP}: {Uniform} {Manifold}
  {Approximation} and {Projection}. \emph{Journal of Open Source Software}
  \textbf{2018}, \emph{3}, 861\relax
\mciteBstWouldAddEndPuncttrue
\mciteSetBstMidEndSepPunct{\mcitedefaultmidpunct}
{\mcitedefaultendpunct}{\mcitedefaultseppunct}\relax
\EndOfBibitem
\bibitem[noa()]{noauthor_enamine_nodate}
Enamine {Screening} {Collections}.
  \url{https://enamine.net/compound-collections/screening-collection}\relax
\mciteBstWouldAddEndPuncttrue
\mciteSetBstMidEndSepPunct{\mcitedefaultmidpunct}
{\mcitedefaultendpunct}{\mcitedefaultseppunct}\relax
\EndOfBibitem
\bibitem[Adasme \latin{et~al.}(2021)Adasme, Linnemann, Bolz, Kaiser, Salentin,
  Haupt, and Schroeder]{adasme_plip_2021}
Adasme,~M.~F.; Linnemann,~K.~L.; Bolz,~S.~N.; Kaiser,~F.; Salentin,~S.;
  Haupt,~V.; Schroeder,~M. {PLIP} 2021: {Expanding} the {Scope} of the
  {Protein}–{Ligand} {Interaction} {Profiler} to {DNA} and {RNA}.
  \emph{Nucleic Acids Research} \textbf{2021}, \emph{49}, W530--W534\relax
\mciteBstWouldAddEndPuncttrue
\mciteSetBstMidEndSepPunct{\mcitedefaultmidpunct}
{\mcitedefaultendpunct}{\mcitedefaultseppunct}\relax
\EndOfBibitem
\end{mcitethebibliography}

\end{document}


\clearpage 

\section{Model-guided Optimization Algorithms}

\begin{algorithm}[h]
\DontPrintSemicolon
\caption{
    Multi-objective Bayesian optimization using scalarization
}
\label{alg:scalarized-bo}
    \KwIn{Objective function $\mathbf f : x \mapsto \mathbb R^N$,
        weight vector $\boldsymbol \lambda = [\lambda_1, \ldots, \lambda_N]$,
        surrogate model $\hat f$,
        candidate set $\mathcal{X}$,
        acquisition function $\alpha : x \mapsto \mathbb R$,
        initial observation size $b_0$,
        batch size $b$
    } 
    Select random subset of design space:
        $\mathcal X_0 \subset \mathcal X\: : \: |\mathcal{X}_0| = b_0 $ \;
    Initialize dataset:
        $\mathcal D_0 \gets \{\tuple{x}{ \boldsymbol \lambda \cdot \mathbf f(x)} : x \in \mathcal X_0 \}$ \;
    \For{$t \gets 1 \ldots T$}{
        Train $\hat f$ on $\mathcal D_{t-1}$ \;
        Select new batch:
            $\mathcal X_t \gets \argmax\limits_{\mathcal X_t \subset \mathcal X \: : \: |\mathcal X_t| = b}\ \sum\limits_{x \in \mathcal X_t} \alpha(x; \hat f, \mathcal D_{t-1})$ \;
        Update dataset: $\mathcal D_t \gets \mathcal D_{t-1} \cup \setbuilder{\tuple{x}{\boldsymbol{\lambda} \cdot \mathbf f(x)}}{x}{\mathcal X_t}$ \;
    }
    \KwOut{$\argmax\limits_{x \in \mathcal D_t} \boldsymbol \lambda \cdot \mathbf f(x)$}
\end{algorithm}

\begin{algorithm}[h]
\DontPrintSemicolon
\caption{
    Multi-objective Bayesian optimization using Pareto optimization
}
\label{alg:pareto-bo}
    \KwIn{Objective function $\mathbf f : x \mapsto \mathbb R^N$,
        surrogate models $\{\hat f^{(n)}\}_{n=1}^N$,
        candidate set $\mathcal{X}$,
        acquisition function $\alpha : x \mapsto \mathbb R$,
        initial observation size $b_0$,
        batch size $b$
    }
    Select random subset of design space:
        $ \mathcal X_0 \subset \mathcal X\: : \:  |\mathcal{X}_0| = b_0 $ \;
    Initialize dataset:
        $\mathcal D_0 \gets \setbuilder{\tuple{x}{\mathbf f(x)}}{x}{\mathcal X_0}$ \;
    Calculate Pareto front: $\mathcal P_0  \gets \mathtt{pareto\_front}(\mathcal D_0)$ \;
    \For{$t \gets 1 \ldots T$}{
        \For{$n \gets 1 \ldots N$}{
            Train $\hat f^{(n)}$ on
            $\setbuilder{\tuple{x}{f_n}}{x\,,\,\mathbf{f}(x)}{\mathcal D_{t-1}}$ \;
            }
        Select new batch:
            $\mathcal X_t \gets \argmax\limits_{\mathcal X_t \subset \mathcal X\: : \: |\mathcal X_t| = b}\ \sum\limits_{x \in \mathcal X_t} \alpha(x; \{\hat f^{(n)}\}, \mathcal D_{t-1}, \mathcal P_{t-1})$ \;
        Update dataset: $\mathcal D_t \gets \mathcal D_{t-1} \cup \setbuilder{\tuple{x}{\mathbf f(x)}}{x}{\mathcal X_t}$ \;
        Update Pareto front: $\mathcal P_t  \gets \mathtt{pareto\_front}(\mathcal D_t)$ \;
    }
    \KwOut{$\mathcal P_T$}
\end{algorithm}

\clearpage
\section{Performance Metrics}
\begin{figure}[H]
    \centering
    \includegraphics{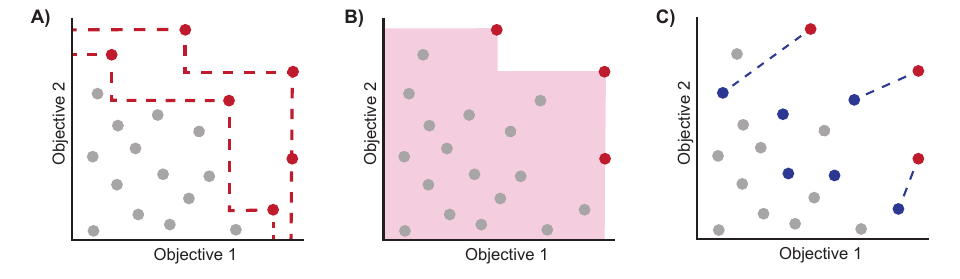}
    \caption{Illustration of the Pareto optimization evaluation metrics considered in this work. (A) Definition of true top-$k$ molecules through non-dominated sorting. Here, the top 30\% are shown in red. (B) Hypervolume metric. (C) Inverted generation distance, which averages the shortest distance between points on the true Pareto front (red) and acquired points (blue). }
    \label{fig:metrics}
\end{figure}

\clearpage

\section{Comparison of Acquisition Functions}

\begin{figure}[H]
    \centering
    \includegraphics{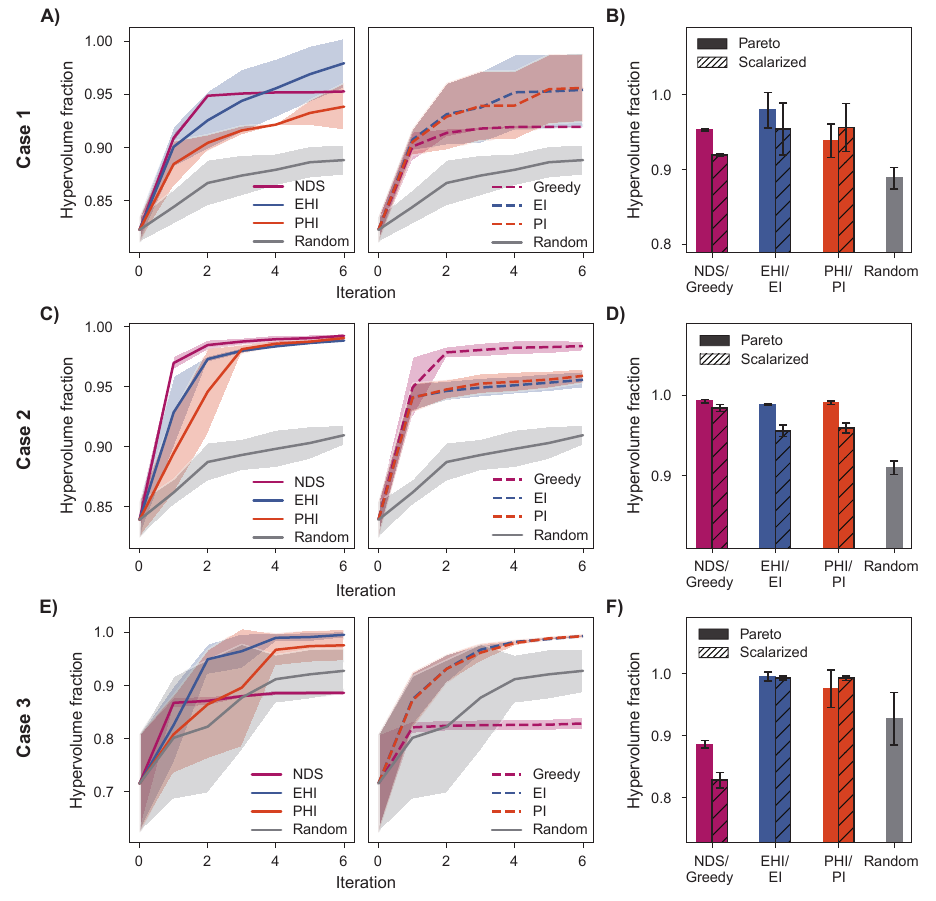}
    \caption{Fraction of hypervolume acquired across iterations (A, C, E) and after six iterations (B, D, F) for three case studies. \casedescriptions \iterationsentence \errorbarsentence{(B, D, F)}{(A, C, E)} }
    \label{fig:hv_si}
\end{figure}

\clearpage
\section{Molecular Diversity Visualization}

\begin{figure}[H]
    \centering
    \includegraphics{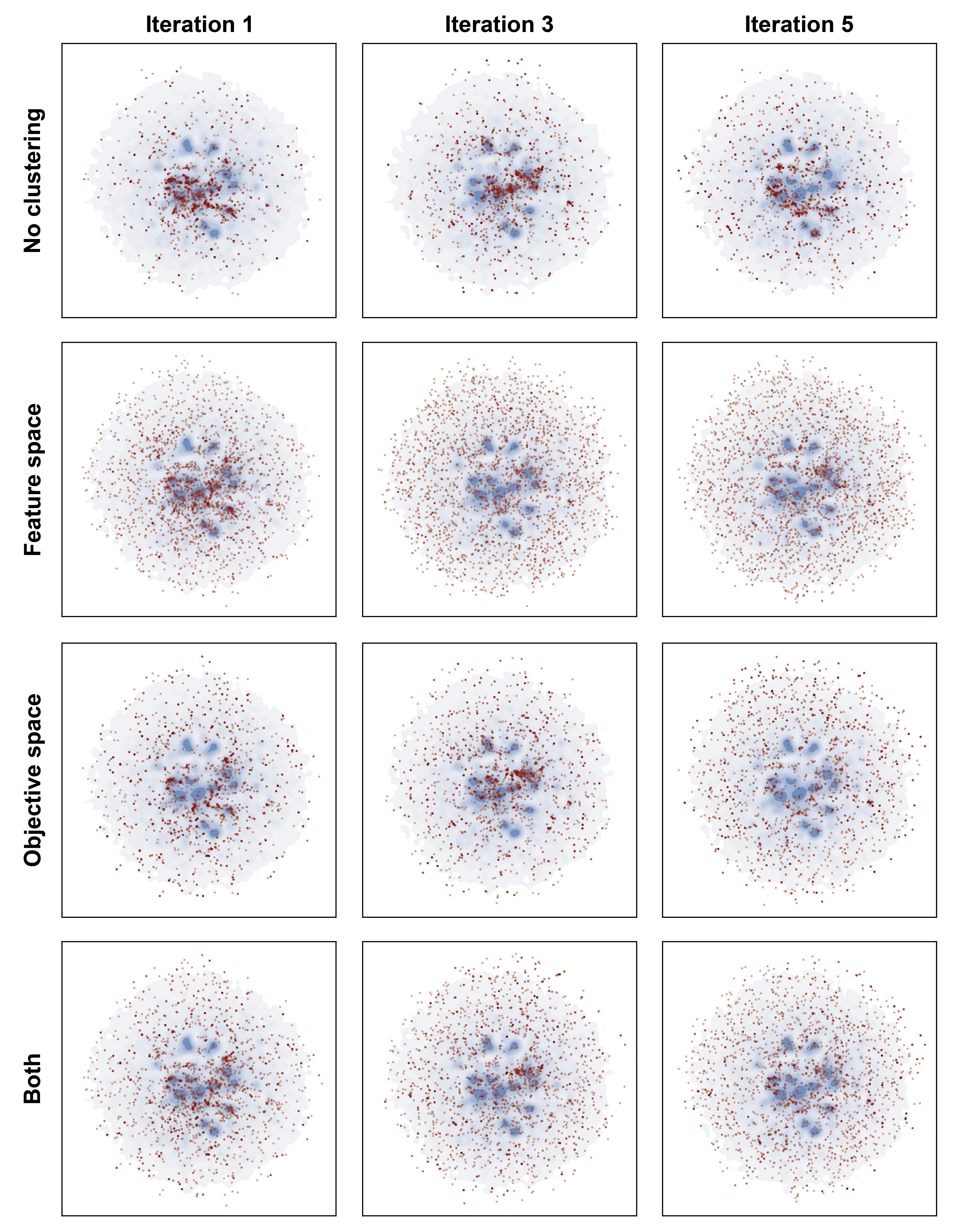}
    \caption{UMAP projections demonstrating molecular diversity of acquired points using standard and diversity-enhanced acquisition. Each row is a single run using the specified diversity-enhanced acquisition strategy with PHI. The runs corresponding to each row were initialized with the same random set of acquired points at iteration 0 and the same model seed. Docking scores computed with DOCKSTRING\cite{garcia-ortegon_dockstring_2022} to IGF1R and CYP3A4 were minimized and maximized, respectively, to identify putative selective inhibitors of IGF1R. Points acquired at iterations 1, 3, and 5 are shown for the four acquisition strategies tested in red. UMAP projections were trained on the entire virtual library (shown as a blue density behind acquired points). }
    \label{fig:umap}
\end{figure}

2-dimensional projections of molecular fingerprints can illustrate molecular diversity in a qualitative sense. We use UMAP projections \cite{mcinnes_umap_2018} to visualize the improvement in molecular diversity of acquired points with diversity-enhanced acquisition strategies. 
Figure \ref{fig:umap} shows UMAP projections of acquired points at iterations 1, 3, and 5 for single experiments using different diversity-enhancing acquisition strategies. These experiments were for the identification of putative IGF1R inhibitors with selectivity over CYP3A4. UMAP embeddings were trained on the entire searched library, shown as blue density plots. Diversity is compared for acquisition that implements no clustering, clustering in the feature space, clustering in the objective space, and clustering in both spaces. The acquired molecules (red points) are qualitatively more dispersed across the chemical space spanned by the library when compared to the points acquired without clustering. While the visualization of molecular diversity through dimensional reduction is qualitative in nature, the difference in the chemical space acquired in the two runs suggests that diversity-enhanced acquisition improves the structural diversity of acquired points.  

\clearpage

\section{Large-Scale Multi-Objective Screen Results}

\begin{figure}[H]
    \centering
    \includegraphics{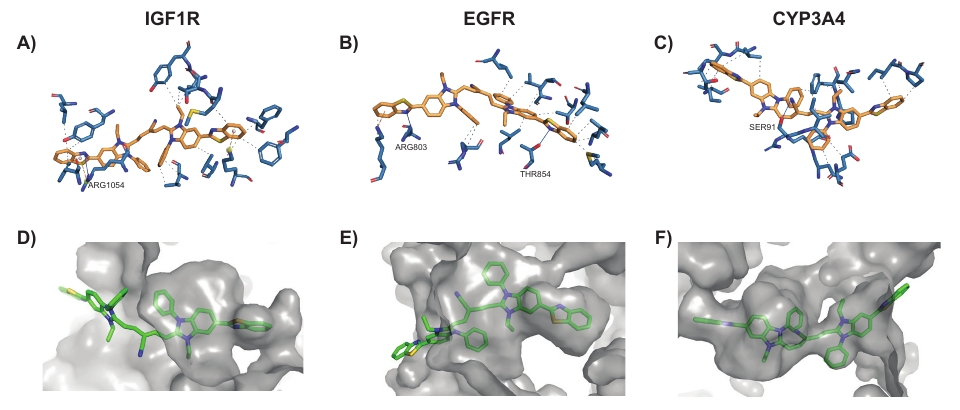}
    \caption{Docking poses of one non-dominated molecule (M2) predicted to selectively bind IGF1R and EGFR over CYP3A4. Docking poses were computed with DOCKSTRING \cite{garcia-ortegon_dockstring_2022}, an AutoDock Vina wrapper with prepared docking settings for IGF1R, EGFR, and CYP3A4. The Enamine screening database \cite{noauthor_enamine_nodate} of over 4M molecules was used as the virtual library. \textbf{(A-C)} Protein-ligand interactions of M2 with IGF1R, EGFR, and CYP3A4 prepared with PLIP \cite{adasme_plip_2021}. \textbf{(D-F)} Space-filling visualization of M5 in the binding pockets of IGF1R, EGFR, and CYP3A4. The docking-based optimization for predicted selectivity favors bulky molecules like M5 that can fit in the binding pockets of targets IGF1R and EGFR but form unresolvable steric clashes with residues that form the pocket of off-target CYP3A4. }
    \label{fig:poses}
\end{figure}

\begin{figure}[H]
    \centering
    \includegraphics[scale=0.9]{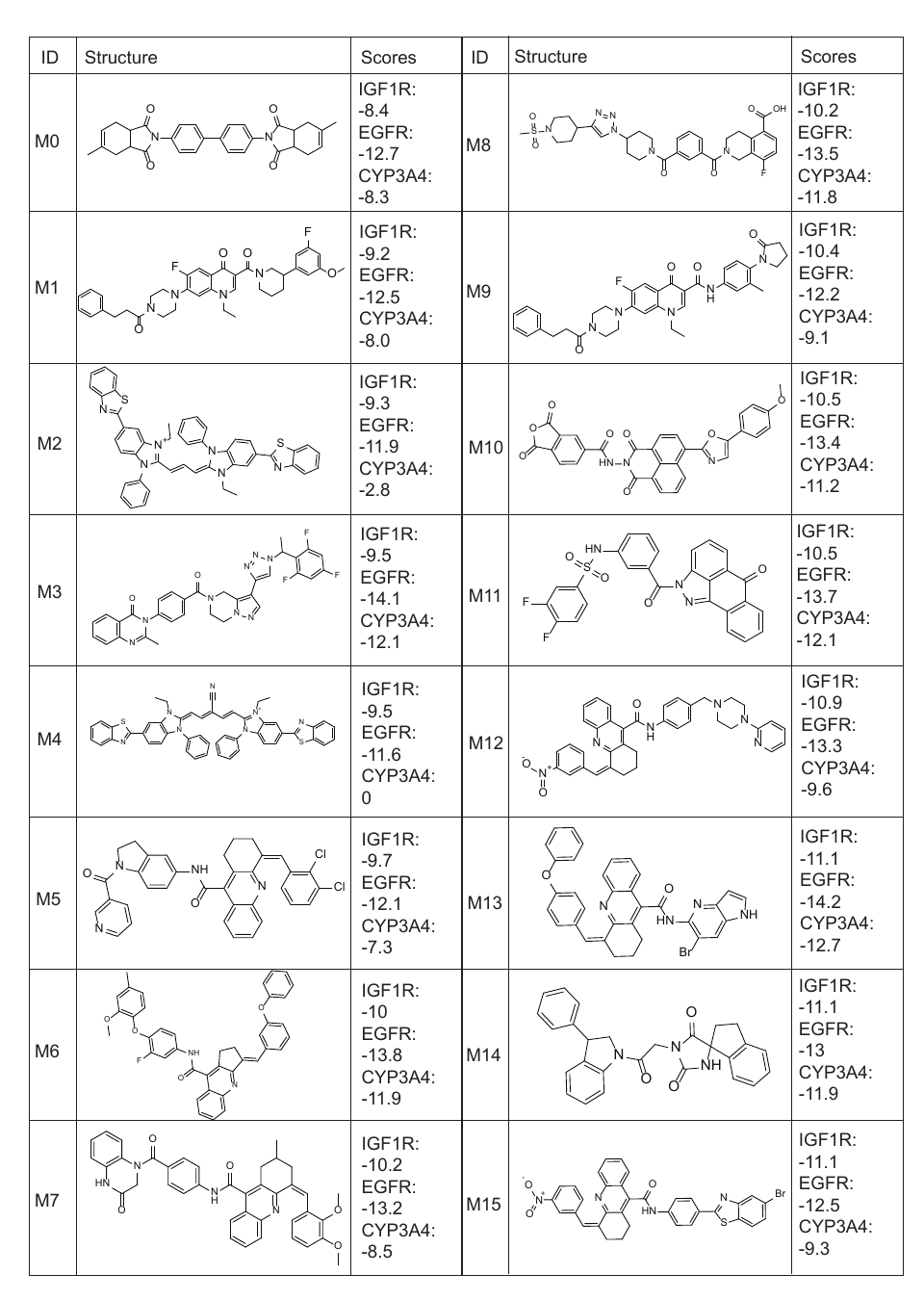}
        \vspace{-20pt}
    \caption{Molecules 0-15 of the 39 non-dominated molecules for an exemplary 3-objective optimization aiming to identify binders of IGF1R and EGFR with selectivity over CYP3A4 from the Enamine screening library of over 4M molecules \cite{noauthor_enamine_nodate}. Docking scores were computed using DOCKSTRING\cite{garcia-ortegon_dockstring_2022}. Docking scores to IGF1R and EGFR were minimized, and scores to the off-target CYP3A4 were maximized.  }
    \label{fig:pf_all_1}
\end{figure}

\begin{figure}[H]
    \centering
    \includegraphics[scale=0.90]{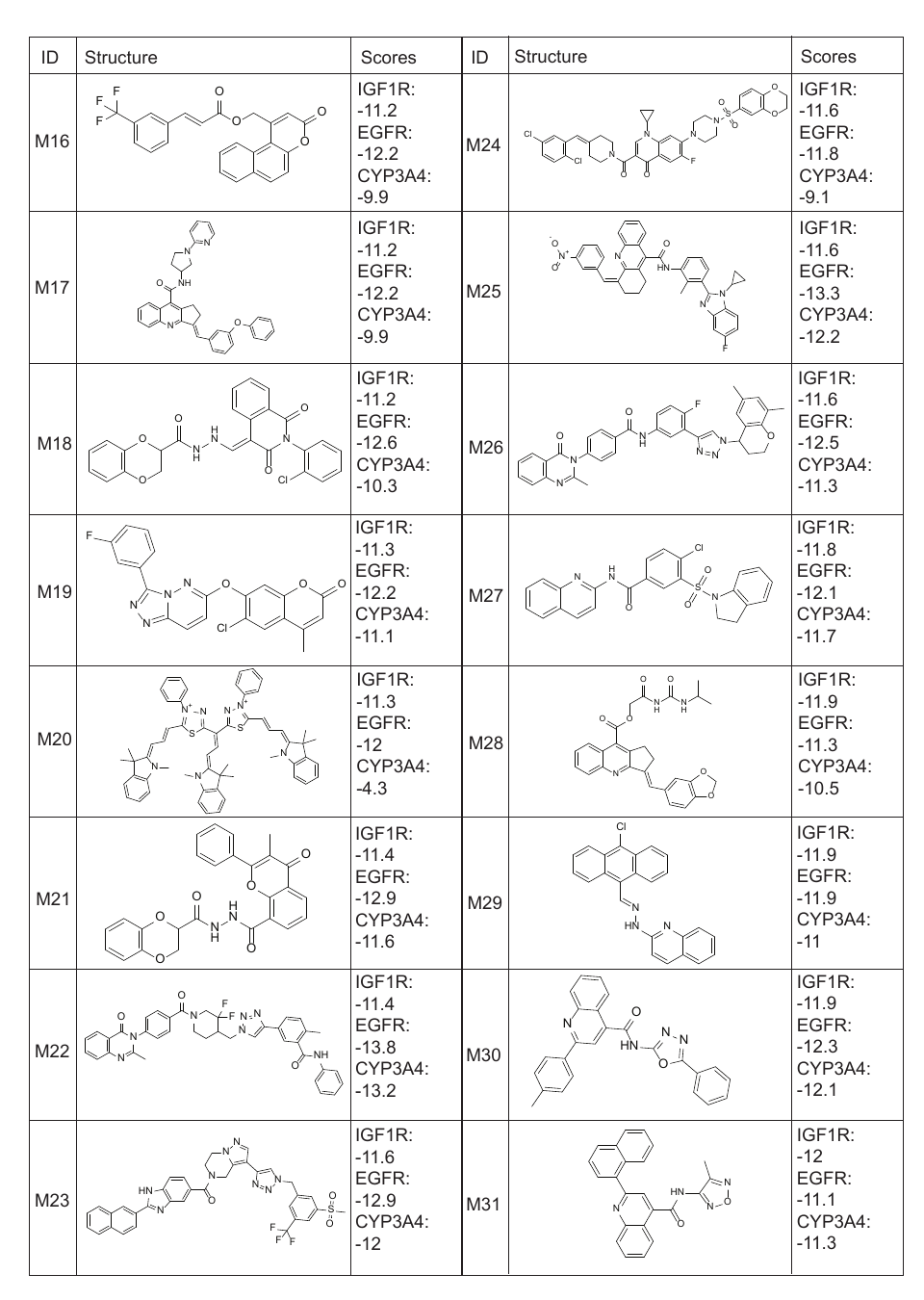}
        \vspace{-20pt}
    \caption{Molecules 16-31 of the 39 non-dominated molecules in the searched library for an exemplary 3-objective optimization aiming to identify binders of IGF1R and EGFR with selectivity over CYP3A4 from the Enamine screening library of over 4M molecules \cite{noauthor_enamine_nodate}. Docking scores were computed using DOCKSTRING\cite{garcia-ortegon_dockstring_2022}. Docking scores to IGF1R and EGFR were minimized, and scores to the off-target CYP3A4 were maximized. }
    \label{fig:pf_all_2}
\end{figure}

\begin{figure}[H]
    \centering
    \includegraphics[scale=0.90]{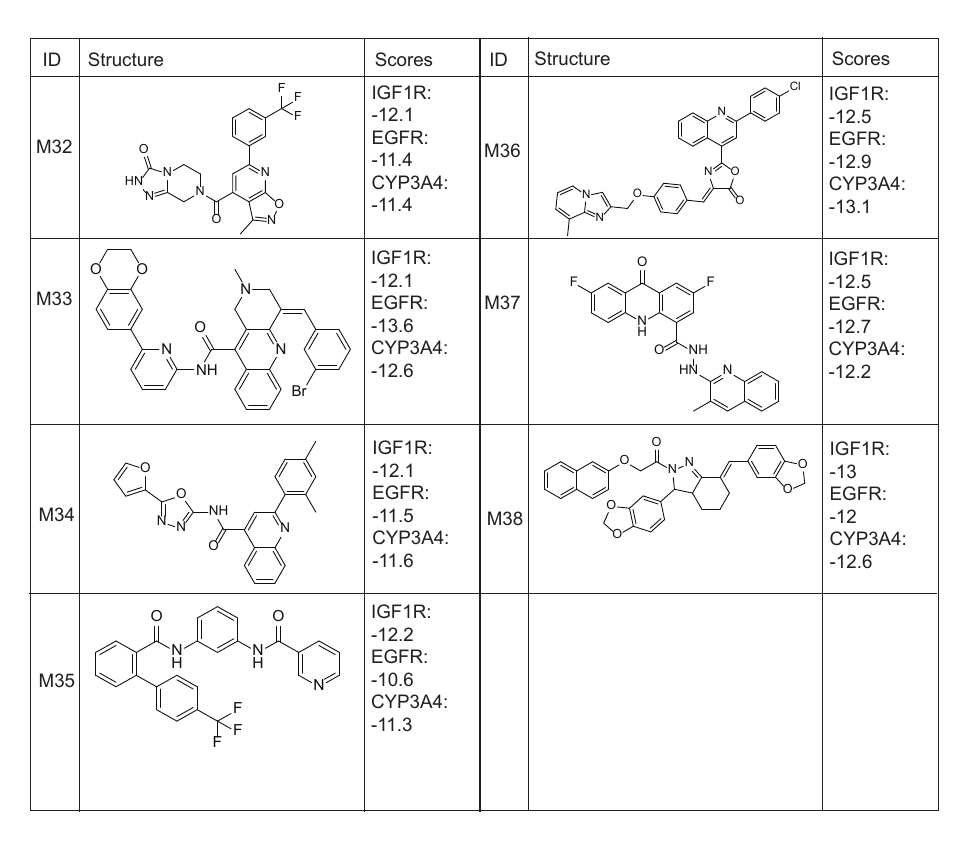}
    \vspace{-20pt}
    \caption{Molecules 32-38 of the 39 non-dominated molecules in the searched library for an exemplary 3-objective optimization aiming to identify binders of IGF1R and EGFR with selectivity over CYP3A4 from the Enamine screening library of over 4M molecules \cite{noauthor_enamine_nodate}. Docking scores were computed using DOCKSTRING\cite{garcia-ortegon_dockstring_2022}. Docking scores to IGF1R and EGFR were minimized, and scores to the off-target CYP3A4 were maximized.  }
    \label{fig:pf_all_3}
\end{figure}

\clearpage
\section{Tables}

\begin{table}[H]
    \hspace*{-1.5cm} 
    \centering
    \footnotesize	    
    \begin{tabular}{rcccccccc}
    \toprule
     Case & Acquisition &  \multicolumn{7}{c}{Iteration}       \\
     & Function & 0 &              1 &              2 &              3 &              4 &              5 &              6 \\
    \midrule  
1 & EHI & $0.01 \pm 0.00$ & $0.15 \pm 0.01$ & $0.21 \pm 0.03$ & $0.25 \pm 0.03$ & $0.30 \pm 0.02$ & $0.33 \pm 0.02$ & $0.36 \pm 0.02$ \\
1 & EI & $0.01 \pm 0.00$ & $0.14 \pm 0.02$ & $0.17 \pm 0.02$ & $0.20 \pm 0.02$ & $0.23 \pm 0.02$ & $0.25 \pm 0.02$ & $0.27 \pm 0.02$ \\
1 & Greedy & $0.01 \pm 0.00$ & $0.17 \pm 0.01$ & $0.25 \pm 0.01$ & $0.31 \pm 0.02$ & $0.36 \pm 0.01$ & $0.41 \pm 0.01$ & $0.45 \pm 0.01$ \\
1 & NDS & $0.01 \pm 0.00$ & $0.16 \pm 0.02$ & $0.25 \pm 0.01$ & $0.30 \pm 0.01$ & $0.34 \pm 0.01$ & $0.37 \pm 0.01$ & $0.40 \pm 0.01$ \\
1 & PHI & $0.01 \pm 0.00$ & $0.17 \pm 0.01$ & $0.25 \pm 0.02$ & $0.31 \pm 0.02$ & $0.36 \pm 0.02$ & $0.40 \pm 0.01$ & $0.43 \pm 0.01$ \\
1 & PI & $0.01 \pm 0.00$ & $0.14 \pm 0.02$ & $0.18 \pm 0.02$ & $0.21 \pm 0.01$ & $0.24 \pm 0.02$ & $0.27 \pm 0.02$ & $0.29 \pm 0.02$ \\
1 & Random & $0.01 \pm 0.00$ & $0.02 \pm 0.00$ & $0.03 \pm 0.00$ & $0.04 \pm 0.00$ & $0.05 \pm 0.01$ & $0.06 \pm 0.01$ & $0.07 \pm 0.01$ \\
\midrule
2 & EHI & $0.01 \pm 0.00$ & $0.13 \pm 0.04$ & $0.15 \pm 0.04$ & $0.17 \pm 0.03$ & $0.20 \pm 0.03$ & $0.23 \pm 0.04$ & $0.25 \pm 0.04$ \\
2 & EI & $0.01 \pm 0.00$ & $0.03 \pm 0.00$ & $0.03 \pm 0.00$ & $0.04 \pm 0.00$ & $0.05 \pm 0.00$ & $0.07 \pm 0.00$ & $0.08 \pm 0.00$ \\
2 & Greedy & $0.01 \pm 0.00$ & $0.09 \pm 0.03$ & $0.13 \pm 0.03$ & $0.16 \pm 0.04$ & $0.20 \pm 0.04$ & $0.22 \pm 0.04$ & $0.25 \pm 0.04$ \\
2 & NDS & $0.01 \pm 0.00$ & $0.13 \pm 0.01$ & $0.21 \pm 0.01$ & $0.26 \pm 0.01$ & $0.29 \pm 0.01$ & $0.32 \pm 0.01$ & $0.34 \pm 0.01$ \\
2 & PHI & $0.01 \pm 0.00$ & $0.18 \pm 0.02$ & $0.21 \pm 0.02$ & $0.25 \pm 0.02$ & $0.28 \pm 0.01$ & $0.31 \pm 0.00$ & $0.35 \pm 0.02$ \\
2 & PI & $0.01 \pm 0.00$ & $0.03 \pm 0.00$ & $0.04 \pm 0.00$ & $0.05 \pm 0.00$ & $0.06 \pm 0.00$ & $0.07 \pm 0.01$ & $0.08 \pm 0.00$ \\
2 & Random & $0.01 \pm 0.00$ & $0.02 \pm 0.00$ & $0.03 \pm 0.00$ & $0.04 \pm 0.00$ & $0.05 \pm 0.00$ & $0.06 \pm 0.00$ & $0.07 \pm 0.00$ \\
\midrule
3 & EHI & $0.01 \pm 0.00$ & $0.04 \pm 0.01$ & $0.06 \pm 0.01$ & $0.11 \pm 0.01$ & $0.17 \pm 0.02$ & $0.23 \pm 0.01$ & $0.26 \pm 0.01$ \\
3 & EI & $0.01 \pm 0.00$ & $0.03 \pm 0.01$ & $0.04 \pm 0.01$ & $0.05 \pm 0.01$ & $0.06 \pm 0.01$ & $0.08 \pm 0.01$ & $0.09 \pm 0.01$ \\
3 & Greedy & $0.01 \pm 0.00$ & $0.03 \pm 0.00$ & $0.06 \pm 0.00$ & $0.08 \pm 0.00$ & $0.11 \pm 0.01$ & $0.13 \pm 0.01$ & $0.15 \pm 0.01$ \\
3 & NDS & $0.01 \pm 0.00$ & $0.05 \pm 0.01$ & $0.08 \pm 0.00$ & $0.12 \pm 0.01$ & $0.15 \pm 0.01$ & $0.17 \pm 0.01$ & $0.19 \pm 0.01$ \\
3 & PHI & $0.01 \pm 0.00$ & $0.03 \pm 0.01$ & $0.07 \pm 0.02$ & $0.11 \pm 0.03$ & $0.17 \pm 0.04$ & $0.24 \pm 0.04$ & $0.28 \pm 0.03$ \\
3 & PI & $0.01 \pm 0.00$ & $0.03 \pm 0.01$ & $0.04 \pm 0.01$ & $0.05 \pm 0.01$ & $0.06 \pm 0.01$ & $0.07 \pm 0.00$ & $0.09 \pm 0.00$ \\
3 & Random & $0.01 \pm 0.00$ & $0.02 \pm 0.00$ & $0.03 \pm 0.00$ & $0.04 \pm 0.00$ & $0.05 \pm 0.00$ & $0.06 \pm 0.00$ & $0.07 \pm 0.00$ \\
\bottomrule
\end{tabular}
    \caption{Fraction of top-1\% acquired across iterations for retrospective multi-objective virtual screening experiments in Section \ref{M-section:acquisition_funcs}, shown in Figure \ref{M-fig:top-k-profiles}. \casedescriptions \iterationsentence \pmsentence{5} }
    \label{tab:top-k-profile}
    \hspace*{-1.5cm}
\end{table}

\begin{table}[H]
    \hspace*{-1.5cm} 
    \centering
    \footnotesize	    
    \begin{tabular}{rcccccccc}
    \toprule
     Case & Acquisition &  \multicolumn{7}{c}{Iteration}       \\
     & Function & 0 &              1 &              2 &              3 &              4 &              5 &              6 \\
    \midrule
1 & EHI & $0.82 \pm 0.01$ & $0.90 \pm 0.02$ & $0.93 \pm 0.03$ & $0.94 \pm 0.03$ & $0.96 \pm 0.03$ & $0.97 \pm 0.03$ & $0.98 \pm 0.02$ \\
1 & EI & $0.82 \pm 0.01$ & $0.91 \pm 0.01$ & $0.93 \pm 0.03$ & $0.94 \pm 0.03$ & $0.95 \pm 0.04$ & $0.95 \pm 0.04$ & $0.95 \pm 0.03$ \\
1 & Greedy & $0.82 \pm 0.01$ & $0.90 \pm 0.01$ & $0.91 \pm 0.00$ & $0.92 \pm 0.00$ & $0.92 \pm 0.00$ & $0.92 \pm 0.00$ & $0.92 \pm 0.00$ \\
1 & NDS & $0.82 \pm 0.01$ & $0.91 \pm 0.00$ & $0.95 \pm 0.00$ & $0.95 \pm 0.00$ & $0.95 \pm 0.00$ & $0.95 \pm 0.00$ & $0.95 \pm 0.00$ \\
1 & PHI & $0.82 \pm 0.01$ & $0.88 \pm 0.02$ & $0.90 \pm 0.01$ & $0.92 \pm 0.00$ & $0.92 \pm 0.00$ & $0.93 \pm 0.01$ & $0.94 \pm 0.02$ \\
1 & PI & $0.82 \pm 0.01$ & $0.91 \pm 0.01$ & $0.93 \pm 0.03$ & $0.94 \pm 0.03$ & $0.94 \pm 0.03$ & $0.95 \pm 0.03$ & $0.96 \pm 0.03$ \\
1 & Random & $0.82 \pm 0.01$ & $0.84 \pm 0.02$ & $0.87 \pm 0.02$ & $0.87 \pm 0.02$ & $0.88 \pm 0.02$ & $0.89 \pm 0.02$ & $0.89 \pm 0.01$ \\
\midrule
2 & EHI & $0.84 \pm 0.01$ & $0.93 \pm 0.03$ & $0.97 \pm 0.00$ & $0.98 \pm 0.00$ & $0.98 \pm 0.00$ & $0.99 \pm 0.00$ & $0.99 \pm 0.00$ \\
2 & EI & $0.84 \pm 0.01$ & $0.94 \pm 0.01$ & $0.95 \pm 0.01$ & $0.95 \pm 0.01$ & $0.95 \pm 0.01$ & $0.95 \pm 0.01$ & $0.96 \pm 0.01$ \\
2 & Greedy & $0.84 \pm 0.01$ & $0.95 \pm 0.03$ & $0.98 \pm 0.00$ & $0.98 \pm 0.01$ & $0.98 \pm 0.01$ & $0.98 \pm 0.01$ & $0.98 \pm 0.00$ \\
2 & NDS & $0.84 \pm 0.01$ & $0.97 \pm 0.01$ & $0.98 \pm 0.00$ & $0.99 \pm 0.00$ & $0.99 \pm 0.00$ & $0.99 \pm 0.00$ & $0.99 \pm 0.00$ \\
2 & PHI & $0.84 \pm 0.01$ & $0.89 \pm 0.03$ & $0.95 \pm 0.04$ & $0.98 \pm 0.00$ & $0.99 \pm 0.00$ & $0.99 \pm 0.00$ & $0.99 \pm 0.00$ \\
2 & PI & $0.84 \pm 0.01$ & $0.94 \pm 0.01$ & $0.95 \pm 0.01$ & $0.95 \pm 0.01$ & $0.95 \pm 0.01$ & $0.96 \pm 0.01$ & $0.96 \pm 0.01$ \\
2 & Random & $0.84 \pm 0.01$ & $0.86 \pm 0.01$ & $0.89 \pm 0.02$ & $0.89 \pm 0.01$ & $0.90 \pm 0.02$ & $0.90 \pm 0.01$ & $0.91 \pm 0.01$ \\
\midrule
3 & EHI & $0.72 \pm 0.09$ & $0.83 \pm 0.07$ & $0.95 \pm 0.03$ & $0.96 \pm 0.03$ & $0.99 \pm 0.01$ & $0.99 \pm 0.01$ & $1.00 \pm 0.01$ \\
3 & EI & $0.72 \pm 0.09$ & $0.87 \pm 0.05$ & $0.93 \pm 0.03$ & $0.97 \pm 0.01$ & $0.98 \pm 0.00$ & $0.99 \pm 0.00$ & $0.99 \pm 0.00$ \\
3 & Greedy & $0.72 \pm 0.09$ & $0.82 \pm 0.01$ & $0.82 \pm 0.01$ & $0.83 \pm 0.01$ & $0.83 \pm 0.01$ & $0.83 \pm 0.01$ & $0.83 \pm 0.01$ \\
3 & NDS & $0.72 \pm 0.09$ & $0.87 \pm 0.01$ & $0.87 \pm 0.01$ & $0.88 \pm 0.00$ & $0.89 \pm 0.01$ & $0.89 \pm 0.01$ & $0.89 \pm 0.01$ \\
3 & PHI & $0.72 \pm 0.09$ & $0.81 \pm 0.07$ & $0.86 \pm 0.10$ & $0.90 \pm 0.11$ & $0.97 \pm 0.03$ & $0.97 \pm 0.03$ & $0.98 \pm 0.03$ \\
3 & PI & $0.72 \pm 0.09$ & $0.87 \pm 0.05$ & $0.93 \pm 0.03$ & $0.96 \pm 0.02$ & $0.98 \pm 0.01$ & $0.99 \pm 0.00$ & $0.99 \pm 0.00$ \\
3 & Random & $0.72 \pm 0.09$ & $0.80 \pm 0.12$ & $0.82 \pm 0.13$ & $0.88 \pm 0.10$ & $0.91 \pm 0.05$ & $0.92 \pm 0.05$ & $0.93 \pm 0.04$ \\
\bottomrule
\end{tabular}
    \caption{Hypervolume profiles for retrospective multi-objective virtual screening experiments in Section \ref{M-section:acquisition_funcs}, shown in Figure \ref{fig:hv_si}. \casedescriptions \iterationsentence \pmsentence{5}}
    \label{tab:hv-profiles}
    \hspace*{-1.5cm}
\end{table}

\begin{table}[H]
    \centering
    \hspace*{-1.5cm} 
    \begin{tabular}{rccccc}
    \toprule
     Case & Acquisition Function &          Top 1\% &               HV &              IGD & Fraction of True Front \\
     \midrule
1 & EHI & $0.36 \pm 0.02$ & $0.98 \pm 0.02$ & $0.11 \pm 0.06$ & $0.68 \pm 0.05$ \\
1 & EI & $0.27 \pm 0.02$ & $0.95 \pm 0.03$ & $0.22 \pm 0.12$ & $0.52 \pm 0.06$ \\
1 & Greedy & $0.45 \pm 0.01$ & $0.92 \pm 0.00$ & $0.33 \pm 0.01$ & $0.51 \pm 0.02$ \\
1 & NDS & $0.40 \pm 0.01$ & $0.95 \pm 0.00$ & $0.20 \pm 0.01$ & $0.49 \pm 0.04$ \\
1 & PHI & $0.43 \pm 0.01$ & $0.94 \pm 0.02$ & $0.23 \pm 0.06$ & $0.61 \pm 0.03$ \\
1 & PI & $0.29 \pm 0.02$ & $0.96 \pm 0.03$ & $0.21 \pm 0.11$ & $0.53 \pm 0.07$ \\
1 & Random & $0.07 \pm 0.01$ & $0.89 \pm 0.01$ & $0.62 \pm 0.04$ & $0.09 \pm 0.04$ \\
\midrule
2 & EHI & $0.25 \pm 0.04$ & $0.99 \pm 0.00$ & $0.10 \pm 0.01$ & $0.41 \pm 0.03$ \\
2 & EI & $0.08 \pm 0.00$ & $0.96 \pm 0.01$ & $0.28 \pm 0.03$ & $0.20 \pm 0.05$ \\
2 & Greedy & $0.25 \pm 0.04$ & $0.98 \pm 0.00$ & $0.10 \pm 0.02$ & $0.38 \pm 0.05$ \\
2 & NDS & $0.34 \pm 0.01$ & $0.99 \pm 0.00$ & $0.07 \pm 0.02$ & $0.48 \pm 0.05$ \\
2 & PHI & $0.35 \pm 0.02$ & $0.99 \pm 0.00$ & $0.08 \pm 0.01$ & $0.47 \pm 0.03$ \\
2 & PI & $0.08 \pm 0.00$ & $0.96 \pm 0.01$ & $0.28 \pm 0.01$ & $0.21 \pm 0.03$ \\
2 & Random & $0.07 \pm 0.00$ & $0.91 \pm 0.01$ & $0.31 \pm 0.03$ & $0.03 \pm 0.03$ \\
\midrule
3 & EHI & $0.26 \pm 0.01$ & $1.00 \pm 0.01$ & $0.07 \pm 0.07$ & $0.82 \pm 0.10$ \\
3 & EI & $0.09 \pm 0.01$ & $0.99 \pm 0.00$ & $0.18 \pm 0.07$ & $0.62 \pm 0.10$ \\
3 & Greedy & $0.15 \pm 0.01$ & $0.83 \pm 0.01$ & $1.25 \pm 0.09$ & $0.02 \pm 0.04$ \\
3 & NDS & $0.19 \pm 0.01$ & $0.89 \pm 0.01$ & $0.74 \pm 0.03$ & $0.24 \pm 0.08$ \\
3 & PHI & $0.28 \pm 0.03$ & $0.98 \pm 0.03$ & $0.19 \pm 0.15$ & $0.72 \pm 0.15$ \\
3 & PI & $0.09 \pm 0.00$ & $0.99 \pm 0.00$ & $0.21 \pm 0.08$ & $0.60 \pm 0.11$ \\
3 & Random & $0.07 \pm 0.00$ & $0.93 \pm 0.04$ & $0.76 \pm 0.22$ & $0.16 \pm 0.12$ \\
    \bottomrule
    \end{tabular}
    \hspace*{-1.5cm} 
    \caption{Comparison of acquisition functions using all four evaluation metrics after a fixed exploration budget of 6 iterations. Top-1\%, hypervolume (HV), inverted generational distance (IGD), and fraction of the true Pareto front are shown. Values are plotted in Figures \ref{M-fig:top-k-profiles}, \ref{M-fig:nd-points}, and \ref{fig:hv_si}. \casedescriptions \iterationsentence \pmsentence{5} }
    \label{tab:end_means_table}
\end{table}

\begin{table}[H]
\hspace*{-1cm} 
\centering
\footnotesize	
\begin{tabular}{cccccccc}
\toprule
\multicolumn{8}{c}{\textbf{Top-1\%}} \\ \toprule
Cluster & \multicolumn{7}{c}{Iteration} \\
 Type & 0 & 1 & 2 & 3 & 4 & 5 & 6 \\
\midrule
Feature & $0.01 \pm 0.00$ & $0.04 \pm 0.01$ & $0.08 \pm 0.02$ & $0.12 \pm 0.02$ & $0.16 \pm 0.02$ & $0.19 \pm 0.02$ & $0.23 \pm 0.02$ \\
Feature + Obj & $0.01 \pm 0.00$ & $0.04 \pm 0.01$ & $0.07 \pm 0.01$ & $0.12 \pm 0.02$ & $0.16 \pm 0.02$ & $0.20 \pm 0.02$ & $0.24 \pm 0.02$ \\
No clustering & $0.01 \pm 0.00$ & $0.03 \pm 0.01$ & $0.07 \pm 0.02$ & $0.11 \pm 0.03$ & $0.17 \pm 0.04$ & $0.24 \pm 0.04$ & $0.28 \pm 0.03$ \\
Obj & $0.01 \pm 0.00$ & $0.03 \pm 0.00$ & $0.05 \pm 0.01$ & $0.09 \pm 0.01$ & $0.13 \pm 0.01$ & $0.17 \pm 0.01$ & $0.20 \pm 0.01$ \\
\toprule
\multicolumn{8}{c}{\textbf{Hypervolume}} \\ \toprule 
Cluster & \multicolumn{7}{c}{Iteration} \\
 Type & 0 & 1 & 2 & 3 & 4 & 5 & 6 \\
\midrule
Feature & $0.72 \pm 0.09$ & $0.86 \pm 0.07$ & $0.97 \pm 0.01$ & $0.97 \pm 0.01$ & $0.98 \pm 0.01$ & $0.98 \pm 0.01$ & $0.99 \pm 0.01$ \\
Feature + Obj & $0.72 \pm 0.09$ & $0.87 \pm 0.04$ & $0.93 \pm 0.04$ & $0.96 \pm 0.04$ & $0.98 \pm 0.02$ & $1.00 \pm 0.00$ & $1.00 \pm 0.00$ \\
No clustering & $0.72 \pm 0.09$ & $0.81 \pm 0.07$ & $0.86 \pm 0.10$ & $0.90 \pm 0.11$ & $0.97 \pm 0.03$ & $0.97 \pm 0.03$ & $0.98 \pm 0.03$ \\
Obj & $0.72 \pm 0.09$ & $0.85 \pm 0.06$ & $0.94 \pm 0.03$ & $0.96 \pm 0.04$ & $0.98 \pm 0.01$ & $0.99 \pm 0.01$ & $0.99 \pm 0.01$ \\
\bottomrule
\end{tabular}
\hspace*{-1.5cm} 
\caption{Top-1\% and hypervolume profiles for Case 3 experiments comparing diversity-enhancing acquisition strategies (Section \ref{M-section:cluster_results}, Figure \ref{M-fig:cluster_metrics}). All runs used PHI for acquisition. Results are shown for top-$k$ batching without clustering and three diversity-enhanced acquisition strategies that apply clustering. Docking scores to IGF1R and CYP3A4 were minimized and maximized, respectively. The virtual library and docking scores were use as published in DOCKSTRING \cite{garcia-ortegon_dockstring_2022}. \iterationsentence \pmsentence{5}}
\label{tab:cluster_topk_hv_profiles}
\end{table}

\begin{table}[H]
\centering
\begin{tabular}{cccccr}
\toprule
\multirow{2}{*}{\parbox{2cm}{\centering Cluster Type}}& \multirow{2}{*}{\parbox{2cm}{\centering Top 1\%}}&
\multirow{2}{*}{\parbox{3cm}{\centering Hypervolume}}&
\multirow{2}{*}{\parbox{2cm}{\centering IGD}}&
\multirow{2}{*}{\parbox{2cm}{\centering Fraction of True Front}}&
\multirow{2}{*}{\parbox{2cm}{\centering Number of Scaffolds}}\\ \\ %
\midrule
Feature & $0.23 \pm 0.02$ & $0.99 \pm 0.01$ & $0.26 \pm 0.08$ & $0.58 \pm 0.07$ & $10605 \pm 146$ \\
Feature + Obj & $0.24 \pm 0.02$ & $1.00 \pm 0.00$ & $0.09 \pm 0.05$ & $0.84 \pm 0.05$ & $9851 \pm \phantom{0}49$ \\
No clustering & $0.28 \pm 0.03$ & $0.98 \pm 0.03$ & $0.19 \pm 0.15$ & $0.72 \pm 0.15$ & $7946 \pm 186$ \\
Obj & $0.20 \pm 0.01$ & $0.99 \pm 0.01$ & $0.25 \pm 0.08$ & $0.64 \pm 0.16$ & $9036 \pm 198$ \\
\bottomrule
\end{tabular}
\caption{Comparison of diversity-enhancing acquisition strategies using all four evaluation metrics after a fixed exploration budget of 6 iterations (Section \ref{M-section:cluster_results}, Figure \ref{M-fig:cluster_metrics}). All runs used PHI for acquisition. Results are shown for top-$k$ batching without clustering and three diversity-enhanced acquisition strategies that apply clustering. Docking scores to IGF1R and CYP3A4 were minimized and maximized, respectively. The virtual library and docking scores were use as published in DOCKSTRING \cite{garcia-ortegon_dockstring_2022}. \iterationsentence \pmsentence{5} }
\label{tab:cluster_end_metrics}
\end{table}

\clearpage

\bibliography{references.bib}